\documentclass[useAMS,usenatbib]{mn2e}

\usepackage{graphicx}
\usepackage{subfigure}
\usepackage{paralist}
\usepackage{amssymb}
\usepackage{bm}
\usepackage{bbding}
\usepackage[fleqn]{amsmath}
\usepackage{multirow}
\usepackage[colorlinks,
            linkcolor=blue, 
            anchorcolor=blue, 
            citecolor=blue, 
            ]{hyperref}
\title[Dynamical structures]{Dynamical structures of retrograde resonances: analytical and numerical studies}
\author[Lei \& Li]{Hanlun Lei$^{1,2}$\thanks{E-mail: leihl@nju.edu.cn}, Jian Li$^{1,2}$\thanks{E-mail: ljian@nju.edu.cn}\\
$^{1}$ School of Astronomy and Space Science, Nanjing University, Nanjing 210023, China\\
$^{2}$ Key Laboratory of Modern Astronomy and Astrophysics in Ministry of Education, Nanjing University, Nanjing 210023, China}

\begin{document}

\date{Accepted. Received; in original form}

\pagerange{\pageref{firstpage}--\pageref{lastpage}} \pubyear{2021}

\maketitle
\label{firstpage}

\begin{abstract}

In this work, retrograde mean motion resonances (MMRs) are investigated by means of analytical and numerical approaches. Initially, we define a new resonant angle to describe the retrograde MMRs and then perform a series of canonical transformations to formulate the resonant model, in which the phase portrait, resonant centre and resonant width can be analytically determined. To validate the analytical developments, the non-perturbative analysis is made by taking advantage of Poincar\'e surfaces of section. Some modifications are introduced in the production of Poincar\'e sections and, in particular, it becomes possible to make direct comparisons between the analytical and numerical results. It is found that there exists an excellent correspondence between the phase portraits and the associated Poincar\'e sections, and the analytical results agree well with the numerical results in terms of the resonant width and the location of resonant centre. Finally, the numerical approach is utilized to determine the resonant widths and resonant centres over the full range of eccentricity. In particular, seven known examples of retrograde asteroids including 2015 BZ509, 2008 SO218, 1999 LE31, 2000 DG8, 2014 AT28, 2016 LS and 2016 JK24 are found inside the libration zones of retrograde MMRs with Jupiter. The results obtained in this work may be helpful for understanding the dynamical evolution for asteroids inside retrograde MMRs.

\end{abstract}


\section{Introduction}
\label{Sect1}

In recent years, a growing number of minor objects have been discovered on retrograde orbits. In particular, among the 1048123 asteroids discovered so far, there are 112 retrograde asteroids in our Solar system\footnote{https://minorplanetcenter.net//iau/MPCORB.html, retrieved 3 February 2020}. \citet{gallardo2019orbital} confirmed that, all along the Solar system, there is a stability stripe around inclination of $\sim$$150^{\circ}$, where the planetary perturbations produce smaller dynamical effects. A remark is that the known examples of retrograde asteroids are concentrated around such an inclination $i$$\sim$$150^{\circ}$ \citep{gallardo2019orbital}. From the viewpoint of stability, the configuration of mean motion resonance (MMR) can provide a protection mechanism against planetary perturbations. Naturally, MMRs play a fundamental role in the long-term stability especially for those high-eccentricity objects in planetary systems. Thus, it becomes of great significance to study dynamical structures of retrograde MMRs which could help to understand the dynamical origin and evolution for those asteroids inside retrograde MMRs.

In our Solar system, an increasing number of asteroids are found inside retrograde MMRs. For example, \citet{morais2013asteroids} identified a set of asteroids among Centaurs and Damocloids inside retrograde MMRs with Jupiter and Saturn, including 2006 BZ8 (in 2/$-5$ resonance with Jupiter), 2008 SO218 (in 1/$-2$ resonance with Jupiter) and  2009 QY6 (in 2/$-3$ resonance with Saturn). As stated by \citet{morais2013asteroids}, these retrograde asteroids are the first examples of Solar system objects in retrograde resonances. Several years later, thanks to an improved orbit determination, \citet{wiegert2017retrograde} confirmed that asteroid 2015 BZ509 is the first asteroid inside a retrograde co-orbital resonance with Jupiter and the authors further predicted that retrograde co-orbital asteroids of Jupiter and other giant planets may be more common than previously expected. Simulations made by \citet{namouni2018coorbital} showed that asteroid 2015 BZ509 sits near the peak of co-orbital capture efficiency. Regarding the transneptunian objects (TNOs), \citet{gladman2009discovery} discovered the first retrograde TNO (2008 KV42) with an orbital inclination of $103^{\circ}$, and then \citet{chen2016discovery} identified a new retrograde TNO 2011 KT19 with an inclination of $110^{\circ}$. One year later, \citet{morais2017first} identified that 2011 KT19 is currently inside a 7:9 polar resonance with Neptune. In a recent survey of retrograde asteroids performed by \citet{li2019survey}, the authors confirmed that 2011 KT19 is currently in a retrograde 7:9 resonance with Neptune.

Recently, \citet{connors2018retrograde} reported that asteroid 2007 VW266 (a retrograde object near Jupiter's orbit) is inside a retrograde 13:14 resonance with Jupiter. \citet{li2018centaurs} numerically explored the resonant behaviors of the clones for the minor bodies among Centaurs and Damocloids and they identified another four Centaurs (including 2006 RJ2, 2006 BZ8, 2017 SV13 and 2012 YE8), which are potential candidates inside retrograde co-orbital resonances with Saturn. Furthermore, \citet{li2019survey} numerically investigated the possible retrograde resonant configurations of minor objects in the Solar system and they identified 38 asteroids to be trapped inside retrograde MMRs with planets.

To understand the dynamical differences of prograde and retrograde MMRs, \citet{morais2012stability} numerically investigated the stability of prograde and retrograde planets and they showed that retrograde planets are stable up to distances closer to the perturber than the prograde counterpart. They pointed out that the enhanced stability of retrograde planets with respect to prograde planets is caused by the difference of dynamical structures between retrograde and prograde resonances: for a certain $p$:$q$ resonance, the strength (or width) of resonance is proportional to $e^{|p-q|}$ for the prograde resonance while the strength is proportional to $e^{p+q}$ for the retrograde resonance. As a continuation, \citet{morais2013retrograde} further studied the dynamics of retrograde resonances in detail by using literal expansion of disturbing function for non-coorbital resonances and using numerical averaging of disturbing function for coorbital resonances, and the numerical technique based on computing surfaces of section is taken to explore phase-space structures in the vicinity of the retrograde 2:1, 1:1 and 1:2 resonances. Recently, in the framework of planar circular restricted three-body problem, \citet{li2020dynamics} studied the dynamics of the exterior retrograde 1:n resonances, and they showed that there is no asymmetric libration for the retrograde 1:n resonances because of the dominant contribution of the first-order harmonic arising in the literal expansion of disturbing function.

The dynamical system theory based on families of periodic orbits in the restricted three-body problem provides a new approach to understand the dynamics of retrograde MMRs. For example, in order to understand the capture mechanism of asteroid 2015 BZ509 and similar co-orbital objects, \citet{morais2019periodic} investigated families of periodic orbits of the retrograde co-orbital problem and studied their stability and bifurcations. Recently, in the Sun--Jupiter system, \citet{kotoulas2020planar} produced families of planar resonant retrograde periodic orbits in the planar circular and elliptic restricted three-body problems for the retrograde 2/1, 3/2, 4/3, 3/1, 5/3, 7/5, 4/1, 5/2 and 7/4 resonances. Focusing on the dynamics of resonant TNOs, \citet{kotoulas2020retrograde} identified planar and three-dimensional retrograde periodic orbits corresponding to the retrograde 1/2, 2/3 and 3/4 resonances with Neptune.

Regarding retrograde MMRs, it is of significance to know (a) the place where the resonance occurs (usually it is different from the nominal resonance location), (b) the width that measures the size of resonance zones, and (c) the dynamical structures in phase space. To this end, in the current work, we investigate retrograde MMRs by means of analytical and numerical approaches.

In the analytical study, a new resonant angle is defined, and then a resonant Hamiltonian model is formulated for retrograde MMRs by performing a series of canonical transformations. Based on the resonant model, it is possible to produce phase portraits, where the resonant centres and resonant widths can be identified. In the numerical study, the non-perturbative approach based on computing Poincar\'e sections is adopted. The technique of Poincar\'e sections has been widely used in studying MMRs, e.g. \citet{malhotra1996phase, winter1997resonanceI, winter1997resonanceII, morais2012stability, morais2013retrograde, wang2017mean, malhotra2018neptune} and \citet{malhotra2020divergence}. About the choice of Poincar\'e sections, \citet{wang2017mean} made an important improvement: recording the states of test particles at every successive perihelion passage. As stated by \citet{malhotra2020divergence}, such an improvement yields a more direct visualization and physical interpretation about the resonance zones arising in Poincar\'e sections. In the present work, we further introduce two slight modifications in the production of Poincar\'e sections: (a) recording the states of test particles when the `faster' angular variable is equal to zero and (b) taking the motion integral $\Gamma_2$ ($\Gamma_2$ is the motion integral of the resonant model) as the conserved quantity. These two minor changes make it possible to establish a correspondence between the phase portraits in the resonant model and the Poincar\'e sections in the non-averaged model. Most importantly, it is possible for us to make direct comparisons between analytical and numerical results for a certain resonance. This comparison is important in terms of validating the analytical model formulated in this work and also understanding the structures arising in Poincar\'e sections. As an extension, the numerical approach based on Poincar\'e sections is utilized to identify resonant widths over the full range of eccentricity, so the results provide global dynamical structures for the retrograde MMRs of interest. In particular, seven known examples of asteroids inside retrograde MMRs with Jupiter (2015 BZ509, 2008 SO218, 1999 LE31, 2000 DG8, 2014 AT28, 2016 LS and 2016 JK24) are considered and it is found that all of them are located inside the libration zones determined by analyzing Poincar\'e sections.

The remaining part of this work is organized as follows. In Section \ref{Sect2}, the Hamiltonian function for the planar circular restricted three-body problem is briefly introduced, and in Section \ref{Sect3}, the Hamiltonian model is formulated for retrograde MMRs. In Section \ref{Sect4}, the resonant model is applied to the retrograde 2:1 and 1:2 resonances with Jupiter and analytical results are presented. In Section \ref{Sect5}, the non-perturbative technique based on computing Poincar\'e sections is introduced and applied to produce numerical results for retrograde 2:1 and 1:2 resonances with Jupiter. Direct comparisons between analytical and numerical results are given in Section \ref{Sect6}. In Section \ref{Sect7}, the numerical approach is applied to identifying resonance widths over the full range of eccentricity for the retrograde 2:1, 1:1, 1:2, 1:3 and 1:4 resonances with Jupiter. At last, the summary and discussion are provided in Section \ref{Sect8}.

\begin{figure}
\centering
\includegraphics[width=0.48\textwidth]{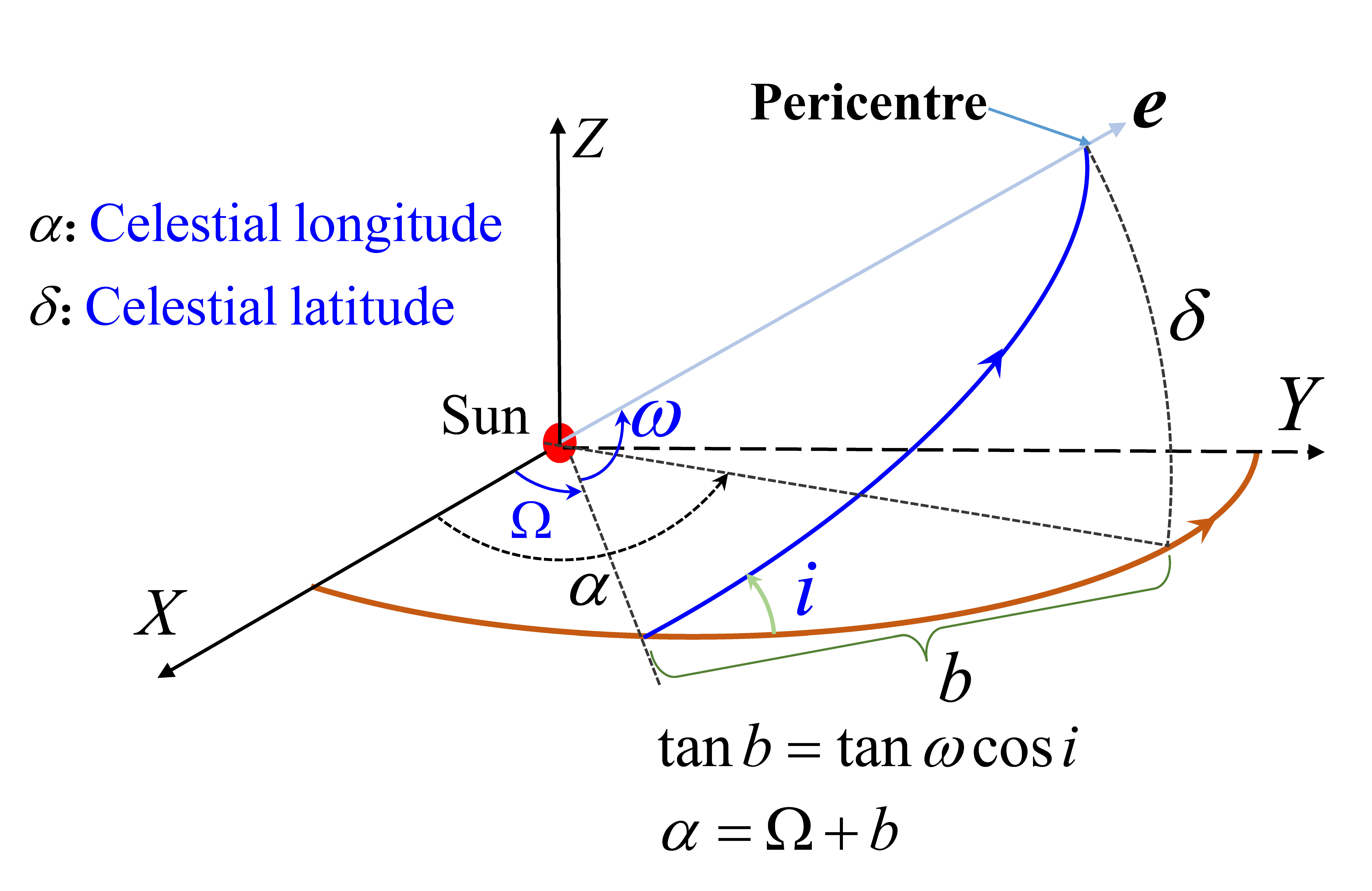}
\caption{Schematic diagram for celestial longitude and latitude of the test particle at the pericentre.}
\label{Fig_1}
\end{figure}

\section{Hamiltonian function}
\label{Sect2}

In this work, we focus on the retrograde motion of asteroids in the planar circular restricted three-body problem with the Sun and a planet as the massive and secondary primaries. In this dynamical model, the primaries move around their barycentre on circular orbits and the asteroid is regarded as a test particle which has no gravitational influence upon the motion of the primaries. The central body (i.e., the Sun) holds the mass of $m_0$ and the planet with mass of $m_p$ plays the role of a perturber. All the objects are moving on the same plane and, in particular, the perturber moves around the Sun in counterclockwise direction and the test particle moves around the Sun in clockwise direction (i.e., the asteroid is moving on a retrograde orbit with inclination of $180^{\circ}$).

For the sake of computational accuracy, it is usual to normalize time and space variables by taking the total mass of the Sun and planet as the unit of mass, their distance as the unit of length, the orbital period of the planet divided by $2\pi$ as the unit of time. Under such a system of normalized units, the universal gravitational constant $\cal G$, the mean motion frequency of the planet $n_p$ and the radius of the planet's orbit $r_p = a_p$ are all unitary. In practical simulations, the Sun--Jupiter system is taken as the fundamental model, in which the length unit is 5.2 $\rm{au}$ and the time unit is 688.955 $\rm{d}$. The normalized mass of the Sun is $m_0 = 0.9990461188$ and that of the Jupiter is $m_p = 1 - m_0$. The system parameters adopted in this work are the same as the ones used in \citet{lei2020multiharmonic}, where the first-order prograde MMRs are investigated.

To describe the orbits of the planet and the test particle, we choose the Sun-centred reference frame with the orbital plane of the planet as the fundamental plane. In the planar problem, the orbit of the planet is described by the semimajor axis $a_p$ ($a_p = r_p = 1$ in normalized unit) and the mean longitude $\lambda_p$, and the orbit of the test particle is characterized by the semimajor axis $a$, eccentricity $e$, the longitude of perihelion $\varpi$ (defined below) and the mean longitude $\lambda = M + \varpi$.

In the Sun-centred reference frame, the motion of the test particle is governed by the planetary disturbing function, given by (the normalized variable $r_p = 1$ is used in the following derivation)
\begin{equation}\label{Eq1}
{\cal R} = {\cal G}{m_p}\left( {\frac{1}{\Delta } - r\cos \psi } \right),
\end{equation}
where $\Delta$ is the mutual distance between the planet and the test particle, given by
\begin{equation*}
\Delta  = \sqrt {1 + {r^2} - 2r\cos \psi }
\end{equation*}
with $\psi$ as the relative angle between the position vectors of the planet and the test particle. In the co-planar retrograde configuration (in this case, the inclination of the test particle's orbit relative the fundamental plane is $i = 180^{\circ}$), the relative angle $\psi$ can be expressed as
\begin{equation*}
\psi  = f + {\lambda _p} - \Omega + \omega = f + {\lambda _p} - \varpi,
\end{equation*}
where $f$ is the true anomaly and $\varpi$ is defined by $\varpi = \Omega - \omega$ in retrograde configurations, which can be found in \citet{shevchenko2016lidov}. In previous studies, $\varpi$ is usually defined by $\varpi = \omega - \Omega$ in the retrograde case \citep{morais2013asteroids, morais2013retrograde}, which has an opposite sign in comparison to our definition.

Regarding the definition of $\varpi (=\Omega - \omega)$ adopted in this work, some discussions are made here. On one hand, in the Sun-centred reference frame, the eccentricity vector can be expressed by
\begin{equation*}
{\bm e} = e\left( \begin{array}{l}
\cos \Omega \cos \omega  - \sin \Omega \sin \omega \cos i\\
\sin \Omega \cos \omega  + \cos \Omega \sin \omega \cos i\\
\sin \omega \sin i
\end{array} \right).
\end{equation*}
Particularly, in the co-planar (prograde or retrograde) configurations, the eccentricity vector becomes ${\bm e} = e \left(\cos{\varpi}, \sin{\varpi}, 0 \right)$ where $\varpi = \Omega + \omega$ in the prograde co-planar case and $\varpi = \Omega - \omega$ in the retrograde co-planar case. On the other hand, when the test particle is located at the perihelion, the celestial longitude is usually defined by $\alpha = \Omega + \arctan(\tan{\omega} \cos{i})$ (please see Fig. \ref{Fig_1} for the definition of $\alpha$). Naturally, it becomes $\alpha = \Omega + \omega$ in the co-planar prograde case (i.e., $i=0$) and $\alpha = \Omega - \omega$ in the co-planar retrograde case (i.e., $i=\pi$). Thus, in the retrograde occasion, the value of $\varpi (= \Omega - \omega)$ measures the relative angle between the test particle's eccentricity vector and the $x$-axis of the defined coordinate system, showing that $\varpi (= \Omega - \omega)$ stands for the longitude of perihelion. Naturally, the definition of $\lambda = M + \varpi (=M+\Omega-\omega)$ is the mean longitude in the retrograde configuration. Thus, we can find the first advantage of the definition $\varpi (=\Omega - \omega)$ adopted in this work: both the angles $\varpi (=\Omega - \omega)$ and $\lambda = M + \varpi (=M+\Omega-\omega)$ have clear physical implications. In the following discussions, we will provide another two advantages for the definition $\varpi (=\Omega - \omega)$ used in this work.

Starting from the co-planar and retrograde assumptions (the assumptions have been used in the expression of $\psi$), we directly follow the procedures given by \citet{murray1999solar} and \citet{ellis2000disturbing} to derive the expansion of planetary disturbing function in a formal series of the orbital elements,
\begin{equation}\label{Eq2}
\begin{aligned}
{\cal R} = & {\cal G}{m_p}\sum\limits_{n = 0}^{N}  {\sum\limits_{j =  - \infty }^\infty  {\sum\limits_{m = 0}^n {\sum\limits_{s =  - \infty }^\infty  {{{\left( { - 1} \right)}^{n - m}}{A_{n,j}}\left( \alpha  \right) {n \choose m}} } } } \\
& \times X_s^{m,j}\left( e \right)\cos \left[ {s \left(\lambda -\varpi\right) + j \left(\lambda _p - \varpi\right)} \right]\\
& - {\cal G}{m_p}a\sum\limits_{s =  - \infty }^\infty  {X_s^{1,1}\left( e \right)\cos \left[ {s \left(\lambda - \varpi\right) + \left(\lambda _p -\varpi\right)} \right]}
\end{aligned}
\end{equation}
where $N$ is the truncated order in the eccentricity (the influence of $N$ upon the accuracy of resonant disturbing function will be reported in Fig. \ref{Fig0}). In equation (\ref{Eq2}), the Hansen coefficients $X_s^{a,b} (e)$ are functions of the eccentricity $e$ and they can be calculated in a recursive manner \citep{hughes1981computation, murray1999solar}, and ${{A_{n,j}}\left( \alpha  \right)}$ is a function of the semimajor axis ratio $\alpha = a/a_p$ ($\alpha = a/a_p = a$ in normalized units), defined by
\begin{equation*}
{A_{n,j}}\left( \alpha  \right) = \frac{{{\alpha ^n}}}{{n!}}\left[ {\frac{{{{\rm d}^n}}}{{{\rm d}{\alpha ^n}}}\frac{1}{2}b_{{1 \mathord{\left/
 {\vphantom {1 2}} \right.
 \kern-\nulldelimiterspace} 2}}^{\left( j \right)}(\alpha )} \right]
\end{equation*}
with $b_{1/2}^j (\alpha)$ as the Laplace coefficients \citep{murray1999solar, ellis2000disturbing}. Since the coefficients $b_{1/2}^j (\alpha)$ are divergent when the semimajor axis ratio is close to unity, the expansion of disturbing function based on Laplace coefficients is not able to deal with co-orbital resonances \citep{murray1999solar, morais2013retrograde}.

It is noted that the literal expansion of disturbing function in the co-planar retrograde case given by equation (\ref{Eq2}) has a similar expression (but not the same) to the one in the co-planar prograde case (please refer to \citet{lei2020multiharmonic} for the explicit expansion of disturbing function in the prograde configuration).

As for the expansion of disturbing function in the retrograde configuration (orbiting in opposite directions), \citet{morais2013retrograde} derived it directly from the standard expansion of the three-dimensional and prograde disturbing function by performing variable substitution: $i^* = 180^{\circ} - i$, $\lambda_p^* = -\lambda_p$, $\omega^* = \omega - \pi$ and $\Omega^* = - \Omega - \pi$. Under the new system of notations, the longitude of perihelion $\varpi$ is transformed into $\varpi^* = \omega - \Omega$ and the mean longitude $\lambda$ into $\lambda^* = M +\omega-\Omega$. As discussed before, $\varpi^*$ and $\lambda^*$ are no longer the longitude of perihelion and mean longitude. The same method utilized to expand the disturbing function in the retrograde configuration can be found in \citet{li2020dynamics}. Although the expansion presented in this work (see equation (\ref{Eq2}) for the explicit expression) and the ones given in previous works \citep{morais2013retrograde, li2020dynamics} are derived from different approaches, we believe they should be equivalent.

Observing the explicit expression of disturbing function given by equation (\ref{Eq2}), we can see that the argument arising in the cosine terms of disturbing function holds
\begin{equation*}
\theta = s (\lambda - \varpi) + j (\lambda_p - \varpi),
\end{equation*}
which can naturally satisfy the d'Alembert rule \citep{murray1999solar, morbidelli2002modern}. This is the second advantage for the definition of $\varpi (=\Omega - \omega)$ adopted in this work.

For convenience, let's denote the independent angles arising in equation (\ref{Eq2}) by
\begin{equation*}
\theta_1 = \lambda - \varpi,\quad \theta_2 = \lambda_p - \varpi,
\end{equation*}
where $\theta_1$ stands for the mean anomaly of the test particle and $\theta_2$ represents the angular separation between the perturber and the test particle's perihelion relative to the Sun. As a result, the expansion of disturbing function given by equation (\ref{Eq2}) can be organized in a compact form as follows:
\begin{equation}\label{Eq3}
{\cal R} = \sum\limits_{{k_1},{k_2}} {{{\cal D}_{{k_1},{k_2}}}\left( {a,e} \right)\cos \left( {{k_1}{\theta _1} + {k_2}{\theta _2}} \right)}, k_1 \in \mathbb{N}, k_2 \in \mathbb{Z}
\end{equation}
where ${\cal D}_{{k_1},{k_2}}$ are the coefficients related to the semimajor axis $a$ and eccentricity $e$ and the explicit expressions can be directly derived from equation (\ref{Eq2}).

Based on the disturbing function, the Hamiltonian function of system, can be written as
\begin{equation}\label{Eq4}
{\cal H} =  - \frac{\mu }{{2a}} + {n_p}{\Lambda _p} - {\cal R}\left( {a,e,\theta_1 ,\theta_2} \right)
\end{equation}
where $\mu$ is the gravitational parameter of the Sun, given by $\mu = {\cal G} m_0$, the mean motion frequency of the planet is $n_p = 1$ and ${\Lambda _p}$ is the momentum conjugated to the mean longitude of the planet $\lambda_p$.

To formulate the dynamical model, let us adopt the modified Delaunay variables as follows:
\begin{equation}\label{Eq5}
\begin{aligned}
&\Lambda  = \sqrt{\mu a},\quad \lambda  = M +\varpi,\\
&P = \sqrt{\mu a} \left(1+ \sqrt{1-e^2}\right),\quad p = -\varpi,\\
&\Lambda _p,\quad \lambda_p,
\end{aligned}
\end{equation}
where the pair $({\Lambda _p}, {\lambda _p})$ is introduced to describe the mean motion of the perturber.

The canonical variables given by equation (\ref{Eq5}) is different from the traditional modified Delaunay variables due to the different definition of $\varpi$ adopted in this work. Please see \citet{morbidelli2002modern} for the classical version of modified Delaunay's variables.

Using a linear transformation, it is possible to introduce the following set of variables,
\begin{equation}\label{Eq6}
\begin{aligned}
&{\Theta _1} = \Lambda ,\quad {\theta _1} = \lambda  + p = \lambda  - \varpi,\\
&{\Theta _2} = P - \Lambda ,\quad {\theta _2} = {\lambda _p} + p = {\lambda _p} - \varpi,\\
&{\Theta _3} = {\Lambda _p} - P + \Lambda ,\quad {\theta _3} = {\lambda _p},
\end{aligned}
\end{equation}
which is a canonical transformation with the generating function,
\begin{equation*}
{\cal S} = \left( {\lambda  + p} \right){\Theta _1} + \left( {{\lambda _p} + p} \right){\Theta _2}  + {\lambda _p}{\Theta _3}.
\end{equation*}
Thus, the Hamiltonian represented by equation (\ref{Eq4}) can be expressed as a function of the new set of canonical variables,
\begin{equation}\label{Eq7}
\begin{aligned}
{\cal H} &=  - \frac{{{\mu ^{\rm{2}}}}}{{2\Theta _1^2}} + {\Theta _2} - {\cal R}\left( {{\Theta _1},{\Theta _2},{\theta _1},{\theta _2}} \right)\\
&= - \frac{{{\mu ^{\rm{2}}}}}{{2\Theta _1^2}} + {\Theta _2} - \sum\limits_{{k_1},{k_2}} {{{\cal D}_{{k_1},{k_2}}}\left( {\Theta_1,\Theta_2} \right)\cos \left( {{k_1}{\theta _1} + {k_2}{\theta _2}} \right)}
\end{aligned}
\end{equation}
where the constant terms have been removed from the Hamiltonian and the equality $n_p= 1$ is used. Evidently, the dynamical model determined by equation (\ref{Eq7}) is of two degrees of freedom with $\theta_1$ and $\theta_2$ as the angular coordinates. The Hamiltonian canonical relations yield the equations of motion as follows:
\begin{equation}\label{Eq8}
\begin{aligned}
\frac{{{\rm d}{\theta _1}}}{{{\rm d}t}} =& \frac{{\partial {\cal H}}}{{\partial {\Theta _1}}},\quad \frac{{{\rm d}{\Theta _1}}}{{{\rm d}t}} =  - \frac{{\partial {\cal H}}}{{\partial {\theta _1}}},\\
\frac{{{\rm d}{\theta _2}}}{{{\rm d}t}} =& \frac{{\partial {\cal H}}}{{\partial {\Theta _2}}},\quad \frac{{{\rm d}{\Theta _2}}}{{{\rm d}t}} =  - \frac{{\partial {\cal H}}}{{\partial {\theta _2}}}.
\end{aligned}
\end{equation}
For convenience, let's denote the state of test particles by
\begin{equation}\label{Eq9}
{\bm X} = \left\{\theta_1,\theta_2, \Theta_1, \Theta_2 \right\}.
\end{equation}
Thus, the equation of motion given by equation (\ref{Eq8}) can be expressed in a vectorial form,
\begin{equation}\label{Eq10}
\dot {\bm X} = {\bm F} \left( {\bm X} \right).
\end{equation}
The dynamical model shown by equations (\ref{Eq9}--\ref{Eq10}) are to be used in defining and producing Poincar\'e surfaces of section, as discussed in Section \ref{Sect5}.

\section{Hamiltonian model of retrograde MMRs}
\label{Sect3}

In this section, we formulate the Hamiltonian model of retrograde MMRs by defining an appropriate resonant angle and performing canonical transformations. For a retrograde $k_p$:$k$ resonance, we define the critical argument as
\begin{equation}\label{Eq11}
\begin{aligned}
\sigma &= \frac{1}{k_{\max}} \varphi \\
&= \frac{1}{k_{\max}} \left[ k \lambda - k_p \lambda_p + (k_p - k)\varpi \right]\\
&= \frac{1}{{{k_{\max }}}}\left( {k (\lambda - \varpi) - {k_p} (\lambda_p - \varpi)} \right)\\
&= \frac{1}{{{k_{\max }}}}\left( {k{\theta _1} - {k_p}{\theta _2}} \right)
\end{aligned}
\end{equation}
where $\varphi$ is the usual critical argument, given by $\varphi = k \lambda - k_p \lambda_p + (k_p - k)\varpi$, and $k_{\max} = \max{\left\{k_p,k\right\}}$ (it holds $k_{\max} = k_p$ for the inner resonances and $k_{\max} = k$ for those exterior resonances). In particular, we have $\sigma  = \frac{1}{2}\left( {{\theta _1} - 2{\theta _2}} \right)$ for the retrograde 2:1 resonance, $\sigma  = \frac{1}{2}\left( {2{\theta _1} - {\theta _2}} \right)$ for the retrograde 1:2 resonance, and the like. Note that the resonant angle defined by equation (\ref{Eq11}) is equal to the one defined by \citet{lei2020multiharmonic} for the inner resonances, but it is different for the outer resonances.

According to the explicit expansion of planetary disturbing function given by equation (\ref{Eq2}), the coefficient of the cosine term associated with the resonant angle, $\cos{(k_{\max}\sigma)}$, holds,
\begin{equation*}
{\cal D}_0 (\alpha) X_k^{m,-k_p} (e),
\end{equation*}
which stands for the magnitude of the resonant term and it is proportional to the power of eccentricity. In particular, at the lowest order, the magnitude of the resonant term is proportional to $e^{k_p + k}$, as pointed out by \citet{morais2012stability}. For the prograde counterpart, it is known that the magnitude of the resonant term, at the lowest order, is proportional to $e^{|k_p-k|}$ \citep{murray1999solar}. Thus, for a certain $k_p$:$k$ resonance with a given eccentricity, the retrograde case has much weaker resonant strength (or, equivalently, smaller resonant width) compared to the prograde case, showing that the planetary perturbations for retrograde MMRs are generally weaker \citep{morais2013retrograde}. Based on this fact, \citet{morais2013retrograde} explained that retrograde resonances are more stable than prograde resonances is due to the weak perturbations in retrograde configurations where the encounter of two objects occurs at a higher relative velocity.

Observing the expression of $\varphi$, we can find the usual critical argument, given by $\varphi = k \lambda - k_p \lambda_p + (k_p - k)\varpi$, keeps the same expression as that in the prograde configuration \citep{murray1999solar, morbidelli2002modern}. This is the third advantage for the definition of $\varpi (=\Omega - \omega)$ adopted in this work. Based on this advantage, it is possible to unify the resonant models of prograde and retrograde resonances from theoretical point of view (this point is very important to understand the retrograde resonances).

To formulate the resonant model, let us further introduce the following canonical transformations:
\begin{equation}\label{Eq12}
\begin{aligned}
&{\Gamma _1} = \frac{k_{\max}}{k} \Theta_1,\quad {\sigma _1} = \frac{1}{{{k_{\max }}}}\left( k \theta_1 - k_p \theta_2 \right) = \sigma,\\
&{\Gamma_2} = \Theta_2 +\frac{k_p}{k} \Theta_1, \quad {\sigma_2} = \theta_2
\end{aligned}
\end{equation}
with the generating function as
\begin{equation*}
{\cal S} = \frac{\Gamma_1}{k_{\max}}\left( k\theta_1 - k_p\theta_2 \right) + \theta_2 \Gamma_2.
\end{equation*}
Using the defined canonical variables, the Hamiltonian given by equation (\ref{Eq7}) can be written as
\begin{equation}\label{Eq13}
\begin{aligned}
{{\cal H}} =& - \frac{{{\mu ^2}}}{{2{{\left( {\frac{k}{{{k_{\max }}}}{\Gamma _1}} \right)}^2}}} - \frac{{{k_p}}}{{{k_{\max }}}}{\Gamma _1}\\
& - \sum\limits_{{k_1},{k_2}} {{\cal D}_{{k_1},{k_2}}}\left( {\Gamma_1,\Gamma_2} \right)\cos \left[{k_{\max}} \frac{k_1}{k}{\sigma_1} + \left(k_2 + \frac{k_1}{k}\right){\sigma_2} \right].
\end{aligned}
\end{equation}
When the test particle is located inside a retrograde MMR, the resonant angle $\sigma_1 (=\sigma)$ becomes a `long-period' variable, while the angle $\sigma_2$ is a `short-period' variable. Thus, in the expansion of disturbing function those terms involving the angle $\sigma_2$ have short-period effects upon the evolution of test particles. In order to study resonant dynamics, it is usual to filter out those short-period terms from the Hamiltonian function, producing the so-called resonant Hamiltonian as follows:
\begin{equation}\label{Eq14}
\begin{aligned}
{{\cal H}^*} = & \frac{1}{{2k\pi }}\int\limits_0^{2k\pi } {{\cal H}\left( {\Gamma_1, \Gamma_2, {\sigma_1}, {\sigma _2}} \right){\rm d}{\sigma _2}}\\
= & - \frac{{{\mu ^2}}}{{2{{\left( {\frac{k}{{{k_{\max }}}}{\Gamma _1}} \right)}^2}}} - \frac{{{k_p}}}{{{k_{\max }}}}{\Gamma _1}\\
& - \frac{1}{{2k\pi }}\int\limits_0^{2k\pi } {{\cal R}\left( {\Gamma_1, \Gamma_2, {\sigma_1}, {\sigma _2}} \right){\rm d}{\sigma _2}}
\end{aligned}
\end{equation}
which becomes
\begin{equation}\label{Eq15}
{{\cal H}^*} = - \frac{{{\mu ^2}}}{{2{{\left( {\frac{k}{{{k_{\max }}}}{\Gamma _1}} \right)}^2}}} - \frac{{{k_p}}}{{{k_{\max }}}}{\Gamma _1} - \sum\limits_{m = 0}^M {{{\cal C}_m}\cos \left( {m{k_{\max }}{\sigma _1}} \right)}
\end{equation}
where ${{\cal C}_m}$ are the coefficients related to the action variables $\Gamma_1$ and $\Gamma_2$ and the number $M$ stands for the number of harmonics of the angle $k_{\max} \sigma_1$ in the expansion of resonant disturbing function. The last term of equation (\ref{Eq15}) corresponds to the resonant disturbing function, whose explicit expression is provided in Appendix \ref{A1}.

\begin{figure}
\centering
\includegraphics[width=0.48\textwidth]{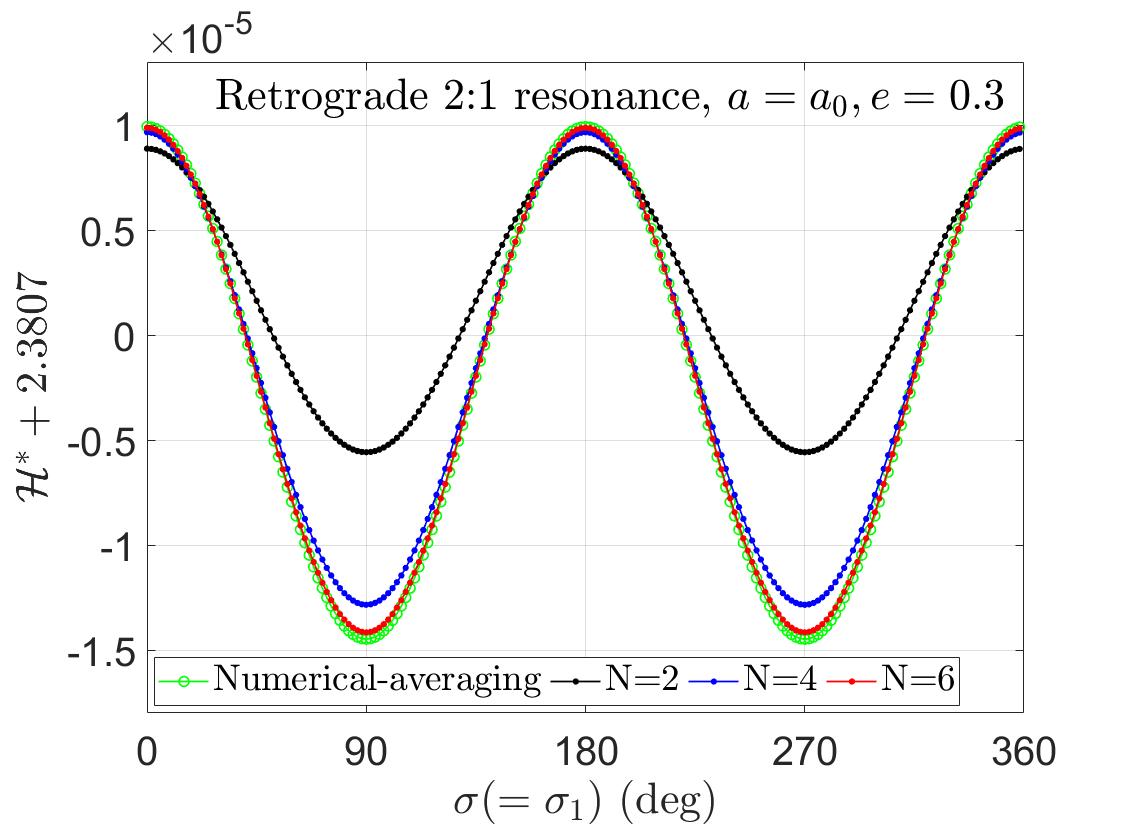}
\caption{Comparisons between the resonant Hamiltonian obtained by means of numerical-averaging technique and series expansions truncated at orders $N=2$, $N=4$ and $N=6$ for the retrograde 2:1 resonance. In simulations, the semimajor axis is taken as the nominal value of resonance (i.e., $a_0=0.62976016$ in normalized units), the value of eccentricity is fixed at $e=0.3$ and the number of harmonics is assumed at $M=2$. For the retrograde 2:1 resonance, the resonant angle is defined by $\sigma = \frac{1}{2} (\theta_1 - 2\theta_2) = \frac{1}{2}\varphi$.}
\label{Fig0}
\end{figure}

Alternatively, the exact value of resonant disturbing function arising in equation (\ref{Eq14}) can be obtained by means of numerical-averaging approach (i.e., the unexpanded disturbing function is numerically averaged with respect to the fast angle $\sigma_2$ over $k$ periods), which has been widely used in previous works \citep{beauge1994asymmetric, gallardo2006occurrence, gallardo2006atlas, morais2013retrograde, li2014studya, li2014studyb, li2020study, huang2018dynamic, gallardo2019strength, gallardo2020three, li2020dynamics}. Please refer to Fig. \ref{Fig0} for the comparisons between the resonant Hamiltonian obtained by means of numerical-averaging technique and series expansions truncated at orders $N=2$, $N=4$ and $N=6$ for the retrograde 2:1 resonance with $e=0.3$ (the semimajor axis is taken as the nominal value of resonance at $a_0=0.62976016$ in normalized units). It shows that the curve corresponding to a larger $N$ is closer to the reference curve obtained by means of numerical averaging, indicating that the series expansion truncated at a higher order has a better accuracy. This is expected by us.

The resonant Hamiltonian given by equation (\ref{Eq15}) determines a dynamical model with a single degree of freedom. Since the angle $\sigma_2$ is absent from the Hamiltonian, its conjugate momentum becomes the motion integral in the resonant model, given by
\begin{equation}\label{Eq16}
\begin{aligned}
{\Gamma _2} &= \Theta_2 +\frac{k_p}{k} \Theta_1\\
&= \sqrt {\mu a} \left( {\frac{{{k_p}}}{k} + \sqrt {1 - {e^2}} } \right) = {\rm const},
\end{aligned}
\end{equation}
which means that, in the resonant model, the semimajor axis and eccentricity exchange with each other in the long-term evolution. For convenience, the motion integral $\Gamma_2$ can be specified by the minimum semimajor axis $a_{\min}$ in the following manner:
\begin{equation*}
{\Gamma _2} =\sqrt {\mu {a_{\min }}} \left( {\frac{{{k_p}}}{k} + 1} \right),
\end{equation*}
where $a_{\min}$ is the magnitude of semimajor axis when the eccentricity is assumed at zero.

\begin{figure}
\centering
\includegraphics[width=0.42\textwidth]{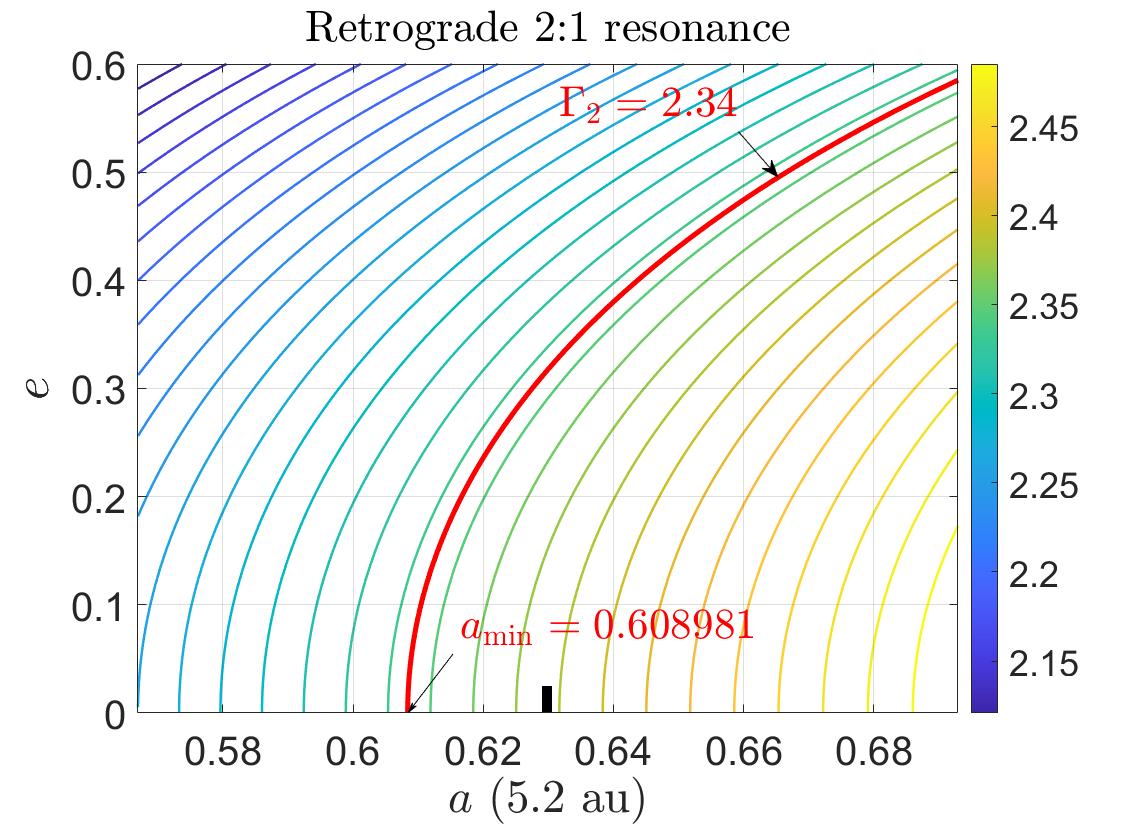}
\includegraphics[width=0.42\textwidth]{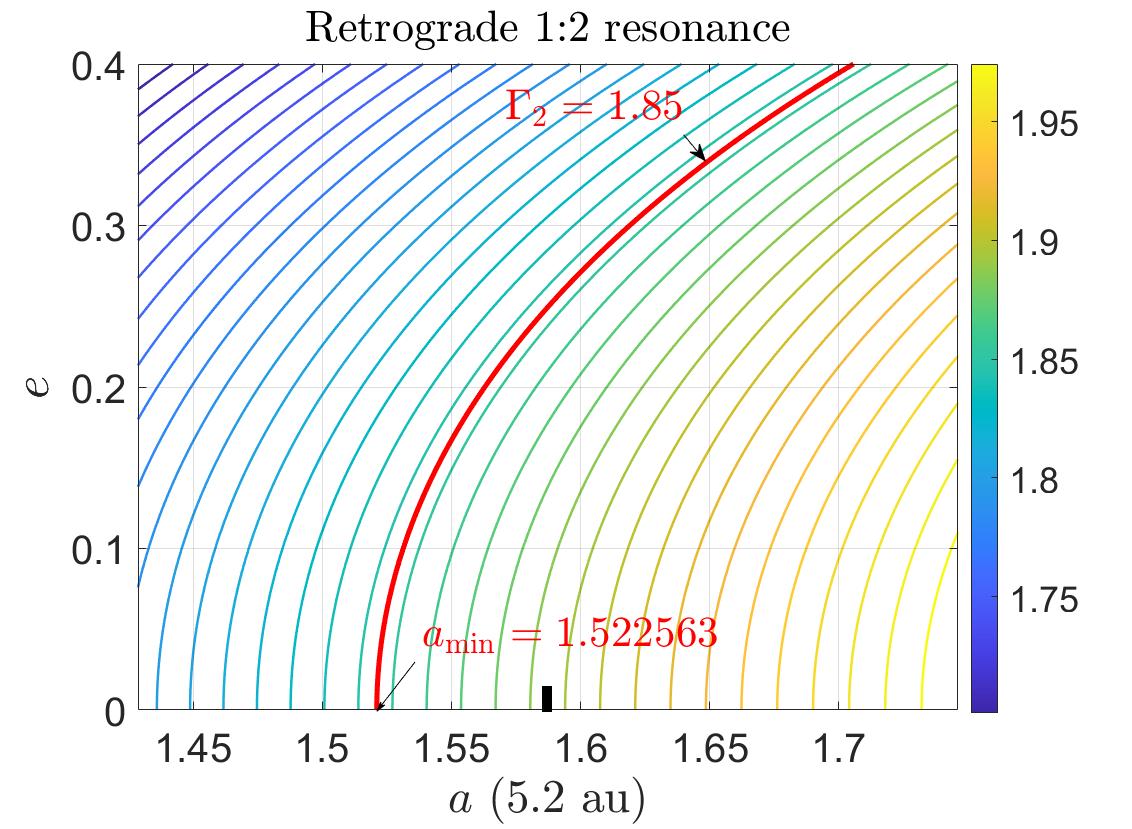}
\caption{Level curves of the motion integral ${\Gamma _2}$ shown in the $(a,e)$ space for the retrograde 2:1 resonance (\emph{upper panel}) and the retrograde 1:2 resonance (\emph{bottom panel}). The motion integral is defined by $\Gamma_2 = \sqrt{\mu a} (\frac{k_p}{k} + \sqrt{1-e^2})$, which can be equivalently represented by $a_{\min}$ in the form of $\Gamma_2 = \sqrt{\mu a_{\min}} (\frac{k_p}{k} + 1)$ ($a_{\min}$ is the semimajor axis when the eccentricity is assumed at zero). The red lines marked in both panels correspond to $\Gamma_2 = 2.34$ (or $a_{\min} = 0.608981$) for the retrograde 2:1 resonance and $\Gamma_2 = 1.85$ (or $a_{\min} = 1.522563$) for the retrograde 1:2 resonance, which are to be used as typical examples in the following simulations. The nominal location of resonance is marked by a short and black vertical line.}
\label{Fig1}
\end{figure}

For the retrograde 2:1 and 1:2 resonances, the level curves of the motion integral $\Gamma_2$ are plotted in the $(a,e)$ space, as shown in Fig. \ref{Fig1}. In particular, two representatives with $\Gamma_2 = 2.34$ (or $a_{\min} = 0.608981$) and $\Gamma_2 = 1.85$ (or $a_{\min} = 1.522563$) are marked by red lines.

\begin{figure}
\centering
\includegraphics[width=0.48\textwidth]{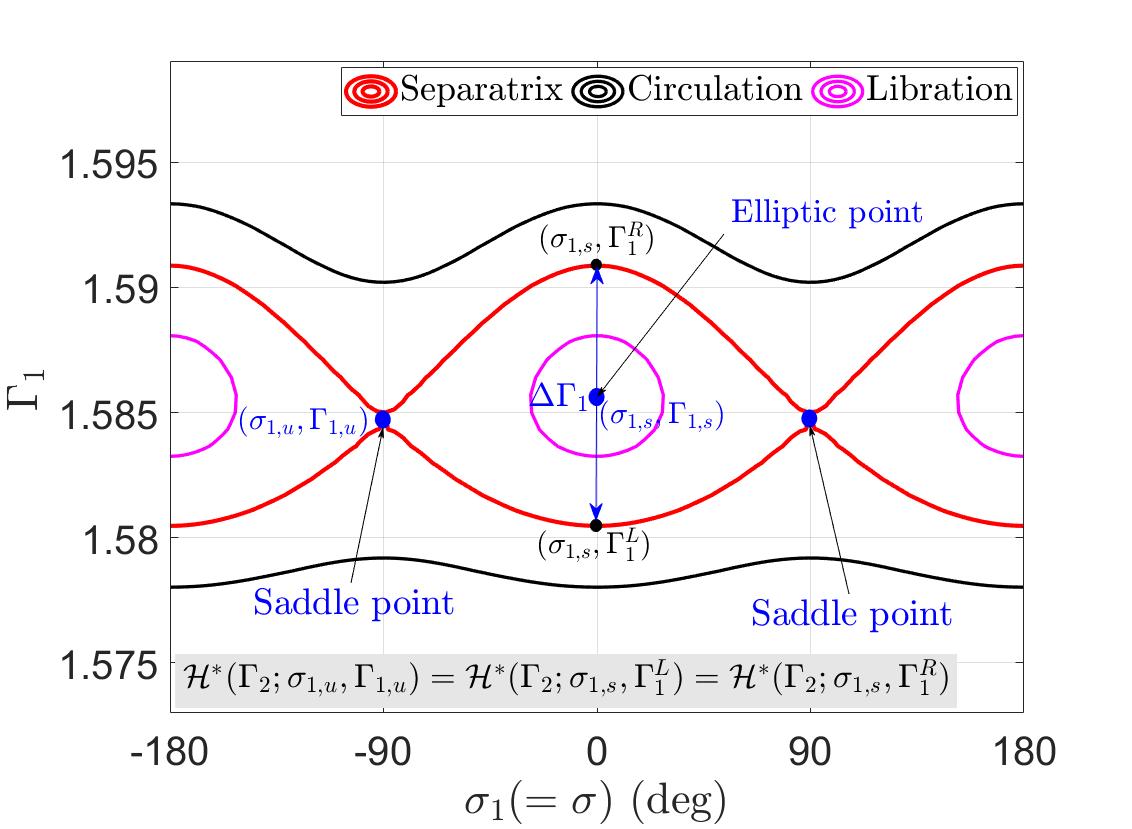}
\caption{Schematic diagram for illustrating elliptic points (i.e., resonant centres), saddle points, dynamical separatrices, curves of libration and circulation, and resonant width measured from the phase portrait in the $(\sigma_1, \Gamma_1)$ space. Here, we take the 2:1 resonance as an example to show the diagram.}
\label{Fig1-1}
\end{figure}

\section{Analytical study}
\label{Sect4}

In the previous section, we have formulated the Hamiltonian model for retrograde MMRs by introducing the resonant angle $\sigma (= \sigma_1)$ and by performing a series of canonical transformations. The resulting resonant model is of a single degree of freedom, and thus it is integrable. In this section, we will introduce the analytical method on how to determine the resonant centre and the associated resonant width, and then apply it to some retrograde MMRs (in practice, the retrograde 2:1 and 1:2 resonances are taken as examples in simulations).

\subsection{Resonant centre and width}
\label{Sect4-1}

In the resonant model (remind that in the resonant model $\Gamma_2$ is a motion integral), the equilibrium points should satisfy the following stationary conditions:
\begin{equation}\label{Eq17}
\begin{aligned}
{{\dot \sigma }_1} &= \frac{{\partial {{\cal H}^*}}}{{\partial {\Gamma _1}}} = 0,\\
{{\dot \Gamma }_1} &=  - \frac{{\partial {{\cal H}^*}}}{{\partial {\sigma _1}}} = \sum\limits_{m = 1}^M {m{k_{\max }}{{\cal C}_m}\sin \left( {m{k_{\max }}{\sigma _1}} \right)}  = 0.
\end{aligned}
\end{equation}
The second condition of equation (\ref{Eq17}) shows that the equilibrium points are located at
\begin{equation*}
{\sigma _1} = \frac{q \pi}{k_{\max}},\quad  q \in \mathbb{Z}.
\end{equation*}
Replacing the resonant Hamiltonian $\cal H^*$ in the first condition of equation (\ref{Eq17}), we can obtain the equilibrium points satisfy
\begin{equation}\label{Eq17A}
k{\mu ^2}{\left( {\frac{k}{{{k_{\max }}}}{\Gamma _1}} \right)^{ - 3}} = {k_p} + {k_{\max }}\sum\limits_{m = 0}^M {\frac{{\partial {{\cal C}_m}}}{{\partial {\Gamma _1}}}\cos \left( {mq\pi } \right)}
\end{equation}
By numerically solving this equation, the positions of equilibrium points, denoted by $(\sigma_1 = \frac{q \pi}{k_{\max}}, \Gamma_1)$, can be determined.

In particular, if the second term in the right-hand side of equation (\ref{Eq17A}) is ignored, it is possible to obtain the nominal resonant centre at
\begin{equation*}
\Gamma_1 = \left( \frac{\mu^2 k} {k_p} \right)^{1/3} \frac{k_{\max}}{k}
\end{equation*}
which is
\begin{equation*}
a = {\mu}^{1/3} (\frac{k}{k_p})^{2/3}.
\end{equation*}
Thus, the deviation of the exact location of equilibrium points from the nominal resonance centre is due to the rightmost term in equation (\ref{Eq17A}).

For a given motion integral $\Gamma_2$, let us denote the stable equilibrium points (i.e., elliptic points or resonant centres) by $({\sigma _{1,s}}, {\Gamma _{1,s}})$ and the unstable equilibrium points (i.e., saddle points) as $(\sigma _{1,u}, \Gamma _{1,u})$. According to the expression of $\Gamma_1 (=\Gamma_{1,s})$ and $\Gamma_2$, we can identify the location of resonant centres in the $(a,e)$ space, denoted by $(a_0, e_0)$. As discussed by \citep{morbidelli2002modern}, the level curves of resonant Hamiltonian passing through saddle points play the role of dynamical separatrices, dividing the total phase space into regions of libration and circulation. In addition, the distance between adjacent separatrices evaluated at the resonant centre stands for the resonant width (please refer to Fig. \ref{Fig1-1} for the definition of resonant width).

Let's denote the states on the separatrices evaluated at the resonant centre with $\sigma_1 = \sigma _{1,s}$ by $(\sigma _{1,s}, \Gamma _1^L)$ and $(\sigma _{1,s}, \Gamma _1^R)$ (see Fig. \ref{Fig1-1} for details). Here, the separatrices stem from the saddle point at $(\sigma _{1,u}, \Gamma _{1,u})$. According to the definition of resonant width, we can obtain the following relation:
\begin{equation}\label{Eq18}
{\cal H} ^* (\Gamma_2; \sigma_{1,u}, \Gamma_{1,u}) = {\cal H} ^* (\Gamma_2; \sigma_{1,s}, \Gamma_1^L) = {\cal H} ^* (\Gamma_2; \sigma_{1,s}, \Gamma_1^R).
\end{equation}
For a given motion integral $\Gamma_2$, it is possible to obtain the action variables $\Gamma_1^L$ and $\Gamma_1^R$ on the separatrices by numerically solving equation (\ref{Eq18}). Naturally, the resonant width can be represented by $\Delta \Gamma_1 = \Gamma_1^R - \Gamma_1^L$.

In our study, we would like to introduce the variations of semimajor axis and eccentricity ($\Delta a$ and $\Delta e$) to measure the resonant width. To this end, we need to determine the points on the separatrices, shown by $(a_L, e_L)$ and $(a_R, e_R)$. According to the relationship between $\Gamma_1$ and the semimajor axis $a$,
\begin{equation*}
\Gamma_1 = \frac{k_{\max}}{k} \sqrt{\mu a},
\end{equation*}
it is possible to determine the semimajor axes of the points at the boundaries, denoted by $a_L$ and $a_R$. Furthermore, the conservation of motion integral $\Gamma_2$ leads to the following relations \citep{lei2020multiharmonic}:
\begin{equation}\label{Eq19}
\begin{aligned}
\Gamma_2 &= \sqrt {\mu a_0} \left( {\frac{{{k_p}}}{k} + \sqrt {1 - {e_0^2}} } \right)\\
&= \sqrt {\mu a_L} \left( {\frac{{{k_p}}}{k} + \sqrt {1 - {e_L^2}} } \right)\\
&= \sqrt {\mu a_R} \left( {\frac{{{k_p}}}{k} + \sqrt {1 - {e_R^2}} } \right)
\end{aligned}
\end{equation}
where $(a_0,e_0)$ stands for the position of the resonant centre. Solving equation (\ref{Eq19}), we can identify the points on the boundaries in the $(a,e)$ space as $(a_L, e_L)$ and $(a_R, e_R)$. Thus, the resonant width in terms of the variations of semimajor axis and eccentricity can be expressed by
\begin{equation*}
\Delta a = a_R - a_L,\quad \Delta e = e_R - e_L.
\end{equation*}

\subsection{Analytical results}
\label{Sect4-2}

In the resonant model, there is a motion integral, denoted by $\Gamma_2$, which can be specified by $a_{\min}$. For such an integral system, the global dynamics in the phase space can be revealed by phase portraits (i.e., level curves of resonant Hamiltonian) with a given motion integral $\Gamma_2$ (or $a_{\min}$). For intuition, in this work we show the phase portraits in the $(e \cos{\sigma}, e \sin{\sigma})$ and $(\sigma, a)$ spaces.

\begin{figure*}
\centering
\includegraphics[width=0.48\textwidth]{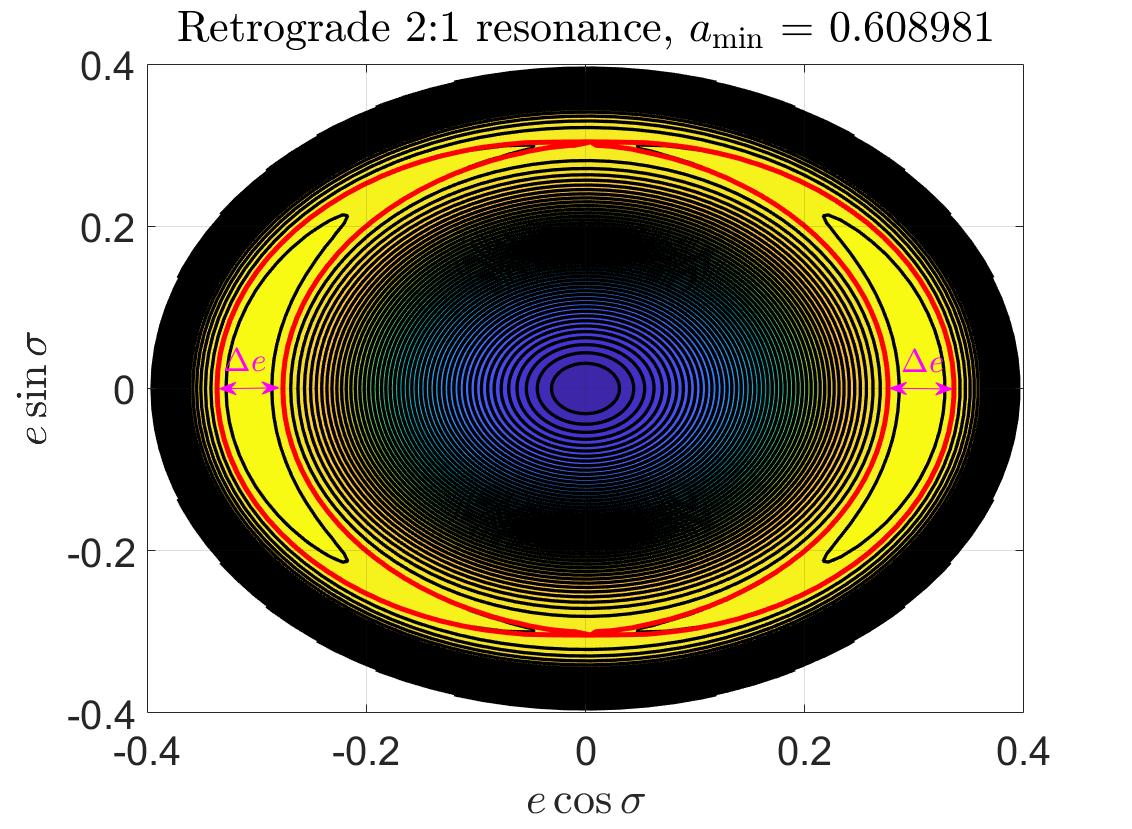}
\includegraphics[width=0.48\textwidth]{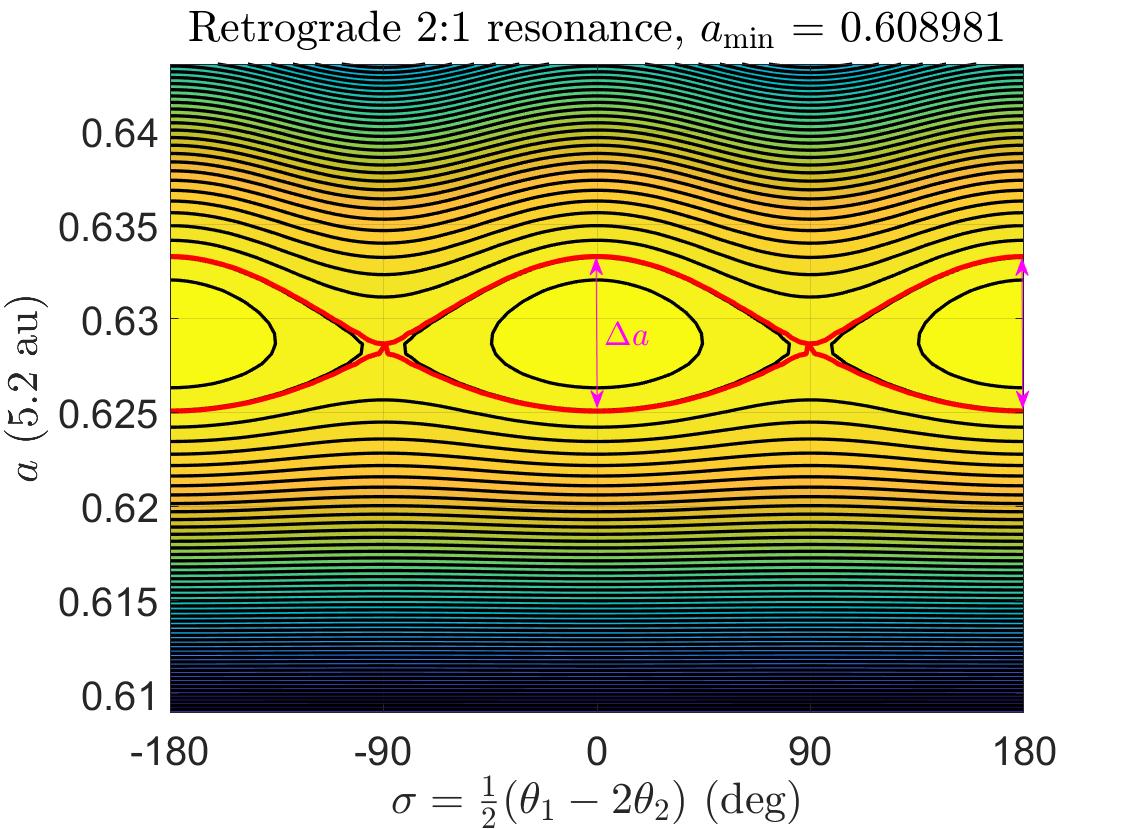}
\caption{The level curves of resonant Hamiltonian (i.e., phase portraits) shown in the $(e \cos{\sigma},e \sin{\sigma})$ space (\emph{left panel}) and in the $(\sigma, a)$ space (\emph{right panel}) for the retrograde 2:1 resonance specified by $a_{\min} = 0.608981$ (or $\Gamma_2 = 2.34$). For the considered resonance, the critical argument is defined as $\sigma = \frac{1}{2} \left(\lambda - 2\lambda_p + \varpi \right) = \frac{1}{2} \left(\theta_1 - 2\theta_2\right)$. The resonant centres are located at $\sigma = 0$ and $\sigma = \pi$ (corresponding to the usual argument at $\varphi = 0$), and the saddle points are located at $\sigma = \pm \pi/2$ (corresponding to $\varphi = \pi$). The level curves passing through the saddle points are marked in red lines, which play the role of dynamical separatrices, dividing the phase space into libration and circulation regions. The distance between two nearby separatrices, evaluated at the resonant centre, measures the resonant width. In particular, the resonant width in terms of the variation of eccentricity $\Delta e$ is marked in the \emph{left panel}, and the one in terms of the variation of semimajor axis $\Delta a$ is marked in the \emph{right panel}.}
\label{Fig2}
\end{figure*}

Figure \ref{Fig2} shows the phase portraits of the retrograde 2:1 resonance specified by $a_{\min} = 0.608981$ (or $\Gamma_2 = 2.34$). The isoline of $\Gamma_2 = 2.34$ is shown in the upper panel of Fig. \ref{Fig1} and, for the retrograde 2:1 resonance, the resonant angle is defined by $\sigma =\frac{1}{2} \left(\theta_1 -2\theta_2\right) = \frac{1}{2} \left(\lambda - 2\lambda_p + \varpi \right) = \frac{1}{2}\varphi$. From Fig. \ref{Fig2}, it is observed that (a) the resonant centres are located at $\sigma = 0$ and $\sigma = \pi$ (corresponding to the usual argument at $\varphi = 0$), (b) the saddle points are placed at $\sigma = \pi/2$ and $\sigma = -\pi/2$ (corresponding to $\varphi = \pi$), (c) the level curves passing through saddle points (marked in red lines) play the role of dynamical separatrices, which divide the entire phase space into libration and circulation regions, and (d) the distance between neighboring separatrices evaluated at the resonant centre can measure the resonant width and, in particular, the resonant width in terms of the variation of eccentricity $\Delta e$ is marked in the \emph{left panel} and the one in terms of the variation of semimajor axis $\Delta a$ is marked in the \emph{right panel}.

In the polar coordinate plane (see the \emph{left panel} of Fig. \ref{Fig2}), the coordinate origin with zero eccentricity is no longer a stationary point. This is different from the prograde case (it is known from \citet{malhotra2020divergence} and \citet{lei2020multiharmonic} that in the prograde case the zero-eccentricity point is always a saddle point in the resonant model).

\begin{figure*}
\centering
\includegraphics[width=0.48\textwidth]{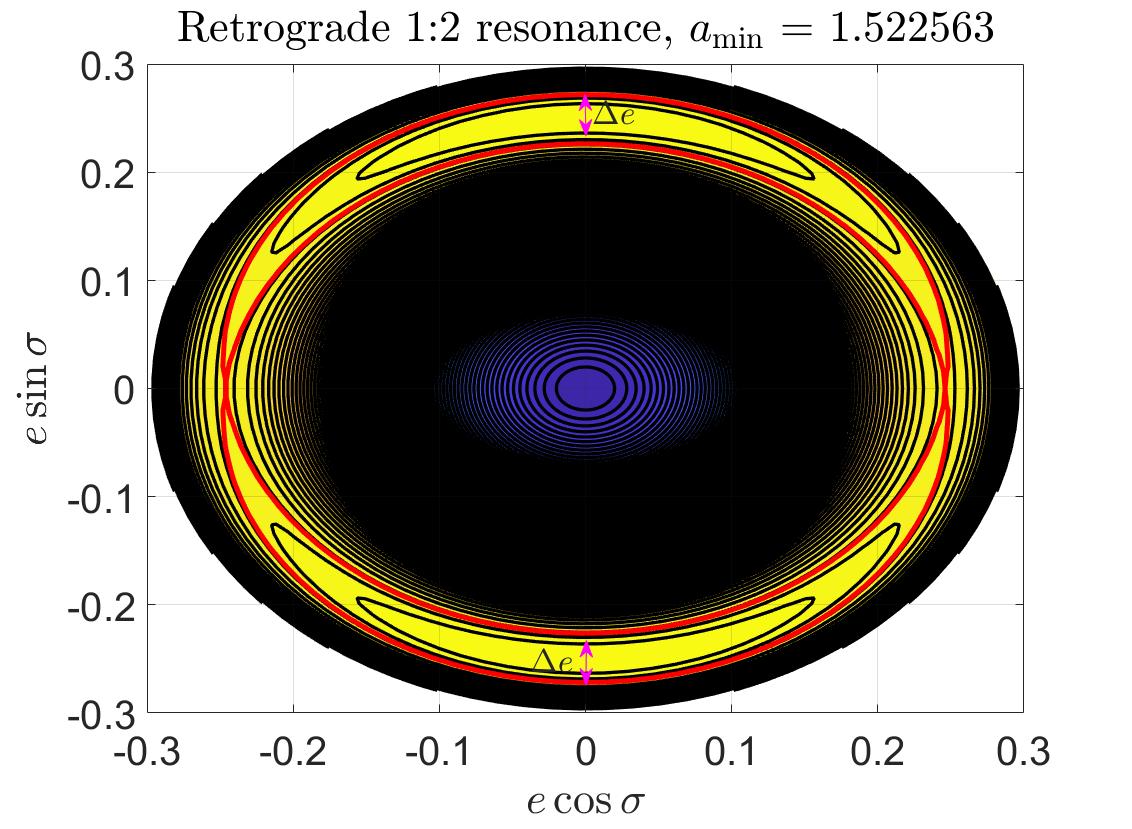}
\includegraphics[width=0.48\textwidth]{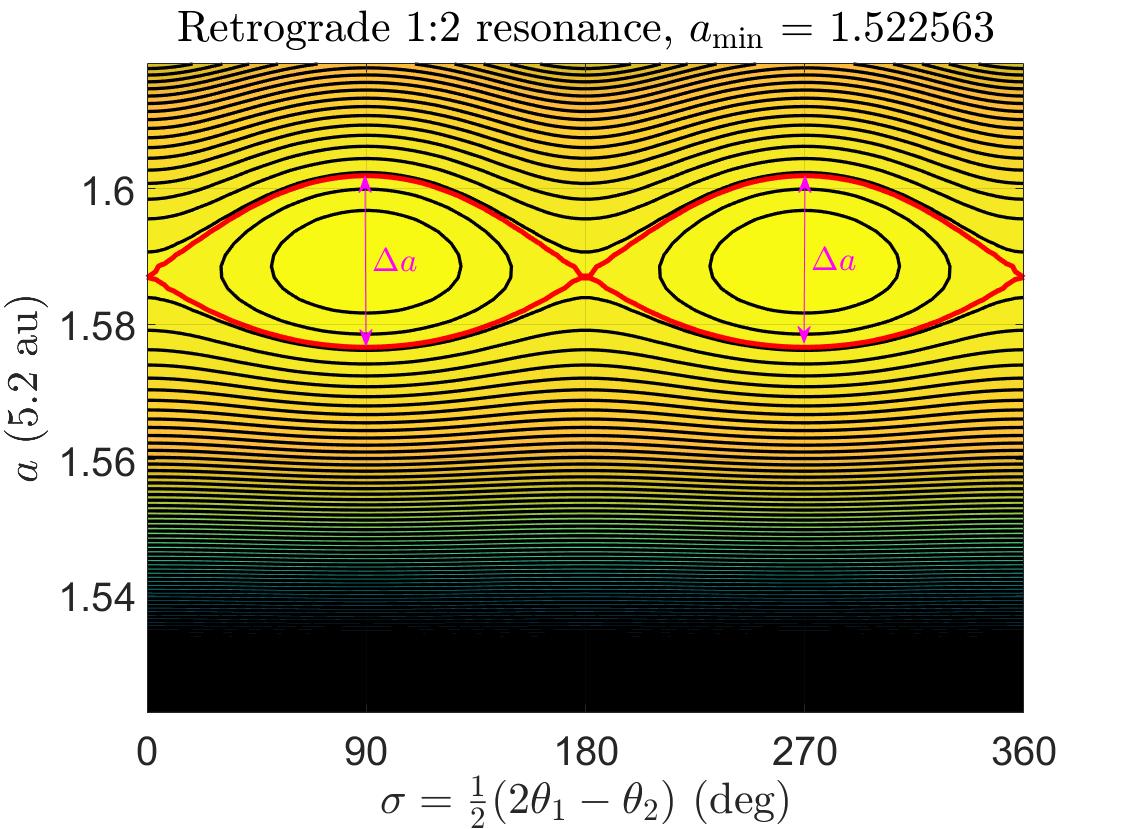}
\caption{Same as Fig. \ref{Fig2} but for the retrograde 1:2 resonance specified by $a_{\min} = 1.522563$ (or $\Gamma_2 = 1.85$). For the considered resonance, the critical argument is defined as $\sigma = \frac{1}{2} \left(2\lambda - \lambda_p - \varpi \right) = \frac{1}{2}\left(2\theta_1 -\theta_2\right)$. The resonant centres are located at $\sigma = \pi/2$ and $\sigma = 3\pi/2$ (corresponding to $\varphi = \pi$), and the saddle points are located at $\sigma = 0$ and $\sigma = \pi$ (corresponding to $\varphi = 0$).}
\label{Fig3}
\end{figure*}

The phase portraits of the retrograde 1:2 resonance specified by $a_{\min} = 1.522563$ (or $\Gamma_2 = 1.85$) are shown in Fig. \ref{Fig3}, where the level curves passing through saddle points are marked in red lines. For the considered resonance, the isoline of $\Gamma_2 = 1.85$ has been marked in the bottom panel of Fig. \ref{Fig1} and the resonant angle is defined by $\sigma = \frac{1}{2} \left(2\theta_1 - \theta_2\right) = \frac{1}{2} \left(2\lambda - \lambda_p - \varpi \right) = \frac{1}{2}\varphi$. From Fig. \ref{Fig3}, it is observed that the resonant centres are located at $\sigma = \pi/2$ and $\sigma = 3\pi/2$ (corresponding to the usual argument at $\varphi = \pi$) and the saddle points are located at $\sigma = 0$ and $\sigma = \pi$ (corresponding to the usual argument at $\varphi = 0$). The distance between neighboring separatrices evaluated at the resonant centre also stands for the resonant width and, in particular, the resonant width in terms of $\Delta a$ and $\Delta e$ is indicated in the phase portraits.

It is known that the prograde 1:$n$ resonances hold asymmetric libration centres with the usual critical argument $\varphi (=k_{\max} \sigma)$ different from zero or $\pi$ \citep{beauge1994asymmetric, morbidelli2002modern}. The appearance of the asymmetric libration centres is because the second-order harmonics dominates the resonant Hamiltonian \citep{morbidelli2002modern}. However, there is no asymmetric libration centres in the phase portraits of the retrograde 1:2 resonance. The difference about the symmetric properties of libration centres between the prograde and retrograde 1:$n$ resonances has been noticed by \citet{li2020dynamics}.

From Figs \ref{Fig2} and \ref{Fig3}, we can also see that the phase portraits shown in the $(e \cos{\sigma},e \sin{\sigma})$ space are symmetric with respect to the lines corresponding to $\sin{\sigma} = 0$ and $\cos{\sigma} = 0$. These symmetric structures indicate that the resonant centres in the analytical model have the same dynamics. Thus, in the following discussions, we take one of them into consideration to discuss the dynamics.

\begin{figure*}
\centering
\includegraphics[width=0.48\textwidth]{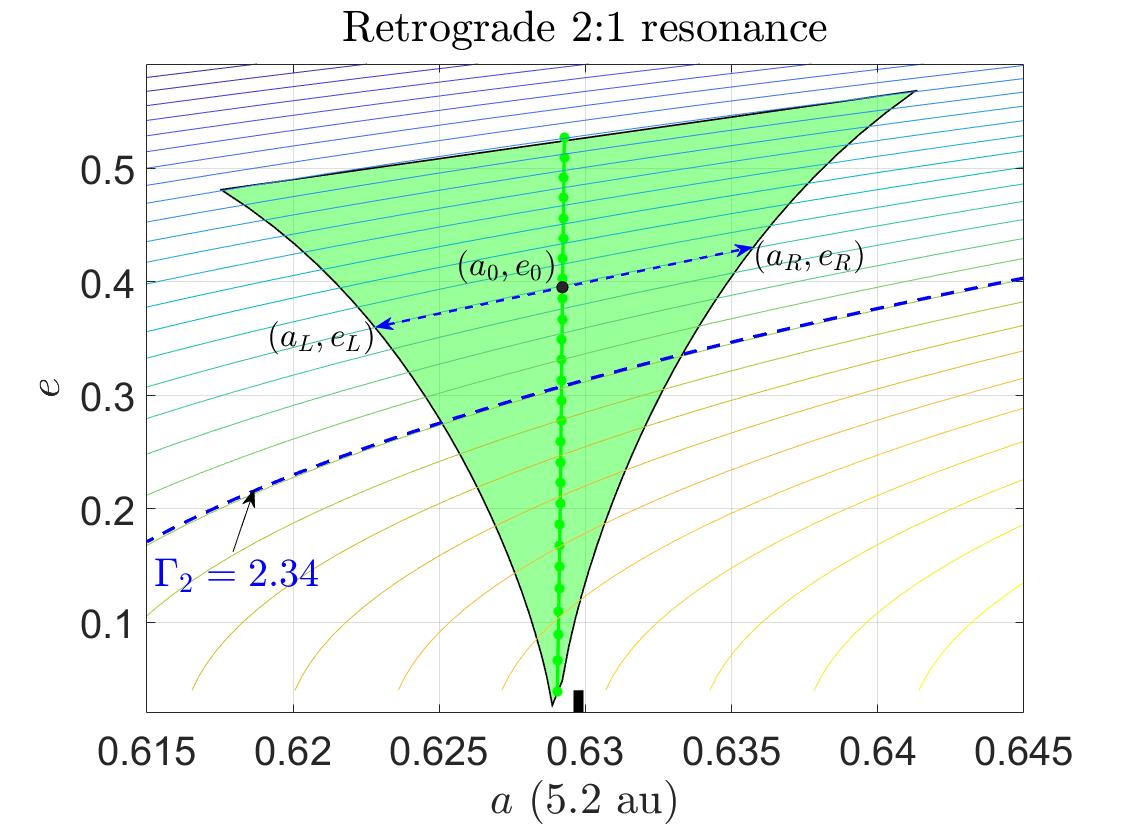}
\includegraphics[width=0.48\textwidth]{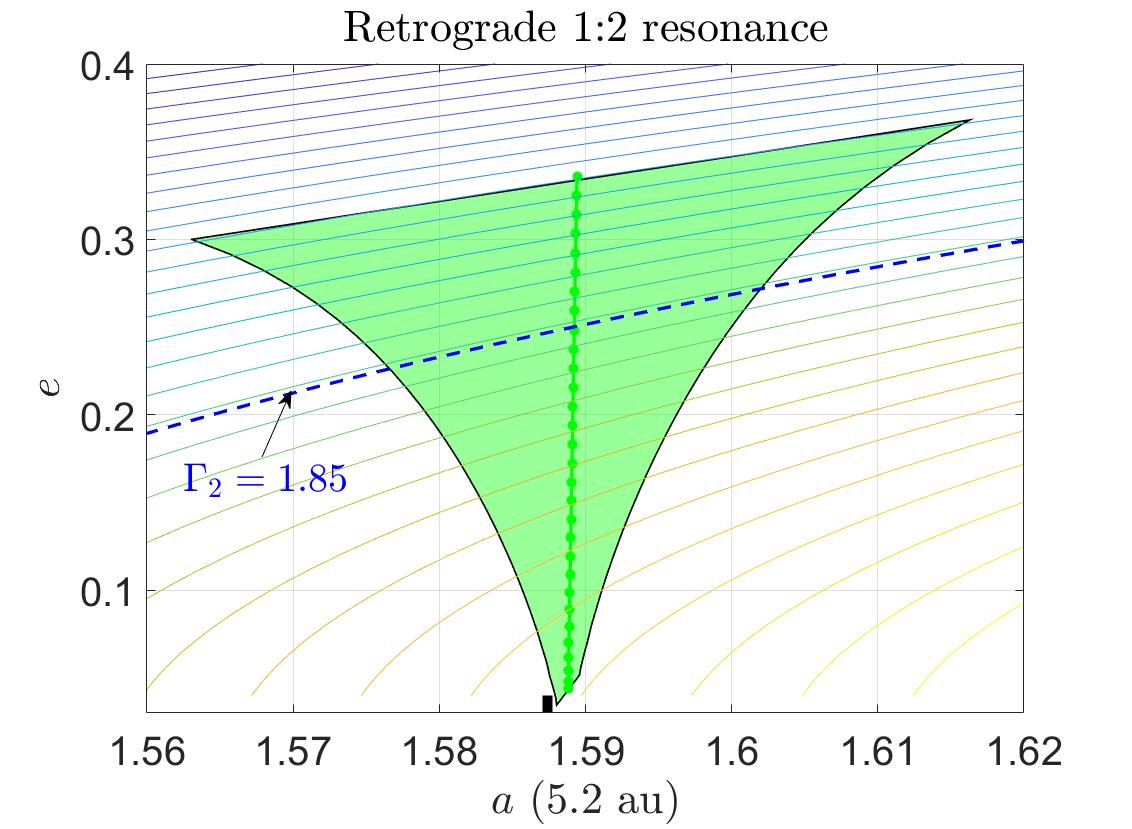}
\caption{Analytical results about the distributions of resonant centre $(a_0,e_0)$ and resonant width $(\Delta a = a_R - a_L, \Delta e = e_R - e_L)$ for the retrograde 2:1 resonance (\emph{left panel}) and for the retrograde 1:2 resonance (\emph{right panel}). For convenience, the level curves of the motion integral $\Gamma_2$ are plotted and, in particular, the isolines of $\Gamma_2 = 2.34$ and $\Gamma_2 = 1.85$ are marked in blue dotted lines. The red lines stand for the location of resonant centres, and the shaded areas represent the libration regions. Note that the resonant width is measured along the isoline of the motion integral $\Gamma_2$, as shown in the \emph{left panel}. The nominal location of resonance is marked by a short and black vertical line at the bottom of each figure.}
\label{Fig4}
\end{figure*}

According to the method presented in Section \ref{Sect4-1}, the location of resonant centre and the boundaries that separate the libration regions from circulation regions can be analytically determined. In Fig. \ref{Fig4}, the resonant centre and resonant width are reported in the $(a,e)$ space for the retrograde 2:1 resonance in the \emph{left panel} and for the retrograde 1:2 resonance in the \emph{right panel}. For convenience, in Fig. \ref{Fig4}, the level curves of the motion integral $\Gamma_2$ are also provided and the nominal location of resonance is marked by a short and black vertical line. The curve of $(a_0,e_0)$ stands for the location of resonant centre, the curve of $(a_L,e_L)$ for the left separatrix, and the curve of $(a_R,e_R)$ for the right separatrix. The shaded areas bounded by the left and right separatrices represent the libration regions in the $(a,e)$ space, and the resonant width is measured along the isoline of $\Gamma_2$. It is observed from Fig. \ref{Fig4} that, in both cases, (a) the left and right separatrices are not symmetric with respect to the nominal resonance location, (b) the curve of resonant centre cannot extend to the extremely low-eccentricity region, (c) the resonant centre deviates from the nominal location of resonance and (d) the resonant width ($\Delta a$ or $\Delta e$) is an increasing function of the eccentricity. In particular, for the retrograde inner 2:1 resonance, the resonant centres are on the left-hand side of the nominal location of resonance and, for the retrograde outer 1:2 resonance, the resonant centres are on the right-hand side (please refer to the short and vertical black lines in Fig. \ref{Fig4} for the nominal locations of resonance).

\section{Numerical study}
\label{Sect5}

To validate the analytical results (including the phase portraits and resonant width), in this section we turn to explore the dynamical structures of retrograde MMRs by analyzing Poincar\'e sections.

\subsection{Poincar\'e surfaces of section}
\label{Sect5-1}

As discussed in Section \ref{Sect2}, the full model represented by equation (\ref{Eq8}) is of two degrees of freedom with $\theta_1$ and $\theta_2$ as angular coordinates. For such a two-degree of freedom system, Poincar\'e surface of section is an efficient technique to explore its global dynamics in the phase space.

\begin{figure*}
\centering
\includegraphics[width=0.9\textwidth]{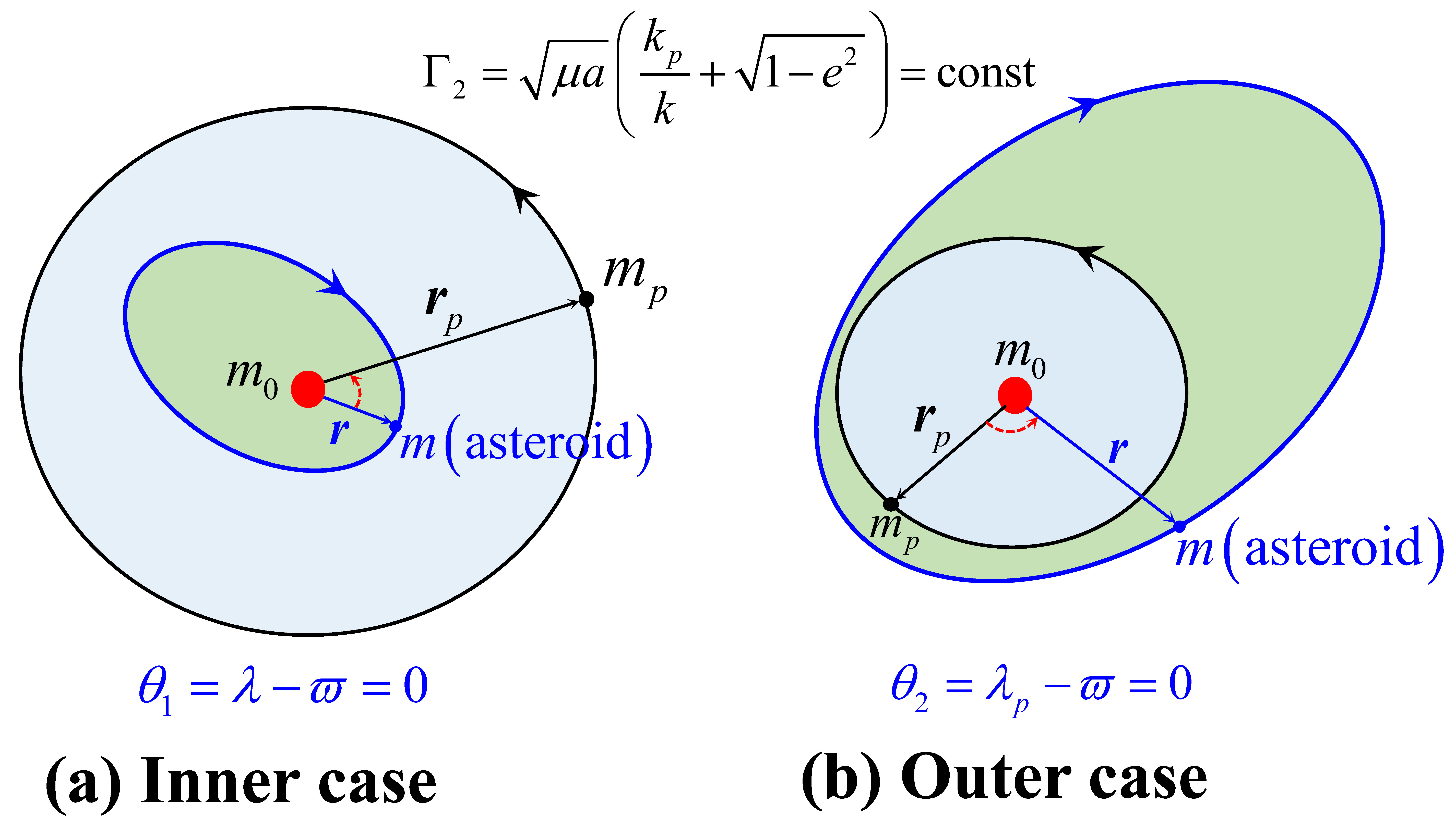}
\caption{Schematic diagrams for the definition of Poincar\'e sections for the inner and outer cases of retrograde MMRs with a given motion integral $\Gamma_2$. In both diagrams, the central body with mass $m_0$ stands for the Sun and the perturber with mass $m_p$ stands for the Jupiter-mass planet. It is known that, in the planar retrograde problem, there are two independent angular variables including $\theta_1 = \lambda - \varpi$ and $\theta_2 = \lambda_p - \varpi$. For the inner case, the angular variable $\theta_1 = \lambda - \varpi$ is faster than $\theta_2 = \lambda_p - \varpi$, so that the section is defined by $\theta_1 = \lambda - \varpi = 0$ (i.e., the particle is at its perihelion), where the angular separation of the perturber from the test particle's perihelion is ${\theta_2} = \lambda_p - \varpi$. For the outer case, the angular variable $\theta_2 = \lambda_p - \varpi$ is faster than $\theta_1 = \lambda - \varpi$, so that the section is defined by $\theta_2 = \lambda_p - \varpi = 0$ (i.e., the perturber is on the line directing from the central body towards the particle's perihelion), where the angular separation between the test particle and its perihelion is $\theta_1 = \lambda - \varpi$. It is noted that the definition in the inner case is similar to the one defined by \citet{malhotra2020divergence}, but the one in the outer case is different.}
\label{Fig5}
\end{figure*}

In this part, we adopt the numerical technique based on Poincar\'e sections to explore the dynamical structures of retrograde MMRs. In particular, we make some modifications for producing Poincar\'e sections in comparison to the traditional versions used by \citet{malhotra1996phase, winter1997resonanceI, winter1997resonanceII, morais2012stability, morais2013retrograde, wang2017mean, malhotra2018neptune} and \citet{malhotra2020divergence} in terms of the following two points.
\begin{itemize}
  \item A modified condition is adopted for computing Poincar\'e sections. It is known from the Hamiltonian given by equation (\ref{Eq7}) that, in the non-averaged model, the particles move in a four-dimensional phase space $(a,e,\theta_1,\theta_2)$, where $\theta_1 = \lambda - \varpi$ and $\theta_2 = \lambda_p - \varpi$. In general, for the inner resonances (i.e., $a < a_p$), the angle $\theta_1$ is faster than $\theta_2$ and, for those outer resonances (i.e., $a > a_p$), the angle $\theta_2$ is faster than $\theta_1$. Based on this, we produce Poincar\'e sections by recording those states of test particles when the short-period variable ($\theta_1$ for the inner case and $\theta_2$ for the outer case) is equal to zero, as illustrated in Fig. \ref{Fig5}. In particular, for the inner case, the section is defined by $\theta_1 = \lambda - \varpi = 0$, meaning that the test particle is located at its perihelion (see the \emph{left panel} of Fig. \ref{Fig5}) and, for the outer case, the section is defined by $\theta_2 = \lambda_p - \varpi = 0$, where the perturber is in the same direction of test particle's eccentricity vector (see the \emph{right panel} of Fig. \ref{Fig5}). It is noted that the choice of Poincar\'e section in the inner case is similar to that taken by \citet{wang2017mean, malhotra2018neptune} and \citet{malhotra2020divergence}, but the one in the outer case is different.
  \item The motion integral $\Gamma_2$ (rather than the approximate Jacobi constant) is taken as the conserved quantity. In the production of Poincar\'e sections adopted by \citet{wang2017mean, malhotra2018neptune} and \citet{malhotra2020divergence}, the approximate Jacobi constant, given by
      \begin{equation*}
      C_J \approx \frac{\mu}{a} + 2\sqrt{\mu a (1-e^2)},
      \end{equation*}
      is taken as the conserved quantity. In this work, in order to compare with the analytical developments presented in Section \ref{Sect3}, we adopt the motion integral, given by
      \begin{equation*}
      \Gamma_2 = \sqrt{\mu a} \left(\frac{k_p}{k} + \sqrt{1-e^2}\right),
      \end{equation*}
      as the conserved quantity in producing Poincar\'e sections. It is noted that the motion integral $\Gamma_2$ is an approximate constant in the non-averaged model and, in the vicinity of resonance, the periodic variation is on the order of the mass ratio between the perturber and the Sun. For a given motion integral $\Gamma_2$, the four-dimensional phase space $(a,e,\theta_1,\theta_2)$ is restrained to a three-dimensional space, i.e., $(a,\theta_1,\theta_2)$ or $(e,\theta_1,\theta_2)$. Furthermore, due to the choice of Poincar\'e section, the three-dimensional space is further reduced to a two-dimensional phase space, i.e., $(a,\theta_2)$ or $(e,\theta_2)$ for the inner case and $(a,\theta_1)$ or $(e,\theta_1)$ for the outer case. Consequently, the Poincar\'e sections can be shown in a two-dimensional parameter space.
\end{itemize}

For the inner case, we can further observe that, in magnitude, the angular separation $\theta_2 = \lambda_p - \varpi$ is equal to $\lambda_p - \lambda$ because it holds $\theta_1 = \lambda - \varpi = 0$ on the Poincar\'e sections. Similarly, for the outer case we can see that the angle $\theta_1 = \lambda - \varpi$ is equal to $\lambda - \lambda_p$ because it holds $\theta_2 = \lambda_p - \varpi = 0$ on the Poincar\'e sections. In other words, for both cases, the angular separation ($\theta_2$ for the outer case and $\theta_1$ for the inner case) on the Poincar\'e section stands for the synodic angle between the test particle and the perturber.

In the production of Poincar\'e sections, the equations of motion of the non-averaged model (i.e., the planar circular restricted three-body problem) are numerically integrated over a long enough time and the states are recorded at every time when test particles pass through the defined sections. The numerical integrator we used in this study is an eighth-order Runge--Kutta algorithm with step-size control at the seventh order \citep{fehlberg1968classical}, and the relative and absolute error tolerances are controlled to be smaller than $1.0 \times 10^{-12}$.

Next, let us consider the relationship between the resonant angle $\sigma$ defined in Section \ref{Sect3} and the angular separation $\theta_2$ (or $\theta_1$) for the inner (or outer) resonances. For the retrograde inner resonances (in this case, it holds $k_p > k$), the resonant angle is given by $\sigma = \frac{1}{k_p} \left(k \theta_1 - k_p \theta_2 \right)$, which is equal to $-\theta_2$ on the Poincar\'e section defined by $\theta_1 = 0$ (see the \emph{left panel} of Fig. \ref{Fig5}). Similarly, for the retrograde outer resonances (in this case, it holds $k_p < k$), the resonant angle $\sigma = \frac{1}{k} \left(k \theta_1 - k_p \theta_2 \right)$ is equal to $\theta_1$ on the Poincar\'e section defined by $\theta_2 = 0$ (see the \emph{right panel} of Fig. \ref{Fig5}). Thus, for a given motion integral $\Gamma_2$, it is possible for us to directly compare the phase portraits obtained in the previous section with the associated Poincar\'e sections for the retrograde $k_p$:$k$ resonance. This comparison is very helpful to understand the structures arising in the Poincar\'e sections and also to validate the analytical developments formulated in Sections \ref{Sect3} and \ref{Sect4}.

\begin{figure*}
\centering
\includegraphics[width=0.48\textwidth]{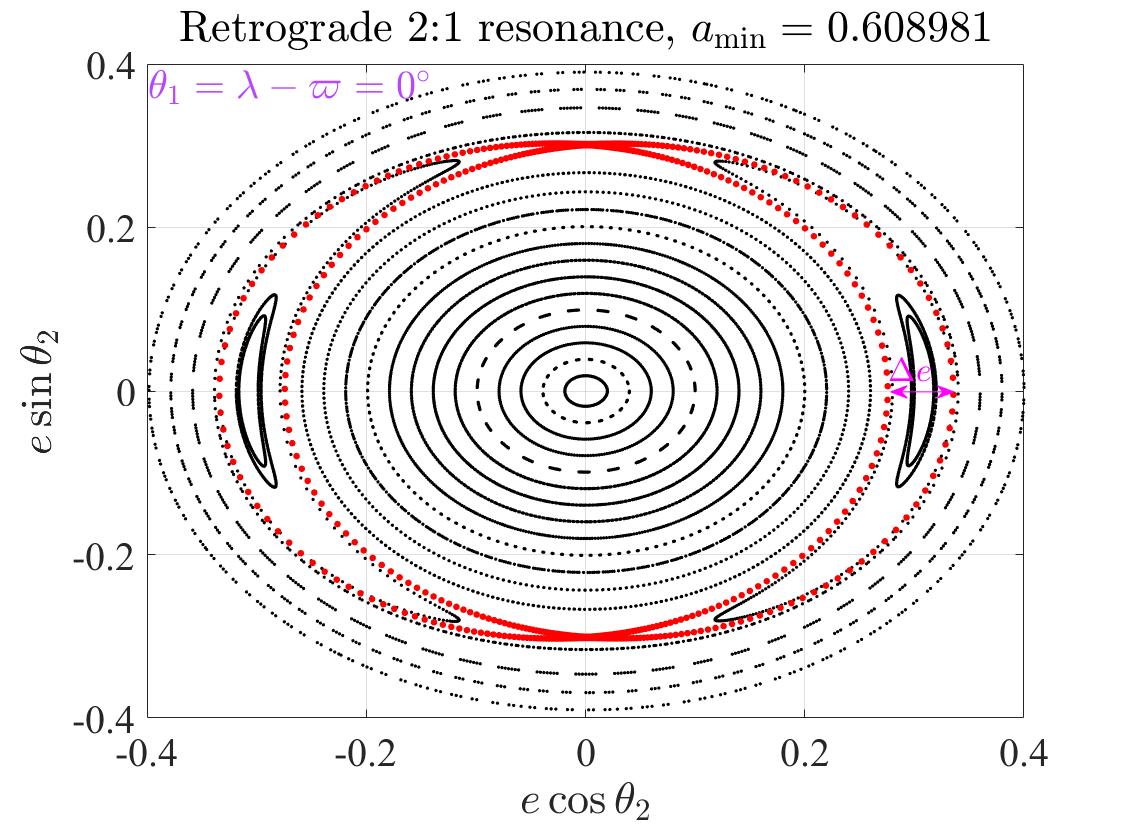}
\includegraphics[width=0.48\textwidth]{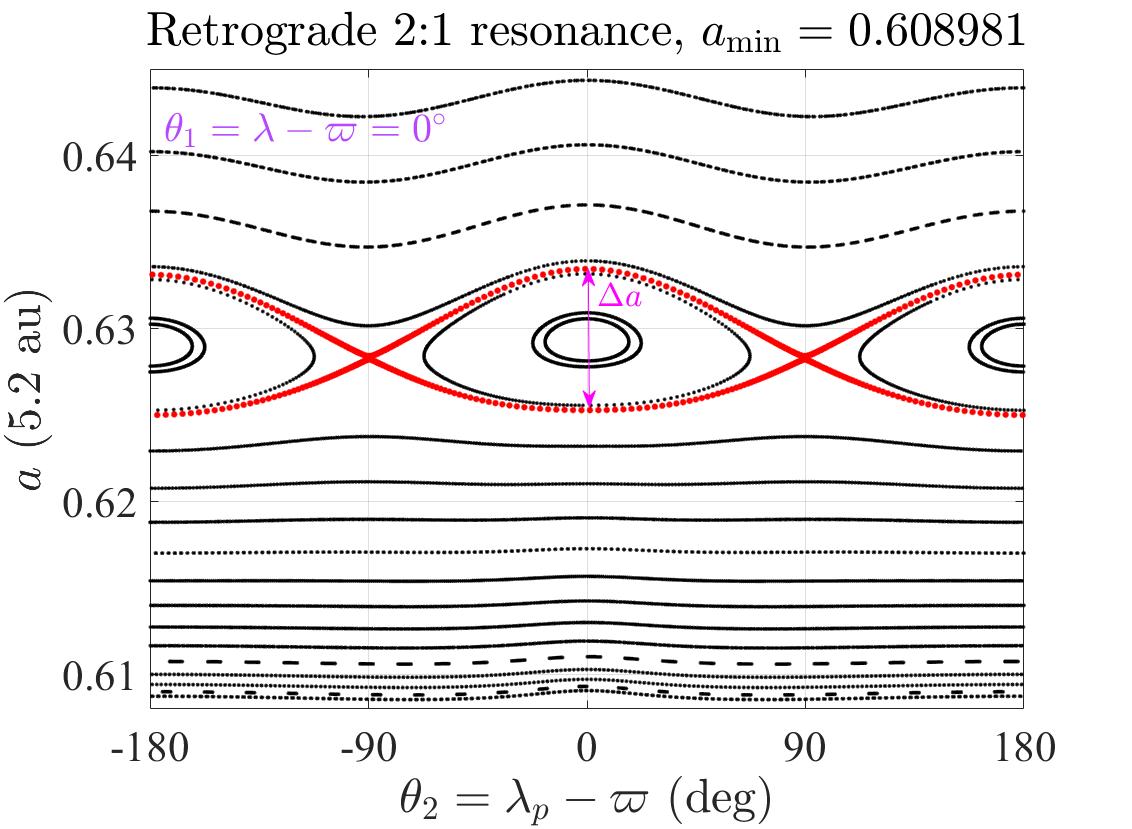}
\caption{Poincar\'e surfaces of section defined by $\theta_1 = \lambda  - \varpi  = 0$ for the retrograde 2:1 resonance specified by $a_{\min} = 0.608981$ (or $\Gamma_2 = 2.34$). At the Poincar\'e section, the critical argument $\sigma = \frac{1}{2} \left(\lambda - 2\lambda_p + \varpi \right) = \frac{1}{2} \left(\theta_1 - 2\theta_2\right)$ becomes $\sigma = \varpi - \lambda_p = - \theta_2$ (because of $\theta_1 = 0^{\circ}$ on the section), measuring the angular separation between the test particle and the perturber when the former is located at its perihelion (i.e., $\theta_1 = \lambda  - \varpi = 0$). In the sections, the curves separating libration regions from circulation regions are marked by red lines. The resonant width in terms of variations of semimajor axis and eccentricity ($\Delta a$ and $\Delta e$) is explicitly marked.}
\label{Fig6}
\end{figure*}

Figure \ref{Fig6} presents the Poincar\'e sections defined by $\theta_1 = \lambda - \varpi = 0$ in the $(e\cos{\theta_2}, e\sin{\theta_2})$ and $(\theta_2, a)$ spaces for the retrograde 2:1 resonance specified by $a_{\min} = 0.608981$ (or $\Gamma_2 = 2.34$). The motion integral adopted here is equal to that used in Fig. \ref{Fig2}. In the Poincar\'e sections, it is known that the scattered points stands for the chaotic motion and the smooth curves represent regular orbits in the full model. In particular, inside the island, the motions are of libration and, outside the island, the motions are of circulation. From Fig. \ref{Fig6}, it is observed that, in the considered phase space, all the orbits are regular (i.e., no chaotic motions are found in the sections) and there are two islands of libration centred at $\theta_2 = \pi/2$ and $\theta_2 = -\pi/2$ and two saddle points located at $\theta_2 = 0$ and $\theta_2 = \pi$. According to the relationship between the resonant angle $\sigma$ and the angular separation $\theta_2$, we have $\theta_2 = -\sigma$ on the sections defined by $\theta_1 = 0$. Thus, we can get that the islands arising in the Poincar\'e sections are centred at $\sigma = -\pi/2$ and $\sigma = \pi/2$ (corresponding to the usual critical argument at $\varphi = \pi$) and the saddle points are located at $\sigma = 0$ and $\sigma = \pi$ (corresponding to $\varphi = 0$). The red curves bounding the islands of libration play the role of dynamical separatrix, which divides the entire phase space into regions of libration and circulation. In particular, in the area bounded by the separatrix, the motions are of libration (i.e., they are quasi-periodic orbits in the full model). In the Poincar\'e sections, the size of libration island can be measured by the distance between two neighboring separatrices evaluated at the centre of libration island \citep{malhotra2020divergence}, as shown by $\Delta e$ in the \emph{left panel} and $\Delta a$ in the \emph{right panel}.

Comparing the Poincar\'e sections shown in Fig. \ref{Fig6} and the phase portraits shown in Fig. \ref{Fig2} (both of them have the same motion integral at $\Gamma_2 = 2.34$), we can observe an excellent correspondence between the structures appearing in the Poincar\'e sections and the ones arising in the phase portraits: (a) the resonant centres and saddle points in both the Poincar\'e sections and phase portraits are in perfect agreement, and (b) the resonant width, measuring the size of libration island, can be determined by analyzing both the Poincar\'e sections and phase portraits (they are called analytical and numerical widths of resonance, respectively).

\begin{figure*}
\centering
\includegraphics[width=0.48\textwidth]{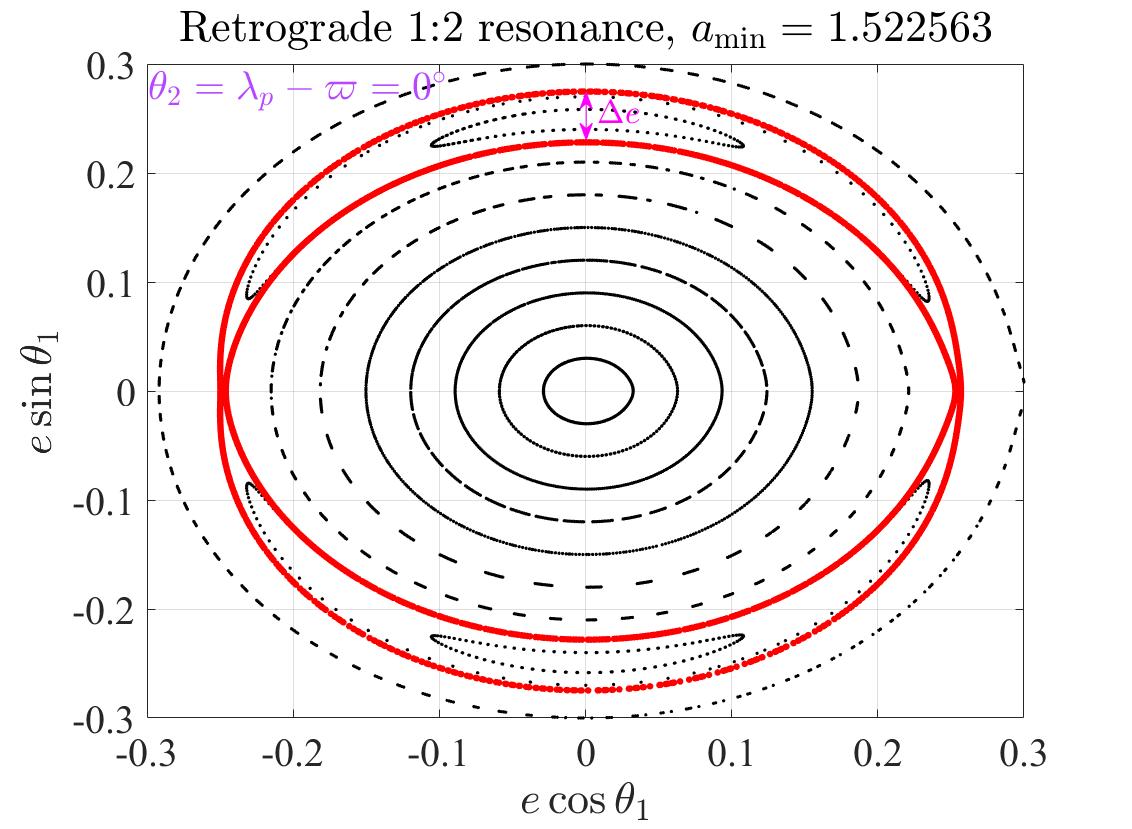}
\includegraphics[width=0.48\textwidth]{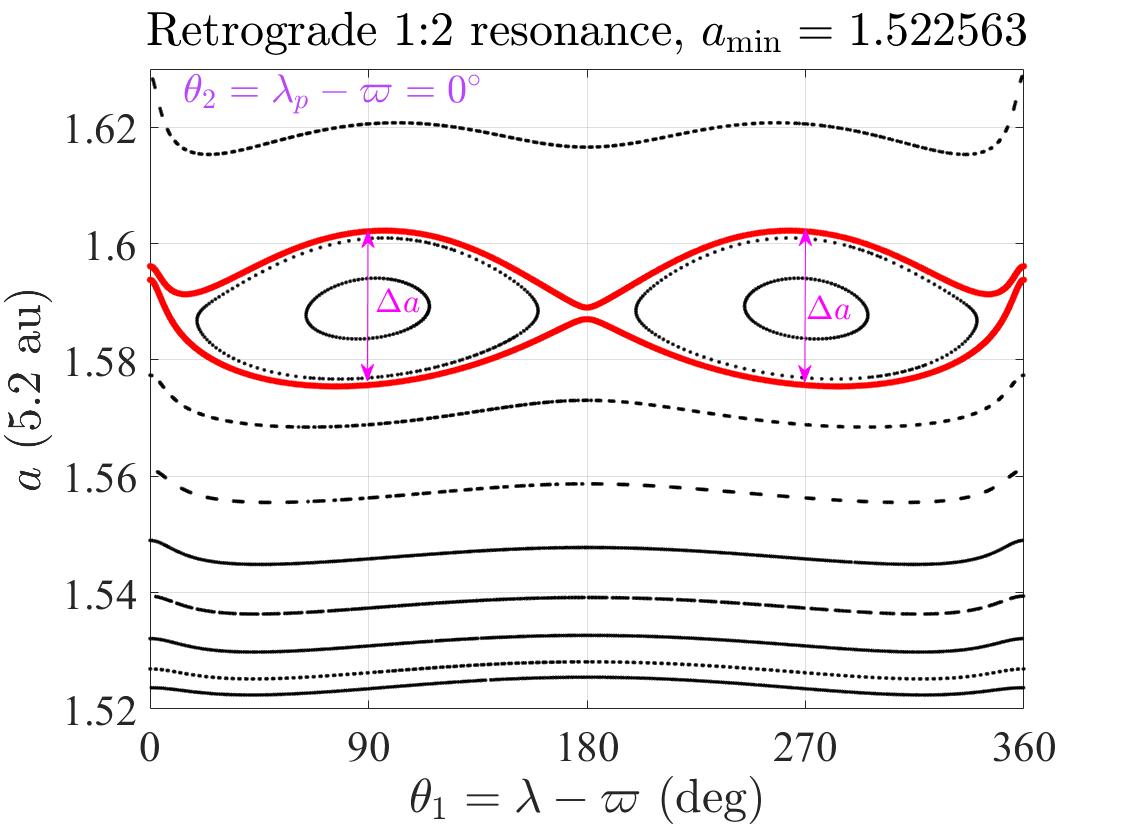}
\caption{Poincar\'e surfaces of section defined by $\theta_2 = \lambda_p  - \varpi  = 0$ for the retrograde 1:2 resonance specified by $a_{\min} = 1.522563$ (or $\Gamma_2 = 1.85$). Regarding the points on the Poincar\'e section, the critical argument $\sigma = \frac{1}{2} \left(2\lambda - \lambda_p - \varpi \right) = \frac{1}{2}\left(2\theta_1 -  \theta_2\right)$ becomes $\sigma = \lambda - \varpi = \theta_1$ (because of $\theta_2 = 0^{\circ}$ on the section), measuring the angular separation between the test particle and its perihelion when the perturber is in the same direction of the particle's perihelion (i.e., $\theta_2 = \lambda_p  - \varpi  = 0$). Similar to Fig. \ref{Fig6}, the curves separating the libration regions from circulation regions are marked by red lines, and the resonant width in terms of the variations of semimajor axis and eccentricity ($\Delta a$ and $\Delta e$) is marked.}
\label{Fig7}
\end{figure*}

Similarly, for the retrograde 1:2 resonance specified by $a_{\min} = 1.522563$ (or $\Gamma_2 = 1.85$), the Poincar\'e surfaces of section are reported in Fig. \ref{Fig7}. The motion integral adopted here is equal to that taken in Fig. \ref{Fig3}. For the considered retrograde outer resonance, we have $\sigma = \theta_1$ on the sections defined by $\theta_2 = 0$. Observing Fig. \ref{Fig7}, we can see that (a) the islands of libration are centred at $\theta_1 = \sigma = \pi/2$ and $\theta_1 = \sigma = 3\pi/2$ (corresponding to $\varphi = \pi$) and (b) the saddle points are located at $\theta_1 = \sigma = 0$ and $\theta_1 = \sigma = \pi$ (corresponding to $\varphi = 0$). The curves passing through the saddle points are marked in red lines and the size of libration island is measured by $\Delta e$ and $\Delta a$, as shown by the left and right panels, respectively. Comparing Fig. \ref{Fig7} with Fig. \ref{Fig3} leads us to the conclusion that, for the retrograde 1:2 resonance, the Poincar\'e sections and phase portraits are in good agreement in terms of the dynamical structures arising in the phase space.

\subsection{Numerical results}
\label{Sect5-2}

As discussed in the previous subsection, the dynamical separatrices can be found in the Poincar\'e sections. Thus, it is possible for us to evaluate their distance at the centre of libration island, which stands for the numerical width of resonance. As shown in Figs \ref{Fig6} and \ref{Fig7}, the width of resonance is described by the variations of semimajor axis and eccentricity, denoted by $\Delta a$ and $\Delta e$. Note that the determination of resonant width by analyzing Poincar\'e sections can be found in a series of publications, e.g. \citet{malhotra1996phase, winter1997resonanceI, wang2017mean, lan2019neptune, malhotra2020divergence}.

\begin{figure*}
\centering
\includegraphics[width=0.48\textwidth]{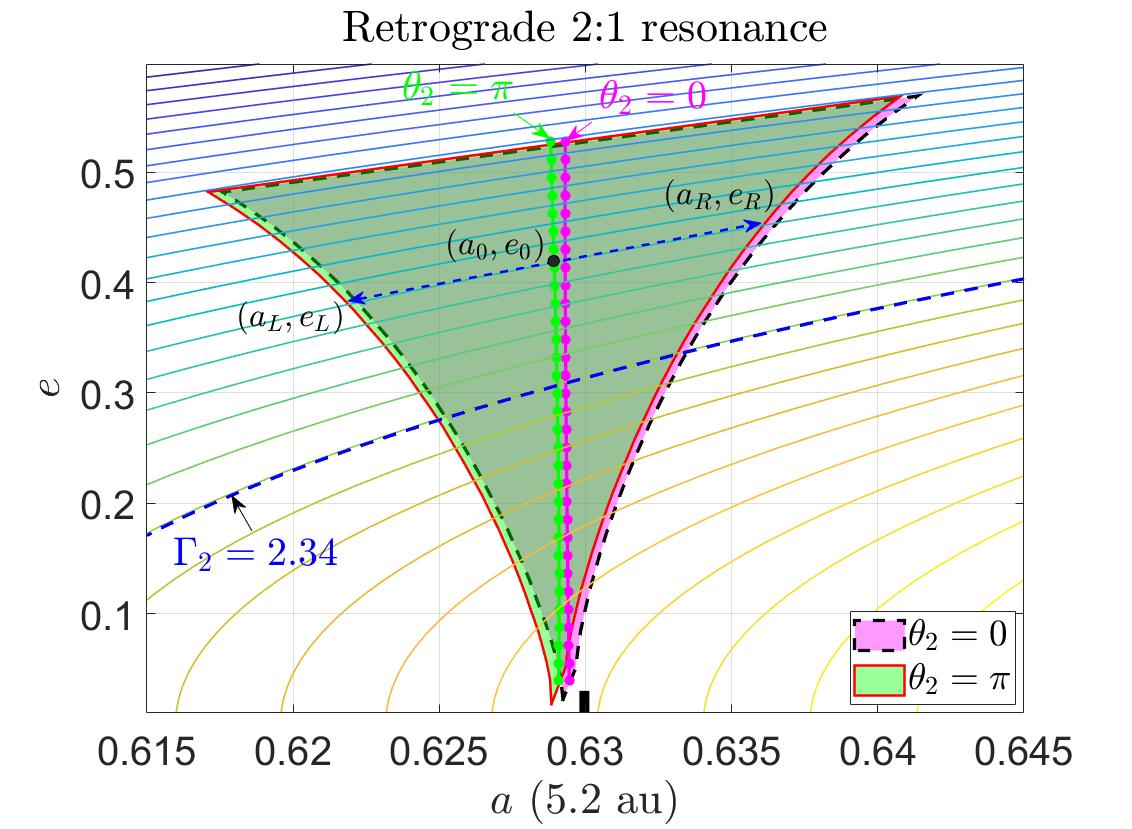}
\includegraphics[width=0.48\textwidth]{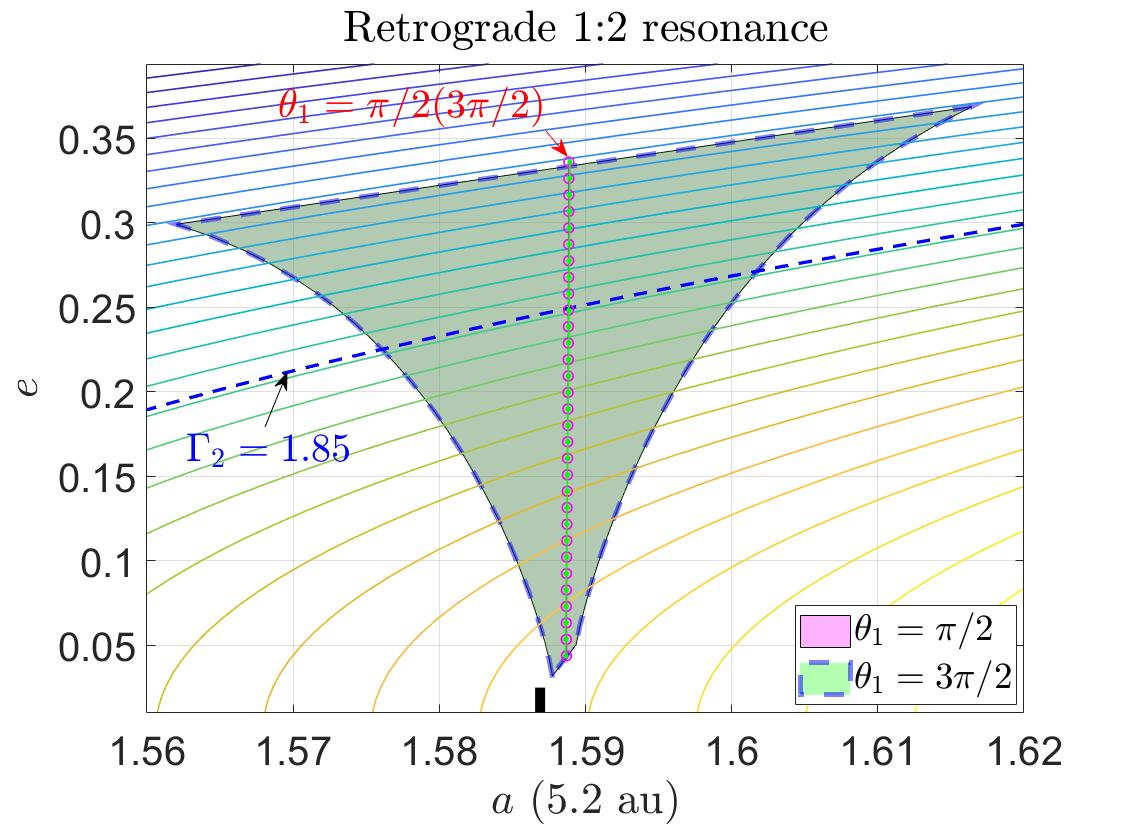}
\caption{Numerical results about the resonant centre $(a_0,e_0)$ and resonant width $(\Delta a = a_R - a_L, \Delta e = e_R - e_L)$ produced by analyzing the Poincar\'e sections for the retrograde 2:1 resonance (\emph{left panel}) and the retrograde 1:2 resonance (\emph{right panel}). For convenience, the level curves of the motion integral $\Gamma_2$ are still plotted and, in particular, the isolines of $\Gamma_2 = 2.34$ and $\Gamma_2 = 1.85$ are marked in blue dotted lines. The resonant width is measured along the isoline of $\Gamma_2$, as shown in the \emph{left panel}. For the retrograde 2:1 resonance, the width is evaluated at $\theta_2 = 0$ and $\theta_2 = \pi$ and, for the retrograde 1:2 resonance, the width is evaluated at $\theta_1 = \pi/2$ and $\theta_1 = 3\pi/2$. It is observed that, for the retrograde 1:2 resonance, the results evaluated at $\theta_1 = \pi/2$ and $\theta_1 = 3\pi/2$ are coincident (this is due to the symmetry between the islands of libration centred at $\theta_1 = \pi/2$ and $\theta_1 = 3\pi/2$). The nominal location of resonance is marked by a short and black vertical line.}
\label{Fig8}
\end{figure*}

For a given motion integral $\Gamma_2$, we can analyze the associated Poincar\'e sections to determine the location of libration centre, denoted by $(a_0,e_0)$, the point on the left boundary, denoted by $(a_L,e_L)$, and the point on the right boundary, denoted by $(a_R,e_R)$. When the motion integral varies in a certain interval, the curve of the libration centre as well as the points on the boundaries can be determined.

In Fig. \ref{Fig8}, the numerical width of resonance and distribution of libration centre are reported for the retrograde 2:1 resonance in the \emph{left panel} and for the retrograde 1:2 resonance in the \emph{right panel}. The shaded areas stand for the libration regions in the $(a,e)$ space. For convenience, the nominal resonance location is marked by a short and vertical black line and the level curves of the motion integral are also presented. Similar to the analytical results shown in Fig. \ref{Fig4}, the resonant width should also be measured along the isoline of the motion integral, i.e., $\Delta a = a_R - a_L$ and $\Delta e = e_R - e_L$, as shown in the \emph{left panel}. Observing Fig. \ref{Fig8}, we can see that (a) the location of libration centre is different from the nominal location of resonance, (b) the left and right separatrices are not symmetric with respect the nominal resonance location and (c) the resonant width is an increasing function of the eccentricity.

In particular, for the retrograde 2:1 resonance, the resonant centres are located on the left-hand side of the nominal resonance location and the resonant widths evaluated at $\theta_2 = 0$ and $\theta_2 = \pi$ have slight difference. The difference can be understood by the asymmetry between the island of libration centred at $\theta_2 = 0$ and the one at $\theta_2 = \pi$ (please see Fig. \ref{Fig6} for the asymmetric structures). For the retrograde 1:2 resonance, the resonant centres are placed on the right-hand side of the nominal location of resonance and it is found that the resonant widths evaluated at $\theta_1=\pi/2$ and $\theta_1=3\pi/2$ are perfectly coincident. This coincidence is due to the symmetry between the islands of libration centred at $\theta_1=\pi/2$ and $\theta_1=3\pi/2$ (please see Fig. \ref{Fig7} for the symmetric structures).

From now on, we denote the outcomes produced from the resonant model as analytical results and the ones obtained by analyzing Poincar\'e sections as numerical results.

\section{Comparison between analytical and numerical results}
\label{Sect6}

In this section, let us compare the analytical results obtained from the resonant model discussed in Sections \ref{Sect3} and \ref{Sect4} with the numerical results produced by analyzing Poincar\'e sections described in Section \ref{Sect5} in order to validate the analytical developments.

\begin{figure*}
\centering
\includegraphics[width=0.48\textwidth]{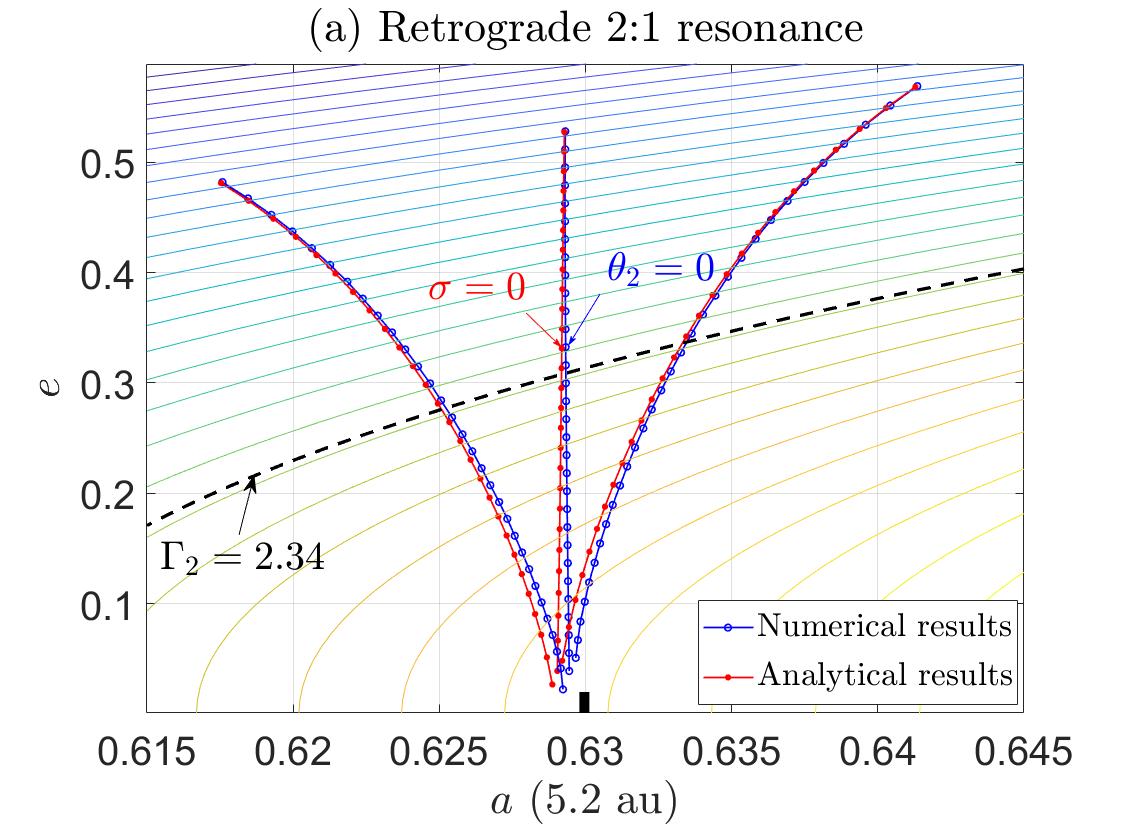}
\includegraphics[width=0.48\textwidth]{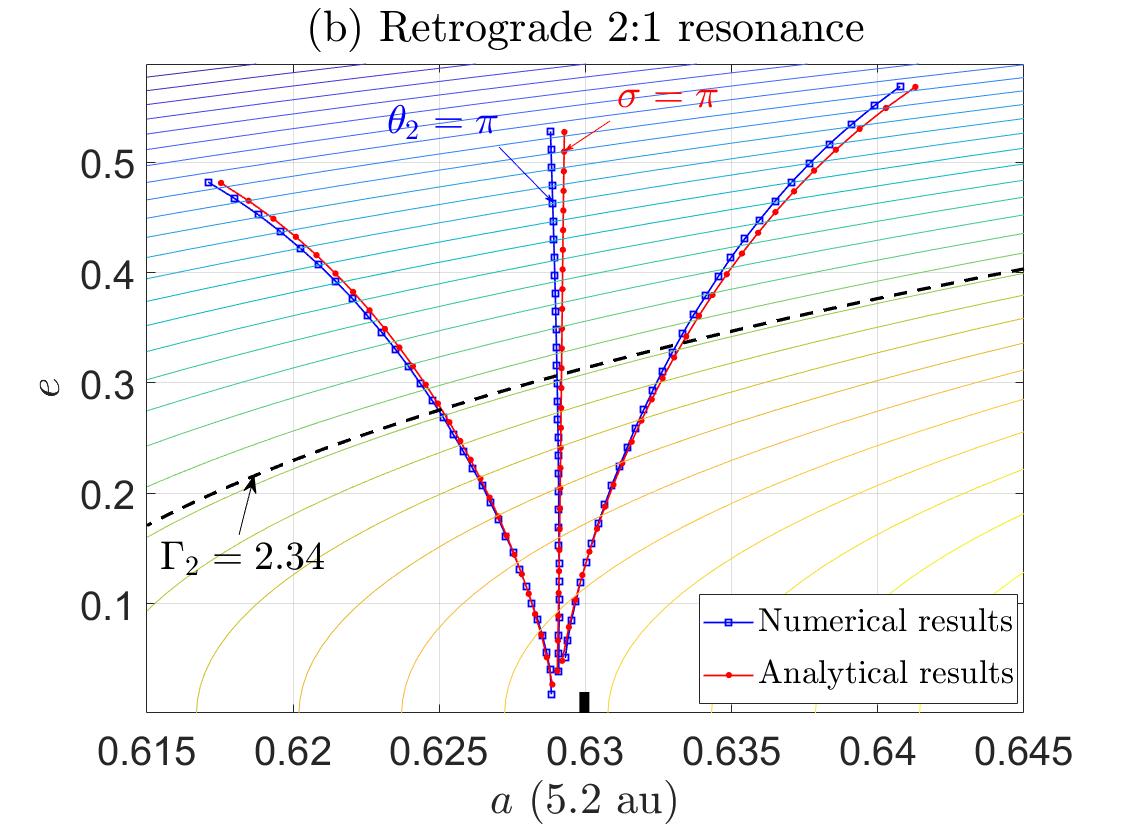}\\
\includegraphics[width=0.48\textwidth]{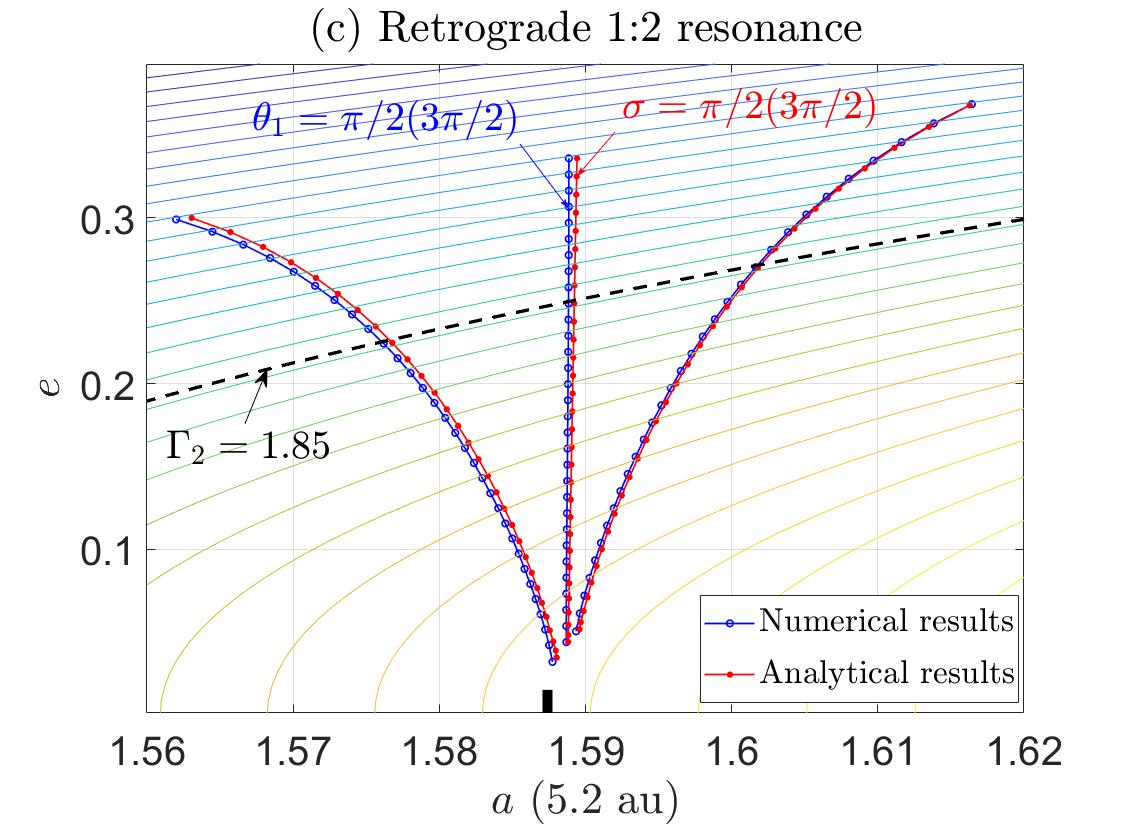}
\caption{Comparisons between the analytical and numerical results of the resonant centre and resonant width. Analytical results are obtained from the resonant Hamiltonian model, while the numerical results are produced by analyzing Poincar\'e sections. The analytical results are the same as the ones shown in Fig. \ref{Fig4} and the numerical results are the same as the ones given in Fig. \ref{Fig8}. In each panel, the nominal location of resonance is indicated by a short and black vertical line.}
\label{Fig9}
\end{figure*}

Figure \ref{Fig9} reports the analytical and numerical results in the $(a,e)$ space for the distribution of resonant centre and resonant width for the retrograde 2:1 resonance in the first two panels and for the retrograde 1:2 resonance in the last panel. The level curves of the motion integral $\Gamma_2$ are plotted and the resonant width should be evaluated along the isoline of $\Gamma_2$. Observing Fig. \ref{Fig9}, we can see that the analytical results are in quite good agreement with the numerical results in terms of the location of resonant centre and resonant width, indicating that our analytical model formulated in Section \ref{Sect3} is accurate to describe the dynamics of retrograde 2:1 and 1:2 resonances.

\begin{figure*}
\centering
\includegraphics[width=0.48\textwidth]{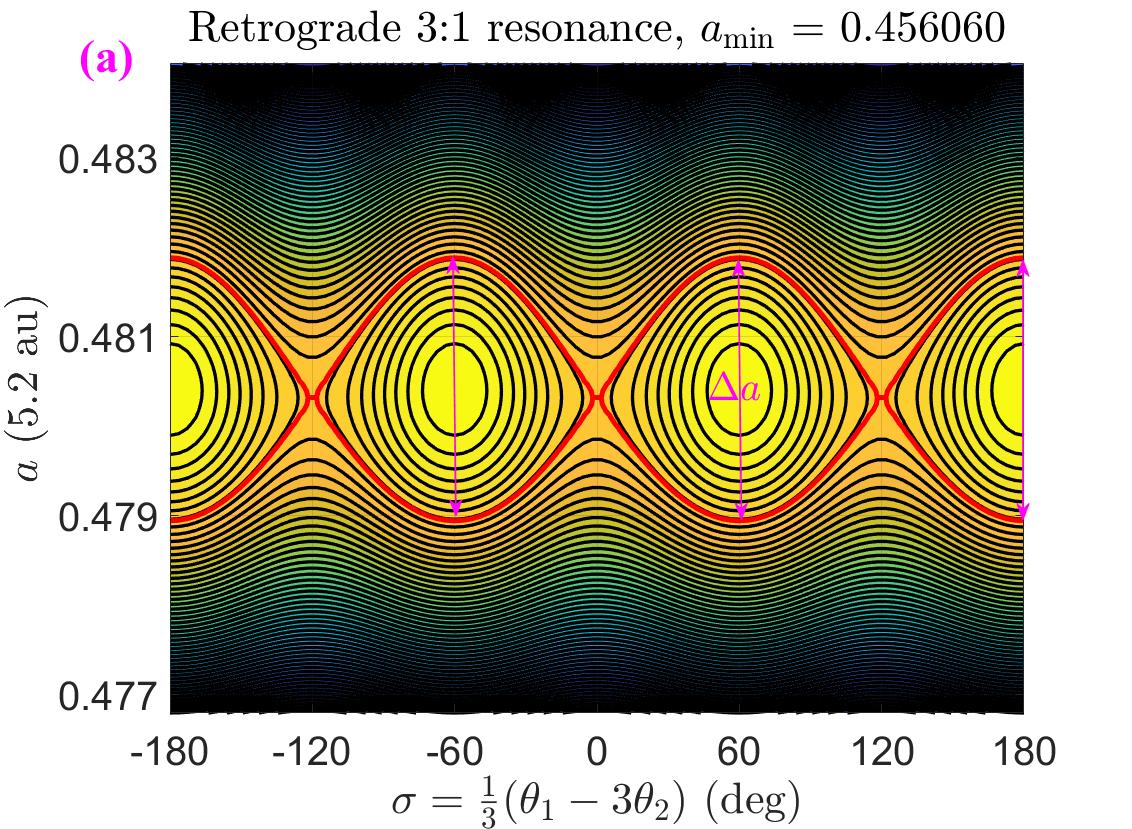}
\includegraphics[width=0.48\textwidth]{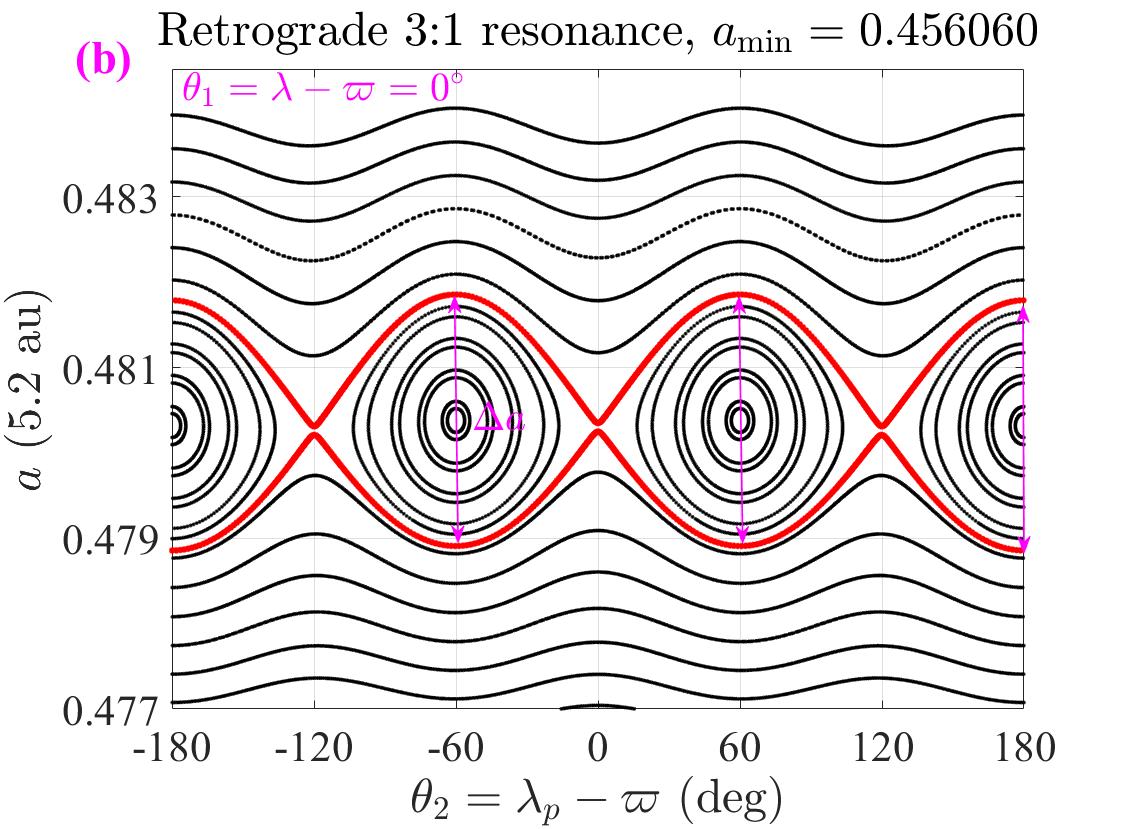}\\
\includegraphics[width=0.48\textwidth]{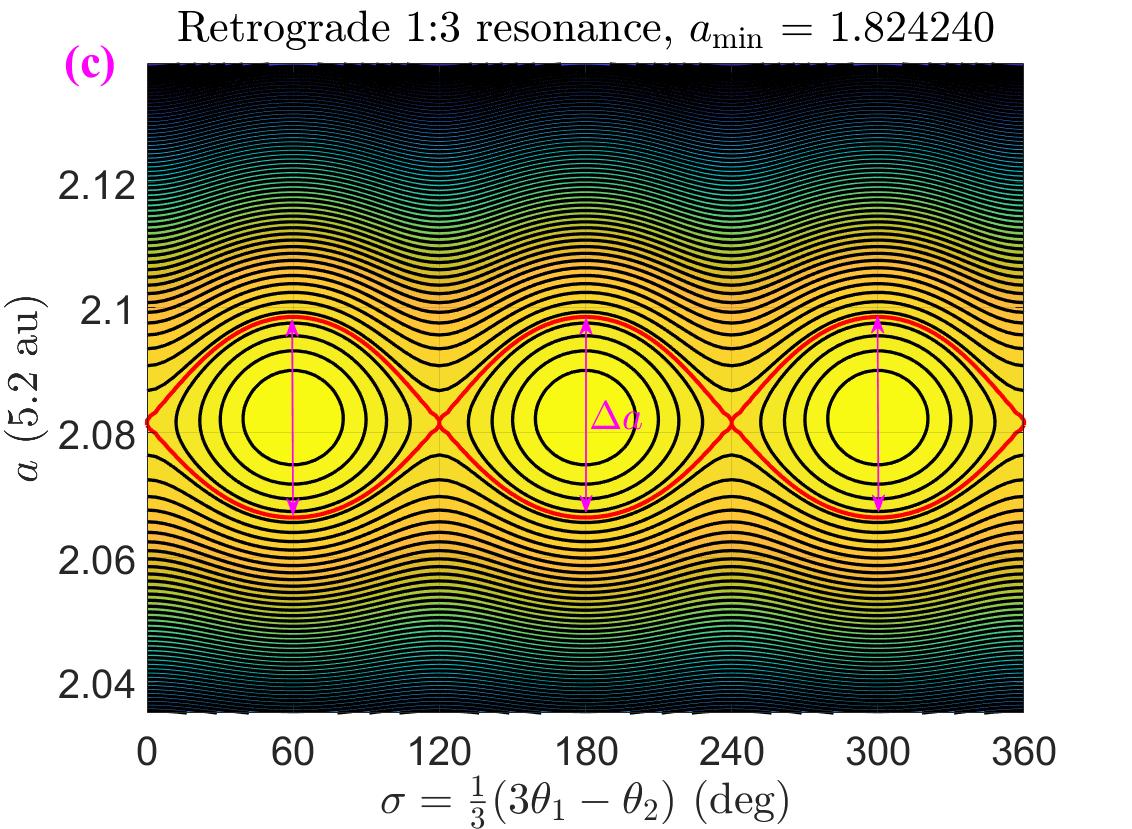}
\includegraphics[width=0.48\textwidth]{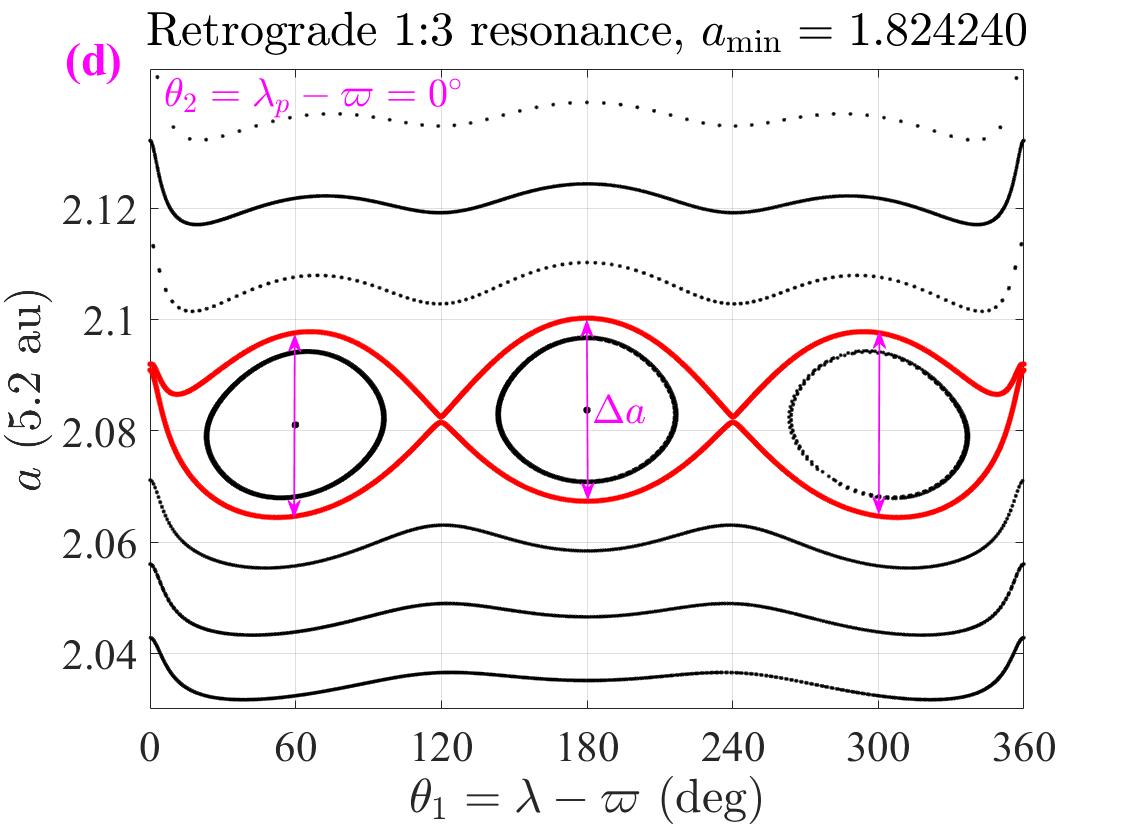}\\
\caption{Phase portraits (\emph{left panels}) as well as the Poincar\'e sections (\emph{right panels}) for the retrograde 3:1 and 1:3 resonances. For the retrograde 3:1 resonance, the resonant argument is $\sigma = \frac{1}{3} (\lambda - 3\lambda_p + 2\varpi) = \frac{1}{3}\left(\theta_1 - 3\theta_2\right)$, the motion integral is taken as $\Gamma_2 = 2.7$ (or $a_{\min} = 0.456060$ in normalized units) and the Poincar\'e section is defined by $\theta_1 = \lambda - \varpi = 0^{\circ}$ (on the sections it holds $\sigma = -\theta_2$). For the retrograde 1:3 resonance, the resonant argument is $\sigma = \frac{1}{3} (3\lambda - \lambda_p - 2\varpi) = \frac{1}{3}\left(3\theta_1 -  \theta_2\right)$, the motion integral is taken as $\Gamma_2 = 1.8$ (or $a_{\min} = 1.824240$ in normalized units) and the Poincar\'e section is defined by $\theta_2 = \lambda_p - \varpi = 0^{\circ}$ (on the sections it holds $\sigma = \theta_1$). The resonant width in terms of the variation of semimajor axis, $\Delta a$, is marked in both the phase portraits and Poincar\'e sections. It is interesting to observe that there is a perfect correspondence between the phase portraits and the associated Poincar\'e sections.}
\label{Fig10}
\end{figure*}

Next, let us apply both the analytical and numerical approaches discussed before to the retrograde 3:1 and 1:3 resonances. In Fig. \ref{Fig10}, their phase portraits and Poincar\'e sections with the same motion integral (or $a_{\min}$) are reported. The \emph{left panels} are for the phase portraits and the right ones are for the Poincar\'e sections. The curves passing through the saddle points are shown in red lines, and the distance $\Delta a$ stands for the resonant width in terms of the variation of semimajor axis.

For the retrograde 3:1 resonance, the resonant angle is $\sigma = \frac{1}{3}(\theta_1 - 3\theta_2)$, which becomes $\sigma = -\theta_2$ on the sections defined by $\theta_1 = \lambda - \varpi = 0$. Observing panels (a) and (b) of Fig. \ref{Fig10} (the motion integral is $\Gamma_2 = 2.7$ or $a_{\min} = 0.456060$), we can see that there are similar structures arising in the phase portrait and in the Poincar\'e section: (i) the resonant centres are located at $\sigma = \pm \pi/3, \pm \pi$ (corresponding to $\varphi = \pi$) in the phase portrait or, equivalently, at $\theta_2 = \pm \pi/3, \pm \pi$ on the Poincar\'e section, (ii) the saddle points are located at $\sigma = 0, \pm \pi$ (corresponding to $\varphi = 0$) in the phase portrait or, equivalently, at $\theta_2 = 0, \pm \pi$ on the Poincar\'e section, (iii) in both plots the red curves passing through saddle points play the role of dynamical separatrices, dividing the entire phase space into regions of libration and circulation, and (iv) the distance between the nearby separatrices evaluated at the resonant centre stands for the resonant width, denoted by $\Delta a$, measuring the size of libration region.

For the retrograde 1:3 resonance, the resonant angle is $\sigma = \frac{1}{3}(3\theta_1 - \theta_2)$, which is equal to $\sigma = \theta_1$ on the section defined by $\theta_2 = \lambda_p - \varpi = 0$. Similarly, comparing panels (c) and (d) of Fig. \ref{Fig10} (the motion integral is $\Gamma_2 = 1.8$ or $a_{\min} = 1.824240$), we can observe similar structures in both the phase portrait and the Poincar\'e section: (i) the resonant centres are located at $\sigma = \pi/3, \pi, 5\pi/3$ (corresponding to $\varphi = \pi$) in the phase portrait or, equivalently, $\theta_1 = \pi/3, \pi, 5\pi/3$ on the section, and (ii) the saddle points are placed at $\sigma = 0, 2\pi/3, 4\pi/3$ (corresponding to $\varphi = 0$) in the phase portrait or, equivalently, $\theta_1 = 0, 2\pi/3, 4\pi/3$ on the section.

\begin{figure*}
\centering
\includegraphics[width=0.48\textwidth]{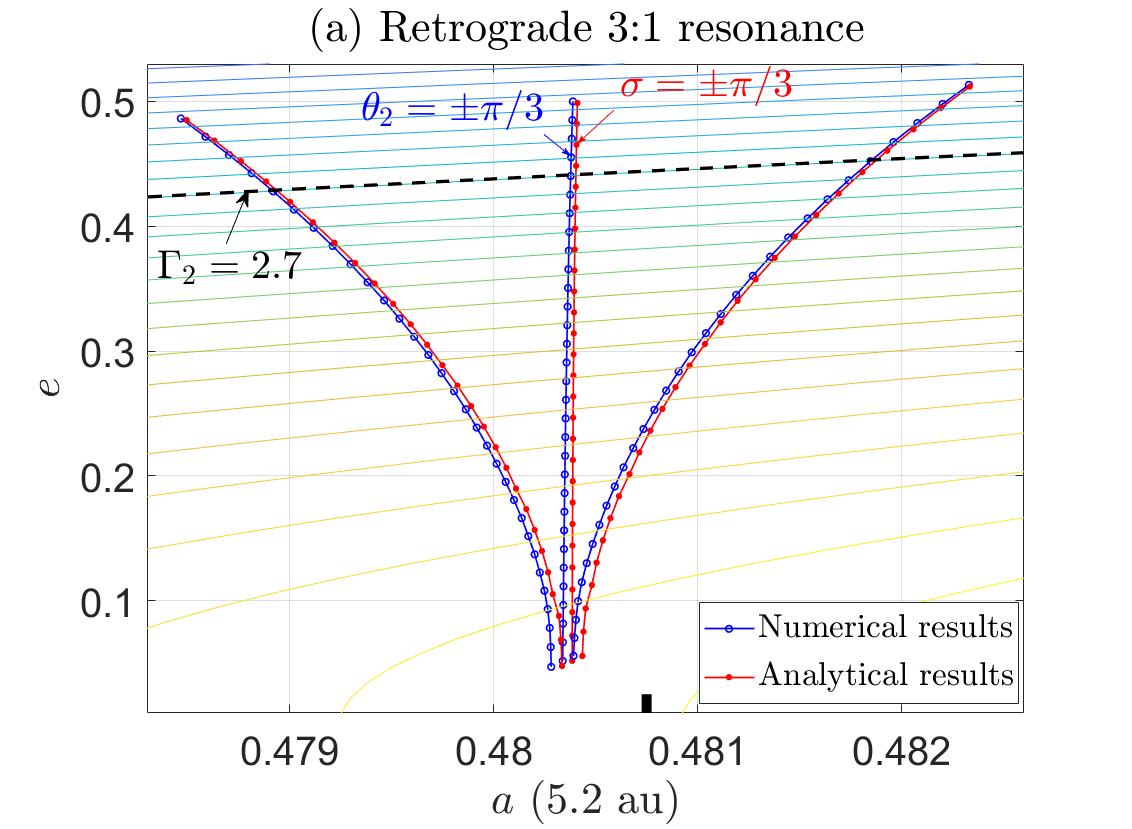}
\includegraphics[width=0.48\textwidth]{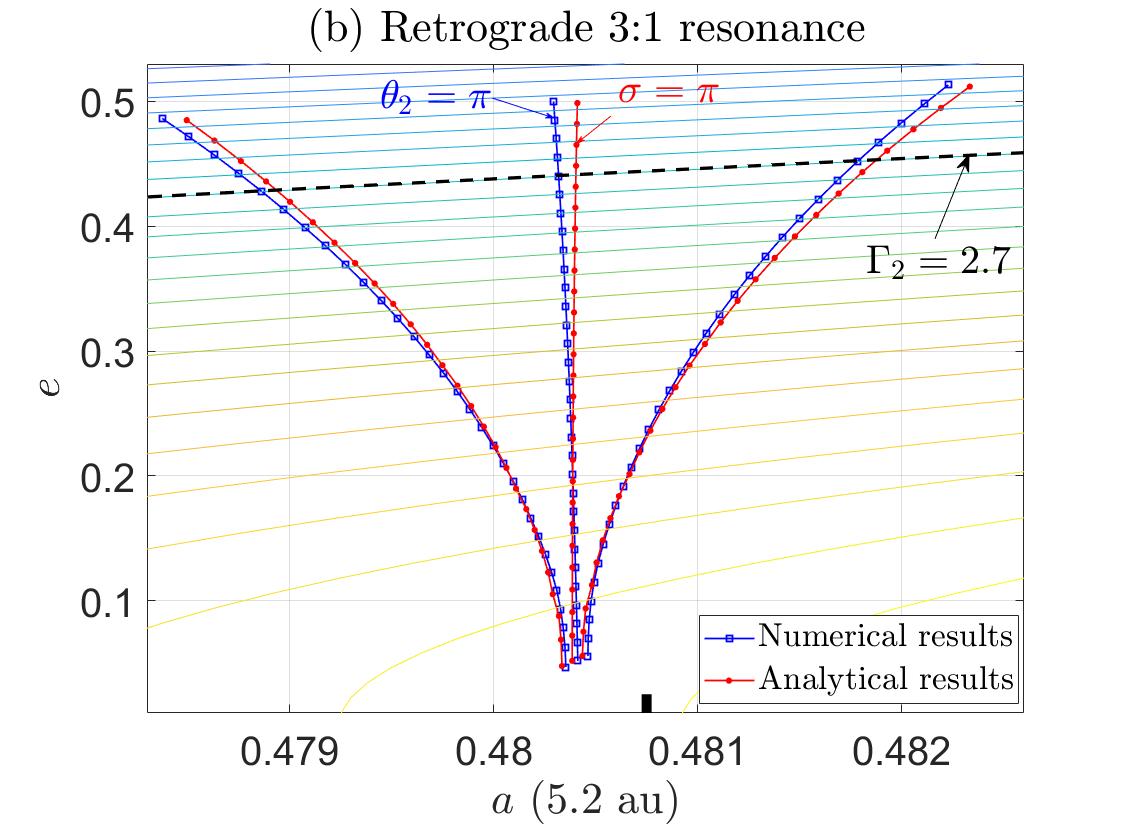}\\
\includegraphics[width=0.48\textwidth]{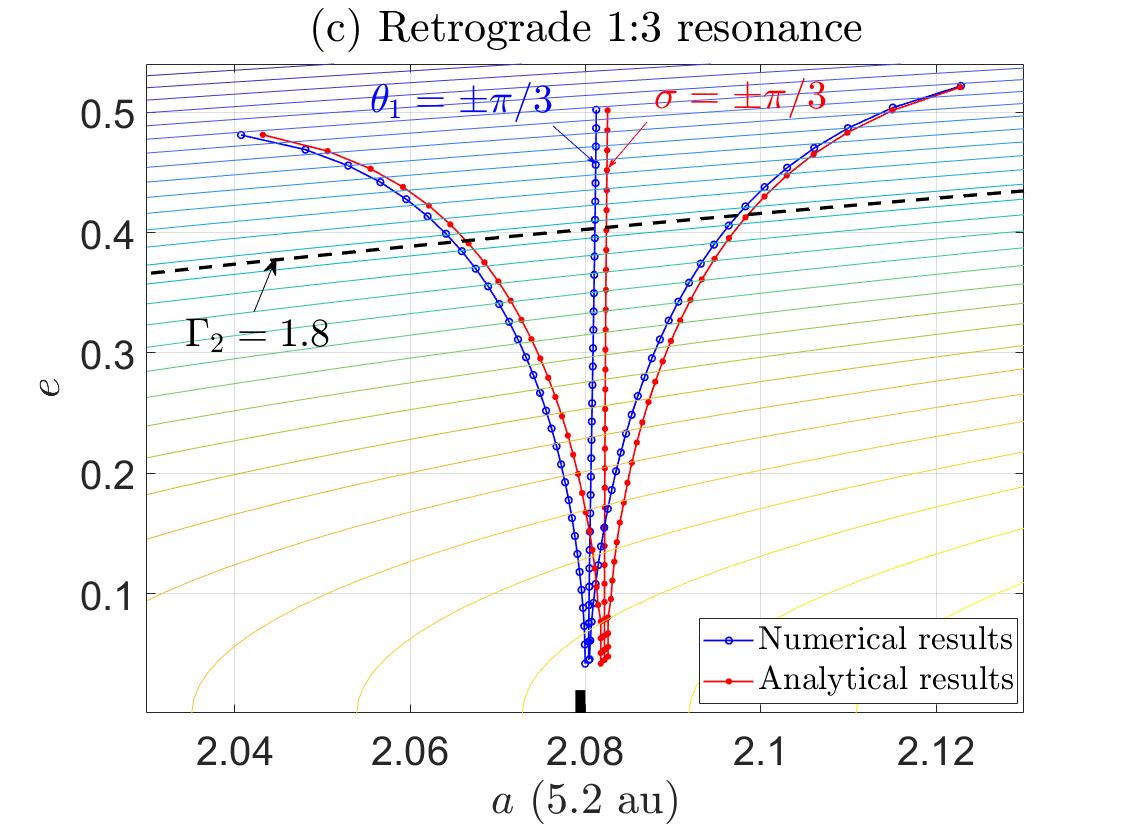}
\includegraphics[width=0.48\textwidth]{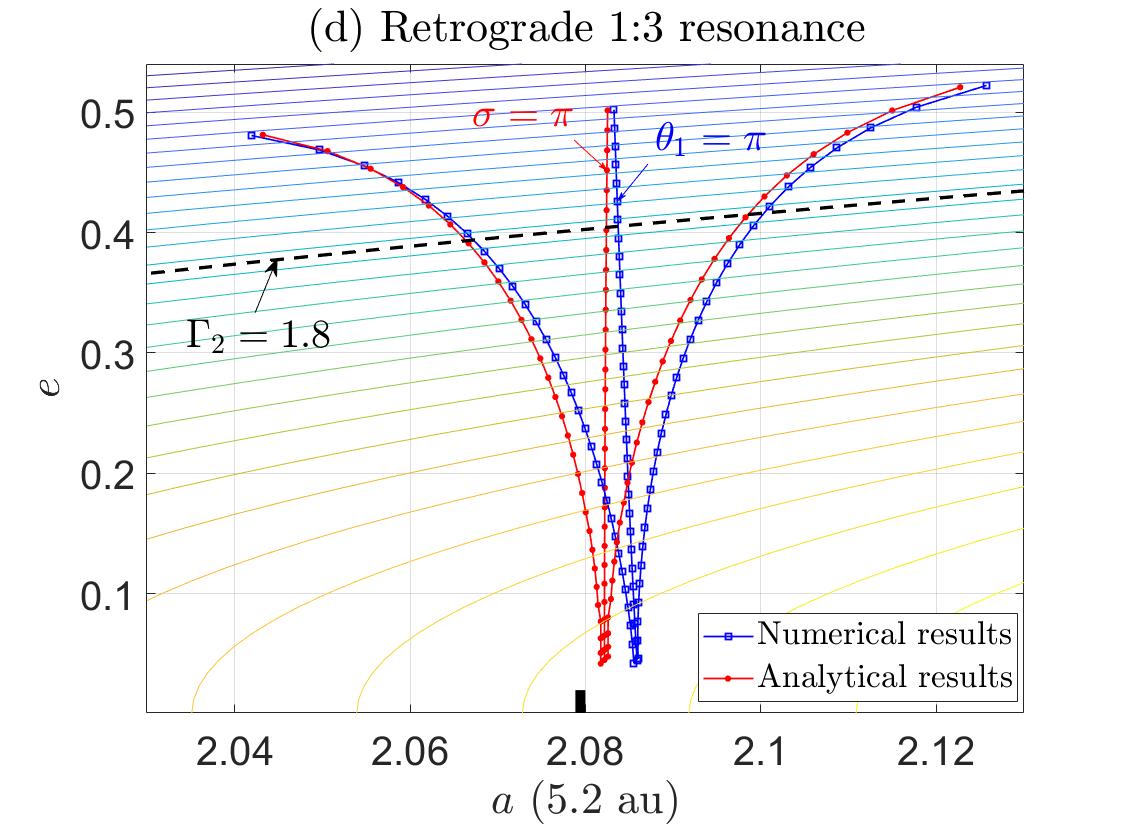}
\caption{Comparisons between the analytical and numerical results in terms of the resonant centre and resonant width. The \emph{upper panels} are for the retrograde 3:1 resonance, and the \emph{bottom panels} are for the retrograde 1:3 resonance. For convenience, the level curves of $\Gamma_2$ are plotted and, in particular, the isolines of $\Gamma_2 = 2.7$ and $\Gamma_2 = 1.8$ are marked (the associated phase portraits and Poincar\'e sections are provided in Fig. \ref{Fig10}). For the retrograde 3:1 resonance, the numerical width of resonance is evaluated at $\theta_2 = \pm \pi/3$ (panel `a') and $\theta_2 = \pi$ (panel `b') and, for the retrograde 1:3 resonance, the numerical width of resonance is evaluated at $\theta_1 = \pm \pi/3$ (panel `c') and $\theta_1 = \pi$ (panel `d'). The nominal location of resonance is indicated by a short and black vertical line.}
\label{Fig11}
\end{figure*}

Figure \ref{Fig11} reports the analytical and numerical results in terms of the distribution of resonant centre and resonant width in the \emph{left panel} for the retrograde 3:1 resonance and in the \emph{right panel} for the retrograde 1:3 resonance. For convenience, the level curves of the motion integral $\Gamma_2$ are plotted in both panels. It is noted that the numerical widths of resonance are evaluated at $\theta_2 = \pm \pi/3, \pi$ for the retrograde 3:1 resonance (see panels `a' and `b' of Fig. \ref{Fig11}) and at $\theta_1 = \pm \pi/3, \pi$ for the retrograde 1:3 resonance (see panels `c' and `d' of Fig. \ref{Fig11}). For the analytical width of resonance, the results have no change when the widths are evaluated at different centres (due to the symmetry of phase portraits). However, the numerical width of resonance will shift slightly if it is evaluated at a different resonant centre, because the islands of libration in the Poincar\'e section are not exactly symmetric (please refer to the Poincar\'e sections shown in Fig. \ref{Fig10} for detailed structures).

According to Fig. \ref{Fig11}, we can observe a good agreement between the analytical and numerical results: (a) both the analytical and numerical widths of resonance ($\Delta a$ or $\Delta e$) increase with the eccentricity, and (b) the location of resonant centre is different from the nominal location. In particular, it is observed from Fig. \ref{Fig11} that the resonant centres are located on the left-hand (right-hand) side of the nominal location of resonance for the retrograde 3:1 (1:3) resonance and their deviation is dependent on the eccentricity. It is not difficult to understand that the slight deviation between analytical and numerical results is caused by the `short-term' effects filtered in the analytical model.

In summary, the comparisons made in Fig. \ref{Fig9} for the retrograde 2:1 and 1:2 resonances and in Figs \ref{Fig10} and \ref{Fig11} for the retrograde 3:1 and 1:3 resonances show that the analytical results could match well with the numerical results in terms of the location of resonant centre and resonant width, indicating that our analytical model is accurate and applicable in predicting the dynamics of retrograde MMRs.

\section{Numerical widths over the full range of eccentricity}
\label{Sect7}

In previous sections, we concentrate on the dynamics of retrograde MMRs with eccentricities smaller than that of the planet-crossing orbit (i.e., $e<e_c$) considering the convergence of series expansion of disturbing function as well as the availability of perturbation theory. In addition, the analytical approach based on series expansion cannot deal with the retrograde co-orbital resonance due to the divergence of Laplace coefficients with $\alpha$ close to unity \citep{murray1999solar}. However, we know that the non-perturbative technique based on computing Poincar\'e sections is not limited by the planet-crossing condition and the co-orbital condition \citep{wang2017mean, malhotra2018neptune, lan2019neptune}, thus it is possible to explore the dynamics of retrograde MMRs (including the interior, co-orbital and exterior resonances) over the entire range of eccentricity $e \in (0,1)$ in order to provide global pictures in the phase space. To this end, we apply the numerical approach described in Section \ref{Sect5} to the retrograde 2:1, 1:1, 1:2, 1:3 and 1:4 resonances.

According to the traditional notations \citep{malhotra2020divergence}, the resonant centres with the usual critical argument at $\varphi = 0$ belong to the pericentric branch and the ones at $\varphi = \pi$ belong to the apocentric branch. From equation (\ref{Eq11}), we have $\varphi = k \theta_1 - k_p \theta_2$. Thus, on the Poincar\'e sections defined by $\theta_1 = 0$ (inner and co-orbital resonances), the resonance centres at $\theta_2 = 0$ (or $\theta_2 = \pm 2\pi/k_p$) belong to the pericentric branch and the ones at $\theta_2 = \pm \pi/k_p$ belong to the apocentric branch. On the sections defined by $\theta_2 = 0$ (outer resonances), the resonance centres at $\theta_1 = 0$ (or $\theta_1 = \pm 2\pi/k$) belong to the pericentric branch and the ones at $\theta_1 = \pm \pi/k$ belong to the apocentric branch.

\begin{figure*}
\centering
\includegraphics[width=0.48\textwidth]{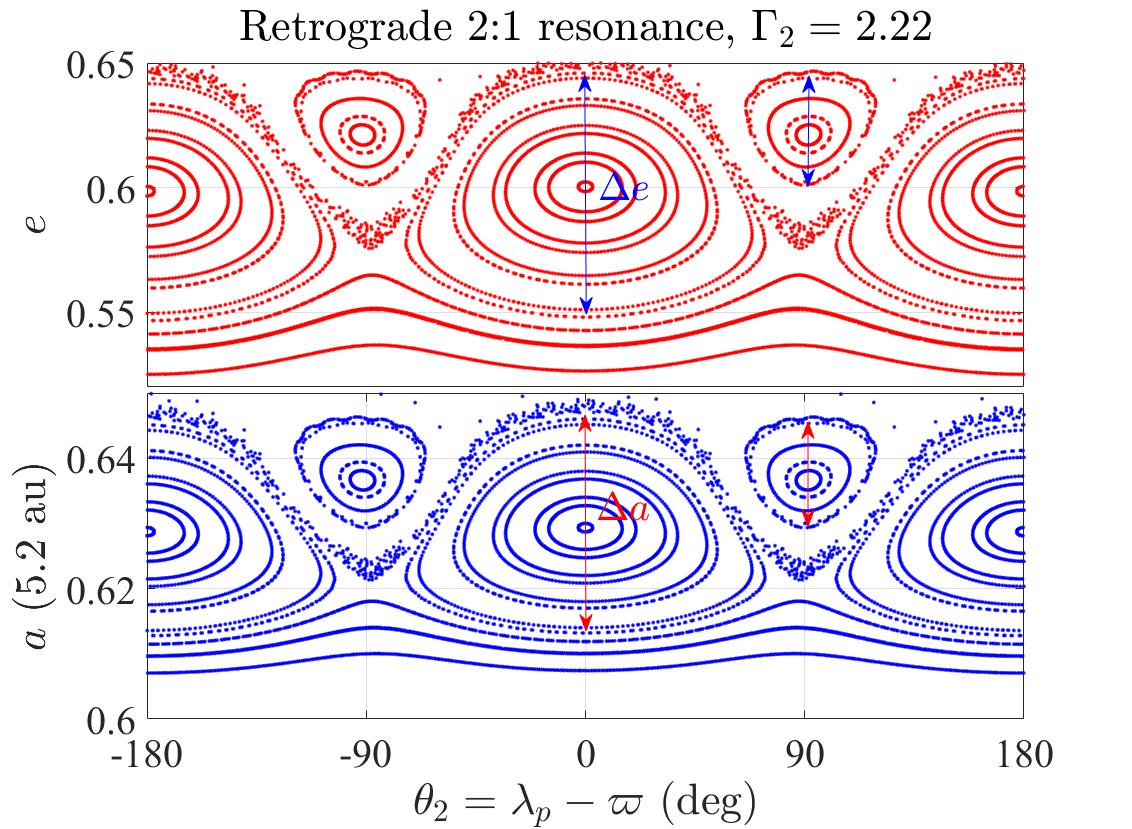}
\includegraphics[width=0.48\textwidth]{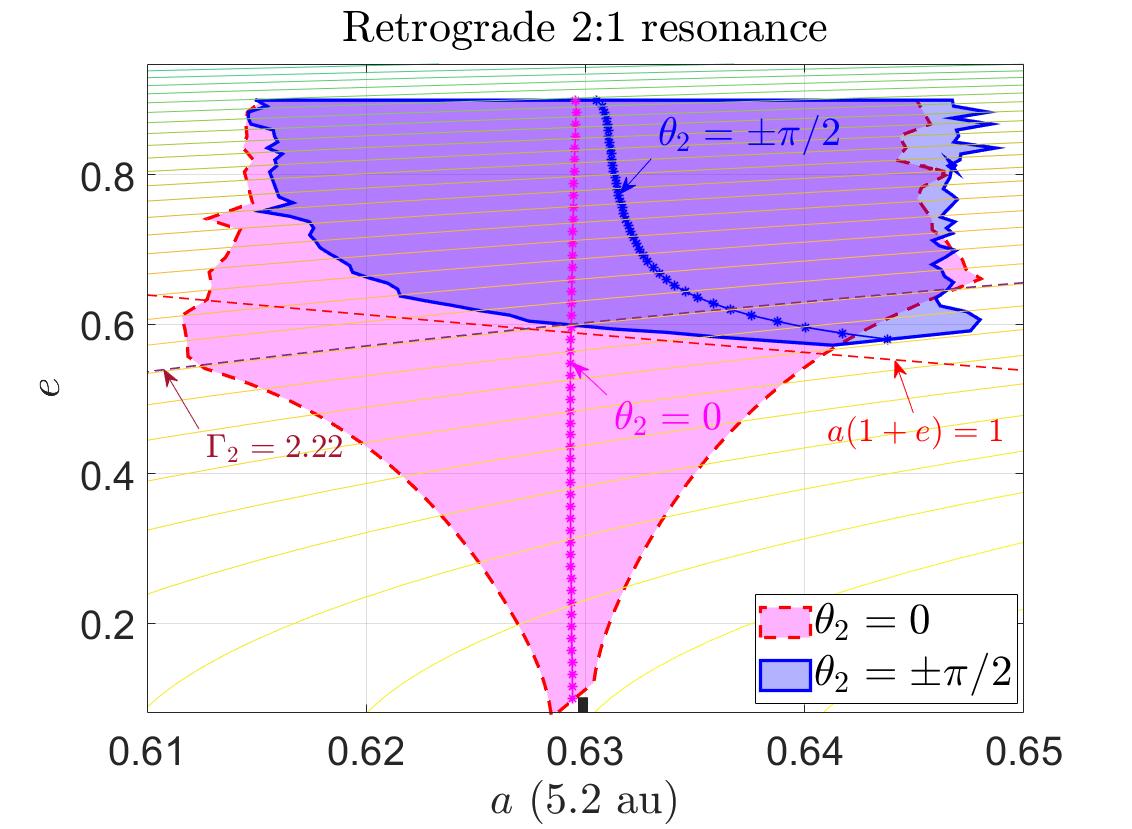}
\caption{Poincar\'e sections of the retrograde 2:1 resonance specified by $\Gamma_2 = 2.22$ (\emph{left panel}) and the numerical widths of resonance over the full range of eccentricity (\emph{right panel}). The numerical widths are evaluated at $\theta_2 = 0$ and $\theta_2 = \pm \pi/2$. In the \emph{right panel}, the level curves of the motion integral $\Gamma_2$ are presented, and the isoline of $\Gamma_2 = 2.22$ and the planet-crossing line corresponding to $a(1+e)=1$ are also plotted. The nominal location of resonance is indicated by a short and vertical black line.}
\label{Fig12}
\end{figure*}

In the \emph{left panel} of Fig. \ref{Fig12}, the Poincar\'e sections for the retrograde 2:1 resonance specified by $\Gamma_2 = 2.22$ are presented in the $(\theta_2,e)$ and $(\theta_2,a)$ spaces. According to the Poincar\'e sections, it is observed that the islands of resonance are centred at $\theta_2 = 0$, $\theta_2 = \pm \pi/2$ and $\theta_2 = \pm \pi$ and those libration islands centred at $\theta_2 = \pm \pi/2$ (or $\theta_2 = \pm \pi$) are symmetric with respect to the line of $\theta_2 = 0$. In addition, the eccentricity ($e$$\sim$0.6) is close to that of planet-crossing orbit, leading to strong perturbation, and thus chaotic layers appear between neighboring islands. According to the Poincar\'e sections, we can see that chaotic layers replace the role of dynamical separatrices. By analyzing Poincar\'e sections, we can numerically identify the resonant widths by evaluating the lower and upper boundaries inside which resonance can always take place. In particular, the resonant width is characterized by the variations of semimajor axis and eccentricity ($\Delta a$ and $\Delta e$), as shown in the \emph{left panel} of Fig. \ref{Fig12}.

The numerical widths of resonance over the full range of eccentricity are reported in the \emph{right panel} of Fig. \ref{Fig12}. It should be noted that the numerical widths are evaluated at $\theta_2=0$ and $\theta_2 = \pm \pi/2$ (the width evaluated at $\theta_2 = \pi$ has a similar behavior to the one evaluated at $\theta_2 = 0$, thus it is not considered here). For convenience, the isoline of $\Gamma_2 = 2.22$ and the planet-crossing critical line corresponding to $a(1+e)=1$ are provided. In the following discussions, we use $e_c (a) = \frac{1}{a} - 1$ to stand for the critical eccentricity of planet-crossing orbit. In the region above the line of $e_c (a)$, close encounter (or even collision) with planet may occur and, in the region below the critical line, no collision can happen. In addition, the nominal resonance location at $a=a_0$ is marked by a short and vertical black line (for the retrograde 2:1 resonance, it holds $a_0 = 0.62976016$). From the \emph{right panel} of Fig. \ref{Fig12}, it is observed that (a) the pericentric libration (with $\theta_2 = 0$) occurs over the entire range of eccentricity while the apocentric libration (with $\theta_2 = \pm \pi/2$) appears only in the region above the line of $e_c (a)$, (b) the resonant centres with $\theta_2 = \pm \pi/2$ (apocentric libration) are located on the right hand of the nominal resonance location (i.e., $a=a_0$) and they diverge away from $a = a_0$ when the eccentricity decreases close to $e_c (a)$, (c) the resonant width evaluated at $\theta_2 = \pm \pi/2$ shrinks as the eccentricity approaches $e_c (a)$, (d) the resonant centres with $\theta_2 = 0$ are located close to $a = a_0$ but on the left-hand side of it, and (e) the resonant width evaluated at $\theta_2 = 0$ reaches the maximum in the vicinity of $e_c (a)$.

\begin{figure*}
\centering
\includegraphics[width=0.48\textwidth]{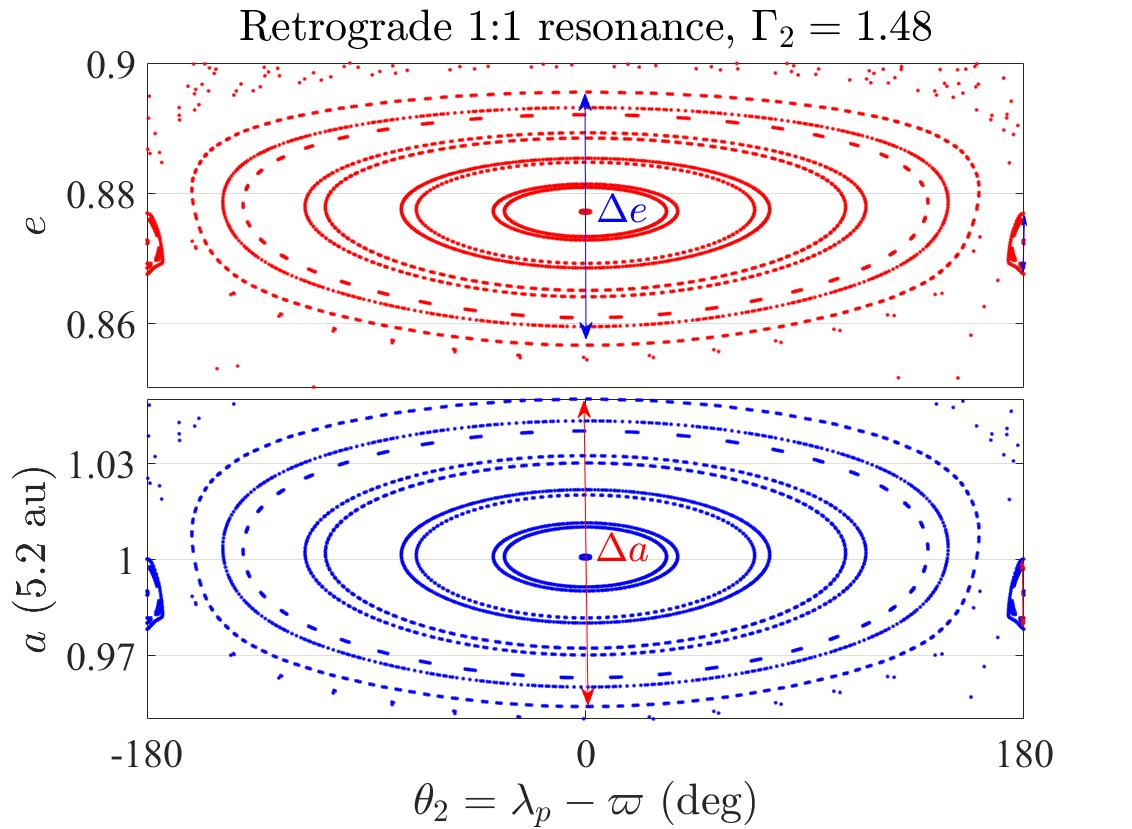}
\includegraphics[width=0.48\textwidth]{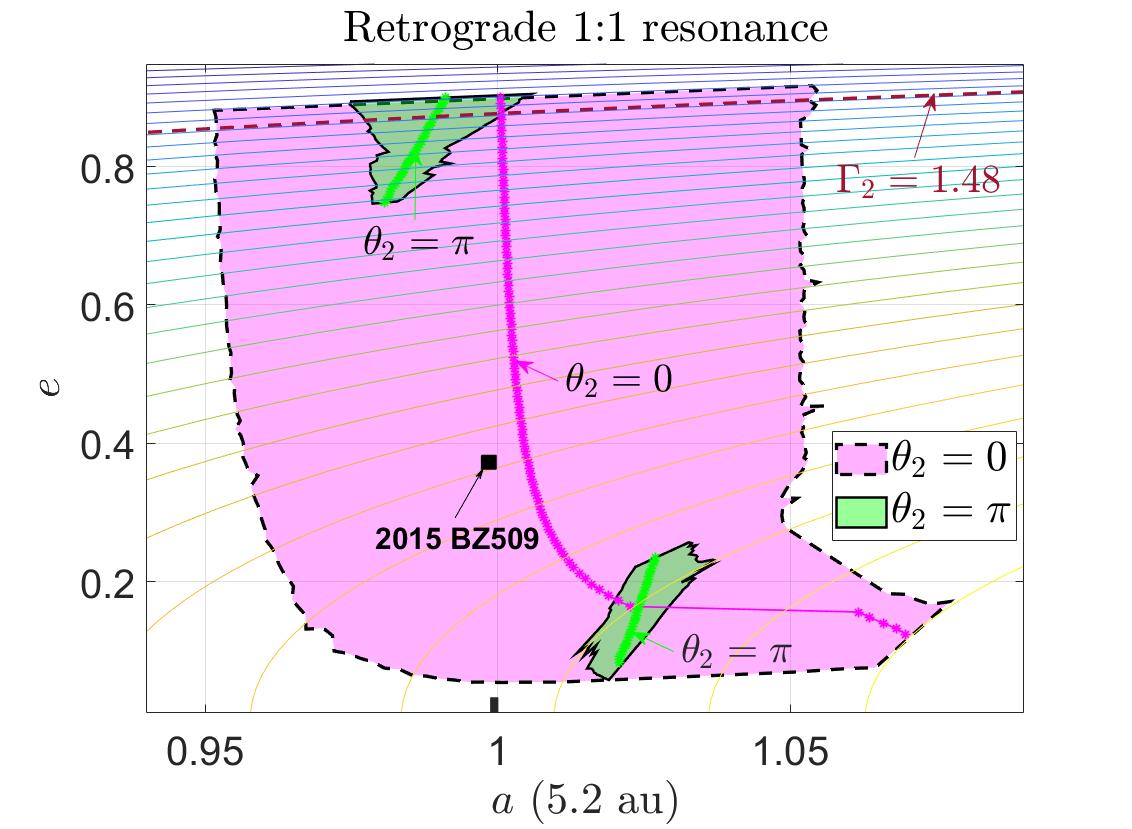}
\caption{Poincar\'e sections of the retrograde 1:1 resonance specified by $\Gamma_2 = 1.48$ (\emph{left panel}) and the numerical widths of resonance in the full range of eccentricity (\emph{right panel}). The location of asteroid 2015 BZ509 is marked by a black square.}
\label{Fig13}
\end{figure*}

For the retrograde co-orbital resonance, the angular variables $\theta_1 = \lambda - \varpi$ and $\theta_2 = \lambda_p - \varpi$ have comparable periods thus, in this case, both conditions illustrated in Fig. \ref{Fig5} can be utilized to produce Poincar\'e sections. In practice, we take the definition of the inner case, i.e., the sections are defined by $\theta_1 =\lambda - \varpi= 0$. As an example, the motion integral is taken at $\Gamma_2 = 1.48$, and the associated Poincar\'e sections are reported in the $(\theta_2, e)$ and $(\theta_2,a)$ spaces, as shown in the \emph{left panel} of Fig. \ref{Fig13}. From the Poincar\'e sections, it is observed that (a) the islands of resonance are centred at $\theta_2=0$ (pericentric libration) and $\theta_2 = \pi$ (apocentric libration), (b) the island centred at $\theta_2 = \pi$ is much smaller than that at $\theta_2= 0$, and (c) chaotic motions can be found in the sections and they fill the space outside the islands.

Similarly, the numerical widths of resonance ($\Delta a$ and $\Delta e$) can be measured from the sections, and they are reported in the \emph{right panel} of Fig. \ref{Fig13}. For convenience, the isoline of $\Gamma_2 = 1.48$ and the nominal resonance location (i.e., $a= a_0$) are marked. It is observed that (a) the islands of resonance centred at $\theta_2 = \pi$ (apocentric libration) occupy two distinct regions (in particular, they disappear in the medium-eccentricity regions), one region with low eccentricities is located on the right-hand side of $a=a_0$ and the other region with high eccentricities occupies on the left-hand side of $a=a_0$, (b) the resonant centres with $\theta_2 = 0$ (pericentric libration) are located on the right-hand side and they diverge away from $a=a_0$ when the eccentricity approaches zero, and (c) the resonant width evaluate at $\theta_2 = 0$ shrinks with eccentricity approaching zero.

As discussed in the introduction, asteroid 2015 BZ509 is the first identified asteroid located inside retrograde co-orbital resonance with Jupiter \citep{wiegert2017retrograde}. For convenience, the location of 2015 BZ509 is marked in the $(a,e)$ space, as shown in Fig. \ref{Fig13}. It is observed that the asteroid is located near the centre of the pericentric libration region.

\begin{figure*}
\centering
\includegraphics[width=0.48\textwidth]{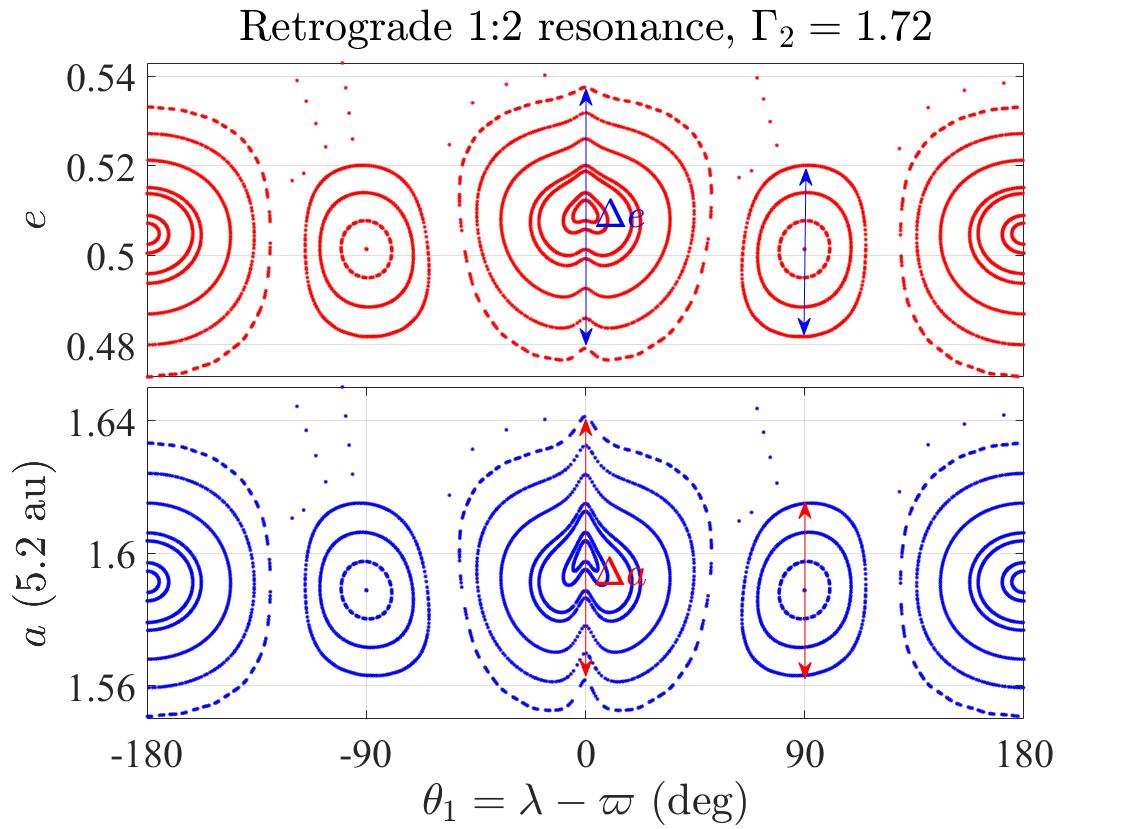}
\includegraphics[width=0.48\textwidth]{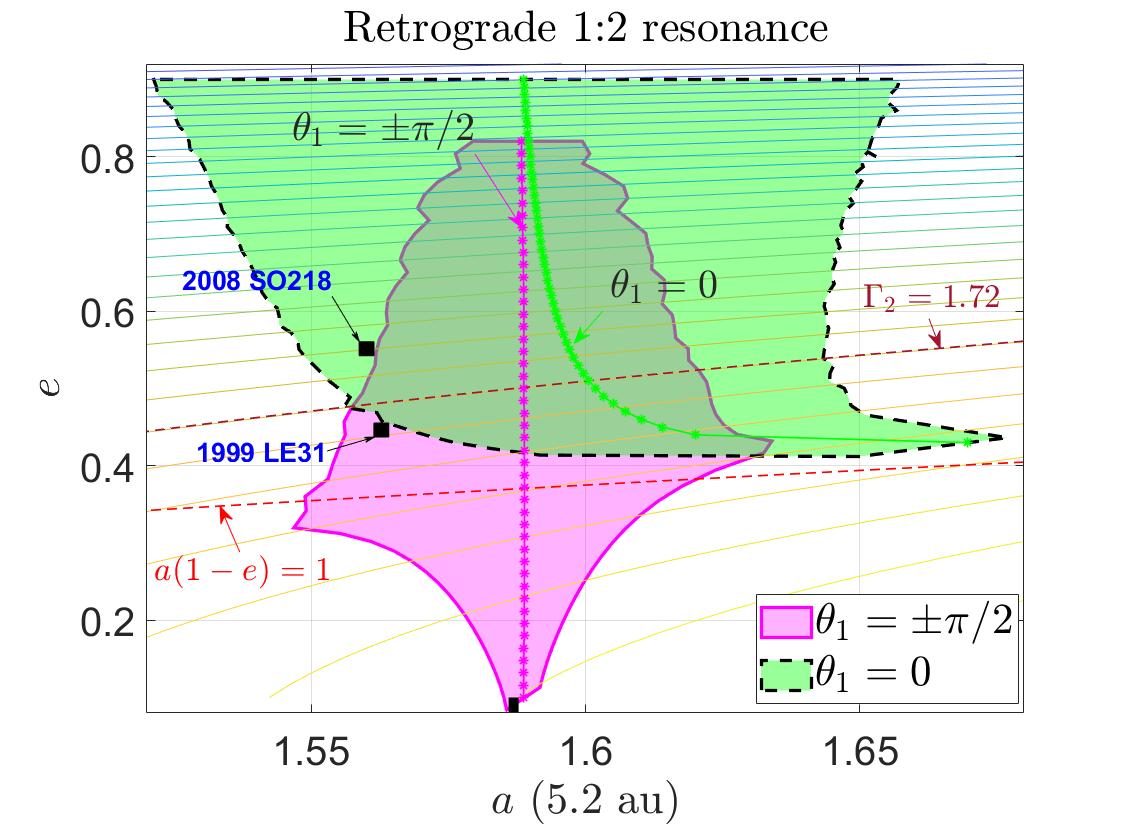}
\caption{Poincar\'e sections of the retrograde 1:2 resonance specified by $\Gamma_2 = 1.72$ (\emph{left panel}) and the numerical widths of resonance in the full range of eccentricity (\emph{right panel}). The locations of asteroids 2008 SO218 and 1999 LE31 are marked by black squares.}
\label{Fig14}
\end{figure*}

Figure \ref{Fig14} presents the Poincar\'e sections and numerical widths for the retrograde 1:2 resonance. For convenience, the nominal resonance location ($a=a_0$) is marked, and the isoline of $\Gamma_2 = 1.72$ and the planet-crossing critical line determined by $a(1-e)=1$ are presented. In the following discussions, we use $e_c (a) = 1-\frac{1}{a}$ to stand for the critical eccentricity of planet-crossing orbit. From the Poincar\'e sections, we can observe that (a) the islands of resonance are located at $\theta_1=0$, $\theta_1=\pm \pi/2$ and $\theta_1=\pm \pi$, (b) the islands centred at $\theta_1=\pm \pi/2$ (or $\theta_1=\pm \pi$) are symmetric with respect to the line of $\theta_1 = 0$ and (c) the separatrices are replaced by chaotic layers. The numerical widths ($\Delta a$ and $\Delta e$) can be measured from the Poincar\'e sections and, in particular, we consider the resonant widths evaluated at $\theta_1=0$ and $\theta_1=\pm \pi/2$, as shown in the \emph{right panel} of Fig. \ref{Fig14}. It is noted that the width evaluated at $\theta_1=\pm \pi$ has a similar behavior to the one evaluated at $\theta_1= 0$, thus it is not considered here. According to the numerical widths of resonance, we can observe that (a) the islands of resonance centred at $\theta_1 = 0$ (pericentric libration) appear in the region above the line of $e_c (a)$, (b) as the eccentricity decreases close to $e_c (a)$, the resonant centres at $\theta_1 = 0$ diverge away from $a=a_0$, (c) the resonant centres at $\theta_1 = \pm \pi/2$ (apocentric libration) stay close to $a=a_0$ but on the right-hand side of it, and (d) the resonant width evaluated at $\theta_1 = \pm \pi/2$ reaches the maximum in the vicinity of $e_c (a)$.

According to \citet{morais2013asteroids} and \citet{li2019survey}, it is known that asteroids 2008 SO218 and 1999 LE31 are currently inside the retrograde 1:2 resonance with Jupiter. For convenience, we mark their locations in the $(a,e)$ space, as shown in Fig. \ref{Fig14}. Evidently, they are located inside the libration regions.

\begin{figure*}
\centering
\includegraphics[width=0.48\textwidth]{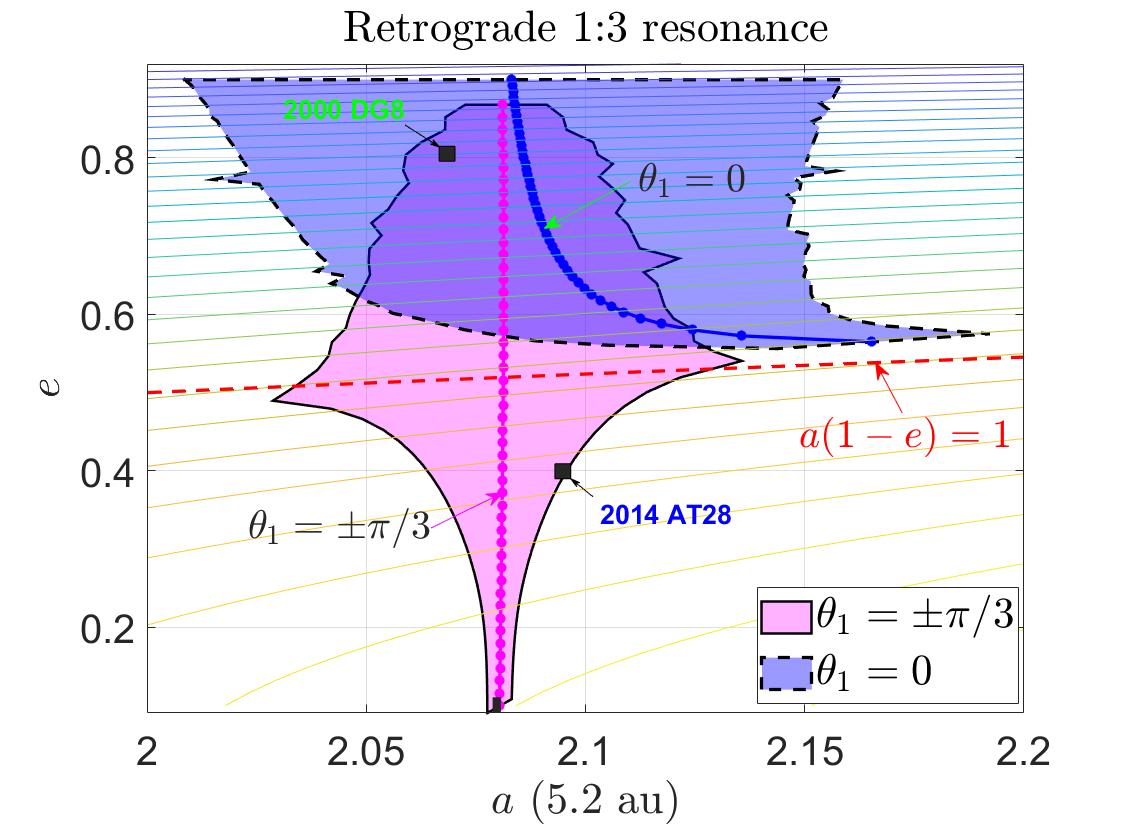}
\includegraphics[width=0.48\textwidth]{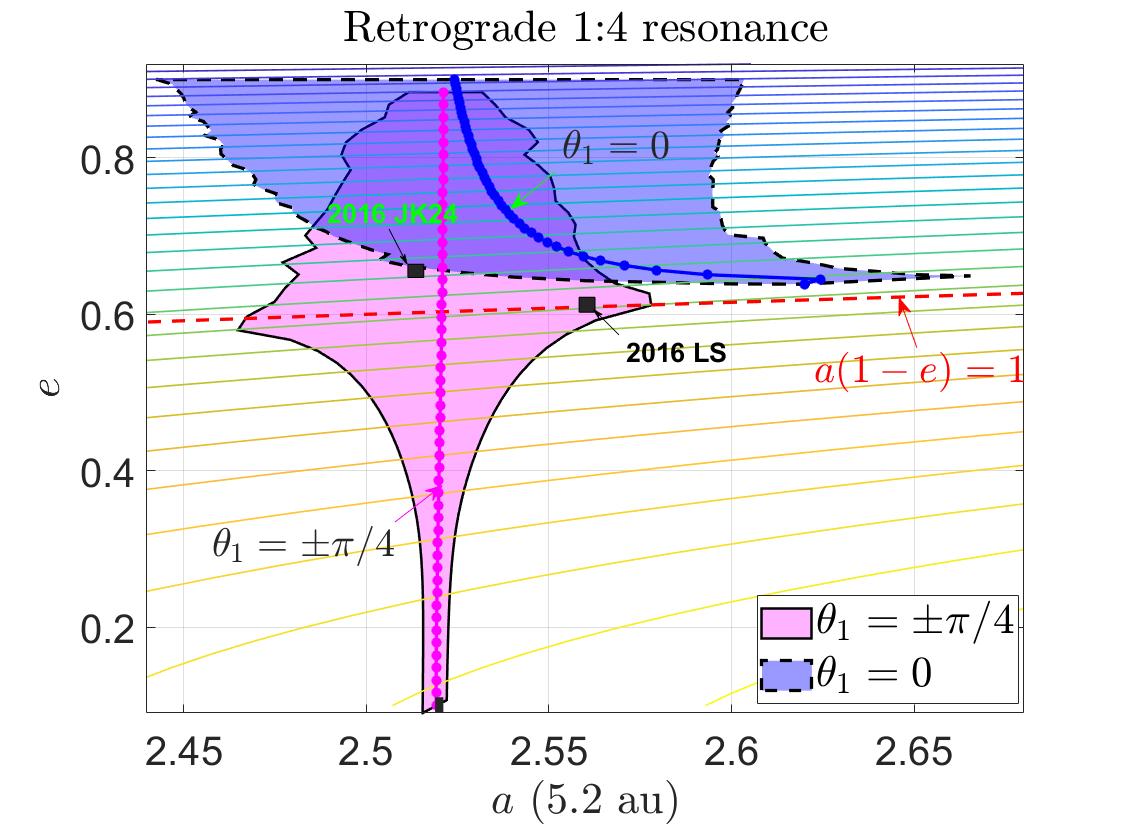}
\caption{The numerical widths of resonance over the full range of eccentricity for the retrograde 1:3 resonance (\emph{left panel}) and for the retrograde 1:4 resonance (\emph{right panel}). The locations of retrograde asteroids 2000 DG8 and 2014 AT28 (in retrograde 1:3 resonance) and 2016 LS and 2016 JK24 (in retrograde 1:4 resonance) are marked by black squares.}
\label{Fig15}
\end{figure*}

In Fig. \ref{Fig15}, the numerical widths of the retrograde 1:3 and 1:4 resonances are reported in the $(a,e)$ space. About these two resonances, only those islands centred at $\theta_1=0$ and $\theta_1=\pm \pi/k$ are taken into consideration. In both panels, the nominal resonant location ($a=a_0$) and the planet-crossing line corresponding to $a(1-e)=1$ are provided (the critical eccentricity is denoted by $e_c (a) = 1 - \frac{1}{a}$). Evidently, the numerical results about the resonant widths for the 1:3 and 1:4 resonances are similar to that of the retrograde 1:2 resonance shown in the \emph{right panel} of Fig. \ref{Fig14}, in terms of the following points: (a) the islands of resonance centred at $\theta_1 = 0$ (pericentric libration) appear in the region above the line of $e_c (a)$, (b) the resonant centres at $\theta_1 = 0$ diverge away from $a=a_0$ when the eccentricity approaches $e_c (a)$, (c) the resonant centres at $\theta_1 = \pm \pi/k$ (apocentric libration) are located close to $a=a_0$, and (d) the resonant width evaluated at $\theta_1 = \pm \pi/k$ reaches the maximum in the vicinity of $e_c (a)$.

According to \citet{li2019survey}, it is known that asteroids 2000 DG8 and 2014 AT28 (or 2016 LS and 2016 JK24) are potentially located inside the retrograde 1:3 (or 1:4) resonance with Jupiter. For convenience, their locations are presented in the $(a,e)$ space, as shown in Fig. \ref{Fig15}. It is observed that asteroids 2000 DG8 and 2016 JK24 sit near the centre of the libration regions, so they will be protected by the resonant configurations for a long enough time. However, asteroid 2016 LS and 2014 AT28 lie close to the boundaries of libration zones, resulting in weak protection effects in the long-term evolution, so the asteroids may exit from the current resonant states. \citet{li2019survey} showed that 2016 LS would be possibly captured in the retrograde 3:5 resonance with Saturn (with possibility of 13.6\%).

In recent years, \citet{huang2018dynamic} discussed the resonant widths by means of semianalytical approach regarding the retrograde 1:1 resonance, and \citet{li2020dynamics} applied the semianalytical method to the dynamics of retrograde 1:2, 1:3, 1:4 and 1:5 resonances. By analyzing the phase portraits, \citet{huang2018dynamic} and \citet{li2020dynamics} measured resonant widths by evaluating the distance of adjacent separatrices that bound libration islands when $e<e_c$ and of the largest boundaries that do not cross the collision curve for $e>e_c$ \citep{morbidelli2002modern}. In comparison to our numerical results produced in this section, some discussions are made as follows.
\begin{itemize}
  \item For the retrograde 1:1 resonance, \citet{huang2018dynamic} noticed that the width evaluated at the usual critical argument $\varphi = 0$ (the pericentric branch) keeps growing with increasing eccentricity (see fig. 2 in their paper). Qualitatively, this behavior of width is in quite good agreement with our numerical results (see the right panel of Fig. \ref{Fig13}). For the apocentric branch, \citet{huang2018dynamic} observed that no apocentric libration with $\varphi = \pi$ exists when the motion integral changes from $-1.94$ to $-1.65$ (corresponding to $\Gamma_2$ used in this work ranging from 1.65 to 1.94). The authors pointed out that the resonant centres in the apocentric branch survives when the eccentricity is small or extremely high. This behavior is reproduced in our numerical results. Please refer to the \emph{right panel} of Fig. \ref{Fig13} for details, which shows that the apocentric libration exists in two distinct regions in the $(a,e)$ space: one with small eccentricities and the other one with high eccentricities (in the medium-eccentricity region the apocentric libration vanishes).
  \item Concerning the retrograde 1:2, 1:3, 1:4 and 1:5 resonances, \citet{li2020dynamics} reported their resonant widths in the $(a,e)$ space and they noticed that all the retrograde 1:$n$ resonances have similar structures (see figs 7 and 10 for details). The authors assumed that the resonance occurs at the nominal resonance location ($a=a_0$), and their results show that the boundaries are symmetric with respect to $a=a_0$. From the numerical viewpoint, we confirm that all the retrograde 1:2, 1:3 and 1:4 resonances considered in this work hold similar structures about resonant widths. However, some discrepancies are observed. Firstly, our numerical results (see Figs \ref{Fig14} and \ref{Fig15}) show that the resonant centres are not located at the nominal resonance locations and, in particular, the libration centres in the pericentric branch ($\theta_1 = 0$) diverge away from the nominal resonance location when the eccentricity is close to $e_c$. Secondly, our results indicate that the boundaries in the $(a,e)$ are not symmetric with respect to the line of $a=a_0$. At last, our results show that the apocentric libration (with $\theta_1 = \pm \pi/k$) vanishes when the eccentricity is higher than a threshold value.
\end{itemize}

\section{Summary and discussion}
\label{Sect8}

In this work, we performed both analytical and numerical studies about the dynamics of retrograde MMRs in the planar circular restricted three-body problem and, in particular, we made direct comparisons between the analytical and numerical results, including the dynamical structures in the phase space, location of resonant centre and resonant width.

Regarding the analytical study, we first presented the explicit expansion of disturbing function for test particles moving on retrograde co-planar orbits and then formulated the Hamiltonian model by (a) introducing a new critical argument as $\sigma = \frac{1}{k_{\max}}\left[k\lambda - k_p\lambda_p + (k_p - k)\varpi\right]$ with $k_{\max} = {\max}\left\{k_p,k\right\}$ and (b) performing a series of canonical transformations. The resulting resonant Hamiltonian determines a single-degree of freedom dynamical model, and the global dynamics of retrograde MMRs can be explored by analyzing phase portraits, where the resonant centre and resonant width can be analytically identified (corresponding to analytical results).

Our numerical investigation is based on the non-perturbative technique by analyzing Poincar\'e sections. We produced Poincar\'e surfaces of section by recording those states of test particles when the `short-period' angular variable is equal to zero under the condition that the motion integral $\Gamma_2 = \sqrt{\mu a} (\frac{k_p}{k} + \sqrt{1-e^2})$ is provided. In particular, the section is defined by $\theta_1 = \lambda - \varpi = 0$ for inner resonances and by $\theta_2 = \lambda_p - \varpi = 0$ for outer resonances. It is found that, on the Poincar\'e sections, the angular separation ($\theta_2$ for the inner resonances or $\theta_1$ for the outer resonances) is equal to the synodic angle between the test particle and the perturber in magnitude. By analyzing the structures arising in the sections, it is possible to numerically determine the location of resonant centre and resonant width (corresponding to numerical results).

According to the definition of sections, there is a relationship between the resonant angle $\sigma$ and the angular separation on the sections: $\sigma = -\theta_2$ for the inner resonances and $\sigma = \theta_1$ for the outer resonances. This relationship makes it possible to compare the phase portraits and Poincar\'e sections for a certain retrograde MMR with a given motion integral $\Gamma_2$. Naturally, we can make a comparison between the analytical and numerical widths of resonance. The main results of the comparative study are summarized as follows.
\begin{itemize}
  \item As for a certain retrograde MMR, there is a perfect correspondence between the phase portraits and Poincar\'e sections with the same motion integral $\Gamma_2$. This correspondence is helpful to understand the structures arising in the Poincar\'e sections and also to validate our analytical developments.
  \item The structures arising in the phase portrait are in good agreement with the ones appearing in the Poincar\'e section, including the islands of libration, dynamical separatrices, resonant centres and saddle points.
  \item For the retrograde MMRs considered in this study, the analytical results (including the resonant width and the location of resonant centre) obtained from the resonant model agree well with the associated numerical results produced by analyzing Poincar\'e sections. It shows that our analytical model formulated in this work is applicable to describe the dynamics of retrograde MMRs in the non-crossing regions.
  \item For the retrograde 2:1 resonance, it is found that the zero-eccentricity point (i.e., the origin in the polar coordinate plane) is no longer a stationary point in both the phase portraits and Poincar\'e sections. This is different from the prograde 2:1 resonance (please refer to \citet{malhotra2020divergence} and \citet{lei2020multiharmonic} for discussions on the prograde 2:1 resonance).
  \item For the outer resonances (i.e., the retrograde 1:2 and 1:3 resonances), both the analytical and numerical results indicate that there are no asymmetric libration centres. This is in agreement with the conclusion given by \citet{li2020dynamics}.
  \item Both the analytical and numerical results show that the width of resonance increases with the eccentricity, the location of resonant centre is dependent on the eccentricity and it deviates from its nominal resonance location (i.e., the Law of Structure exists for retrograde MMRs).
\end{itemize}

The numerical approach based on Poincar\'e sections is not limited to the non-crossing and non-coorbital conditions. Thus, in the last part of this work, we applied the numerical technique to identifying dynamical structures of retrograde MMRs (including the interior, co-orbital and exterior resonances) over the full range of eccentricity. For the retrograde co-orbital resonances, it is found that (i) the apocentric libration occurs in two distinctive regions in the phase space (one with low eccentricities and the other one with high eccentricities) and (ii) the libration centres in the pericentric branch diverge away from the nominal resonance location with eccentricity deceasing close to zero. About the retrograde interior and exterior MMRs, it is found that (a) in the non-crossing regions there is one branch of libration centres while in the planet-crossing regions an additional branch of libration centres appears, (b) chaotic layers replace the role of dynamical separatrices in the planet-crossing regions, making that the boundaries are no longer smooth curves, and (c) the libration centres located in the planet-crossing regions diverge away from the nominal resonance location when the eccentricity decreases close to the critical value $e_c (a)$. In particular, the retrograde asteroids 2015 BZ509 (in retrograde coorbital resonance), 2008 SO218 and 1999 LE31 (in retrograde 1:2 resonance), 2000 DG8 and 2014 AT28 (in retrograde 1:3 resonance), and 2016 LS and 2016 JK24 (in retrograde 1:4 resonance) are located inside the libration regions (please refer to Figs \ref{Fig13}--\ref{Fig15} for details), indicating that these seven asteroids are inside the retrograde resonances with Jupiter.

\begin{figure}
\centering
\includegraphics[width=0.48\textwidth]{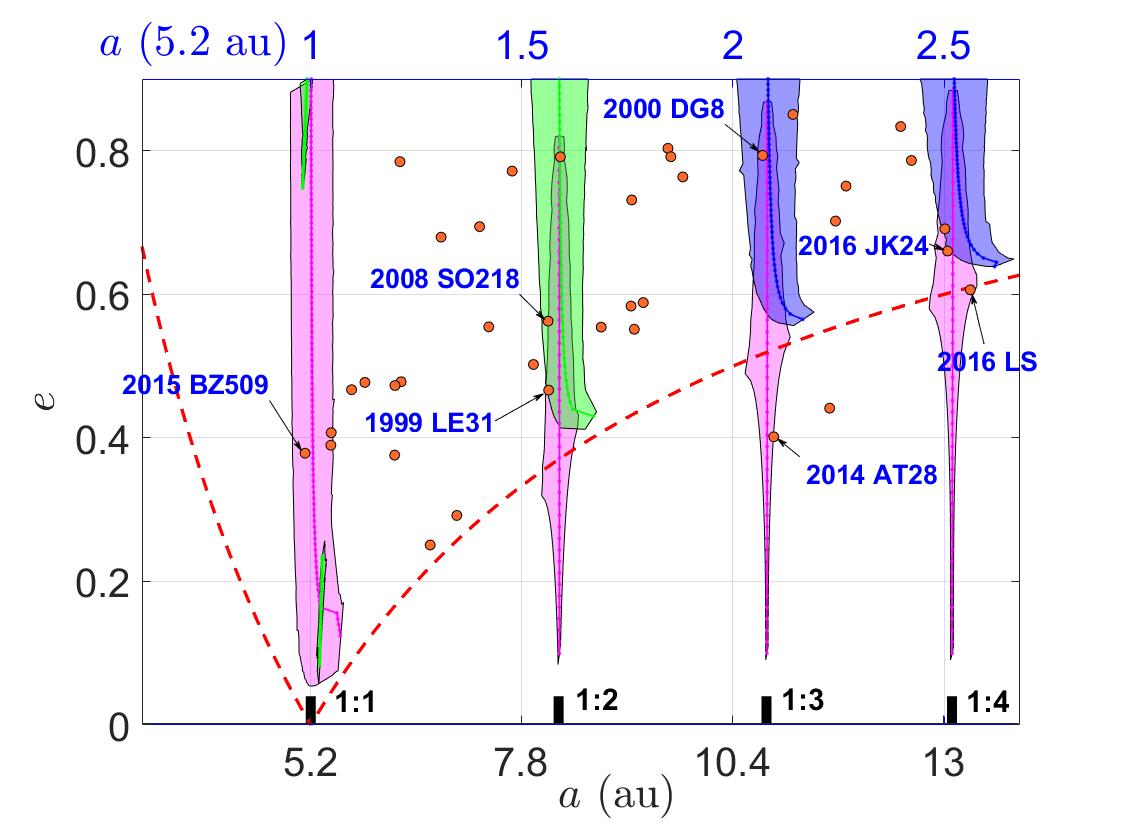}
\caption{The numerical widths of retrograde 1:1, 1:2, 1:3 and 1:4 resonances with Jupiter, together with all the known retrograde asteroids with semimajor axes $a \in (3.12\; \mathrm{au},14.04\; \mathrm{au})$. For convenience, seven potential examples of asteroids inside retrograde MMRs, which are confirmed in previous works, are indicated.}
\label{Fig16}
\end{figure}

Figure \ref{Fig16} reports the numerical widths of resonance together with the retrograde asteroids\footnote{https://minorplanetcenter.net//iau/MPCORB.html, retrieved 3 February 2020} with semimajor axes $a \in (3.12\; \mathrm{au},14.04\; \mathrm{au})$. Evidently, besides the seven known examples of asteroids potentially inside retrograde MMRs with Jupiter, there are additional asteroids which are located inside the libration zones (for example, it is observed that asteroid 2018 TL6 is located at the centre of the libration zone of the retrograde 1:2 resonance). The asteroids inside resonance zones could be considered as new potential candidates to be trapped inside retrograde MMRs. However, we need to notice two points: (a) the PCRTBP adopted in this work is a greatly simplified model for the Solar system, so that the results obtained in such a simplified model could only provide preliminary boundaries for retrograde MMRs, and (b) the inclinations of practical retrograde asteroids are not equal to $180^{\circ}$. Thus, whether these asteroids are really trapped inside resonances or not, it requires numerical integrations to confirm them in the Solar system model. We leave the confirmation to a future study.

\section*{Acknowledgments}
This work is supported by the National Natural Science Foundation of China (Nos 12073011, 11973027, 41774038). This research has made use of data and/or services provided by the International Astronomical Union's Minor Planet Centre.

\section*{Data availability}
The analysis and codes are available upon request.

\bibliographystyle{mn2e}
\bibliography{mybib}

\appendix
\section{Resonant disturbing function}
\label{A1}
For the $k_p$:$k$ resonance, the critical argument is given by $\sigma = \frac{1}{k_{\max}} \left[k\lambda - k_p \lambda_p + (k_p - k)\varpi\right]$. The resonant disturbing function can be written as
\begin{equation}\label{Eq_A1}
\begin{aligned}
{{\cal R}^ * } &= {\cal G}{m_p}\sum\limits_{n = 0}^N {\sum\limits_{\scriptstyle j =  - \infty \hfill\atop
\scriptstyle\bmod (j,{k_p}) = 0\hfill}^\infty  {\sum\limits_{m = 0}^n {{{\left( { - 1} \right)}^{n - m}}{A_{n,j}}\left( \alpha  \right) {n \choose m}}}}\\
&\times X_{ - jk/{k_p}}^{m,j}\left( e \right)\cos \left( {\frac{j}{{{k_p}}}{k_{\max }}\sigma } \right)\;\\
&- {\cal G}{m_p}aX_{ - k}^{1,1}\left( e \right){\delta _{{k_p},1}}\cos \left( {{k_{\max }}\sigma } \right)
\end{aligned}
\end{equation}
where ${\delta _{k_p,1}} =1$ when $k_p = 1$ and ${\delta _{k_p,1}} =0$ when $k_p \ne 1$. For convenience, the resonant disturbing function is denoted by
\begin{equation}\label{Eq_A2}
{{\cal R}^*} = \sum\limits_{q = 0}^Q {{{\cal C}_q}\cos \left( {q{k_{\max }}\sigma } \right)},
\end{equation}
where the coefficients ${\cal C}_q$ are related to the semimajor axis and eccentricity. Based on equation (\ref{Eq_A1}), it is not difficult to obtain
\begin{equation*}
\begin{aligned}
{{\cal C}_q}\left( {a,e} \right) &= {\cal G}{m_p}\sum\limits_{n = 0}^N {\sum\limits_{m = 0}^n {{{\left( { - 1} \right)}^{n - m}}{A_{n,q{k_p}}}\left( \alpha  \right) {n \choose m} X_{ - qk}^{m,q{k_p}}\left( e \right)}} \\
&- {\cal G}{m_p}a X_{-k}^{1,1}\left( e \right){\delta _{k_p,1}} {\delta _{q,1}}.
\end{aligned}
\end{equation*}

\bsp
\label{lastpage}
\end{document}